\documentclass[%
 pre, amsmath,amssymb,showkeys,
 aps,superscriptaddress,onecolumn,a4paper,english]{revtex4}
 \usepackage[T1]{fontenc}
 \usepackage[utf8]{inputenc}
 \usepackage{amsmath}
 \usepackage{amssymb}
 \usepackage{graphicx}
 \usepackage{bm}
 \usepackage{multirow,array}
 \usepackage{mathtools}
 \usepackage{empheq}
 \usepackage{chessfss}
 \usepackage{epsdice}
 \usepackage{pifont}
 \usepackage{amsthm}
 \usepackage[
 citecolor=blue,
 colorlinks,
 linkcolor=blue,
 urlcolor=blue,
 ]{hyperref}
 \usepackage[dvipsnames]{xcolor}
 \usepackage{dcolumn}
 \usepackage{bm}
 \usepackage{etoolbox}  
 \usepackage{soul} 
 \usepackage{cancel}
\begin{document}

	\title{\textcolor{black}{Evolutionarily stable strategy in asymmetric games: Dynamical and information-theoretical perspectives}}

\author{Vikash Kumar Dubey}
\email{vdubey@iitk.ac.in}
\affiliation{Department of Physics, Indian Institute of Technology Kanpur, Uttar Pradesh 208016, India}

\author{Suman Chakraborty}
\email{ph25r001@smail.iitm.ac.in}
\affiliation{\textcolor{black}{Department of Physics, Indian Institute of Technology Madras, Tamil Nadu 600036, India}}

\author{Arunava Patra}
\email{arunava20@iitk.ac.in}
\affiliation{Department of Physics, Indian Institute of Technology Kanpur, Uttar Pradesh 208016, India}

\author{Sagar Chakraborty}
\email{sagarc@iitk.ac.in}
\affiliation{Department of Physics, Indian Institute of Technology Kanpur, Uttar Pradesh 208016, India}

	%

\begin{abstract}
Various extensions of evolutionarily stable strategy (ESS)---the central concept in evolutionary game theory---defined for asymmetric games differ in how they correspond to fixed points of replicator equation which models evolutionary dynamics of frequencies of strategies in a population. Along with reporting interesting new results, this paper is partially \textcolor{black}{intended} as a contextual mini-review of some of the most important definitions of ESS in asymmetric games. We present the definitions coherently and scrutinize them closely while establishing equivalences---some of them hitherto unreported---between them. Since it is desirable that a definition of ESS should correspond to asymptotically stable fixed points of the replicator dynamics, we bring forward the connections between various definitions and their dynamical stabilities. Furthermore, in this context, we use the principle of relative entropy minimization to gain information-theoretic insights into the concept of ESS, thereby establishing a three-fold connection between game theory, dynamical system theory, and information theory.
\end{abstract}
	\maketitle
	

\section{Introduction}

Evolutionarily stable strategy (ESS) is the most fundamental concept of evolutionary game theory~\cite{SMITH1973,Maynard_Smith1982,Thomas1985}, and its exploration has yielded a substantial body of scientific literature~\cite{Maynard_Smith1982,Nowak_2006_book,hofbauer_book,Cressman2003,Weibull1997,Haigh1975,Smith1977,Hines1987,Broom2014,Vickers1987,Eshel2005,Hammerstein1994,Smith1977,Pohley1983,Szidarovszky2008,Broom2016,Traulsen2023}.  In fact, one may remark~\cite{dawkins1989} that this concept is only second to the Darwin's contribution in advancing the evolutionary theory. No wonder that the paper~\cite{MaynardSmith1974} putting forward this concept has been included as one of eighty-eight foundational papers in complexity science~\cite{Krakauer_FP_3_2024} of the 20th century. In a large (mathematically, infinite) population with all individuals adopting a strategy (equivalently, a phenotype) that is evolutionarily stable, a small fraction of alternative strategy (called mutant strategy) cannot invade the population. Essentially, the concept can be understood as follows: ESS either renders a higher expected payoff through its performance against itself or, in case the strategy ties with that of the mutant's, it fetches more expected payoff against the mutant compared to what the mutant would.

The ESS was originally introduced for a population where all the individuals are identical (or have the same role), from the strategic interaction point of view; in such a population, one says that symmetric games are played by the individuals. However, the concept of ESS in asymmetric games can be quite tricky. Asymmetric games refer to games that involve individuals categorized into more than one class---each with a distinct role---and the strategic interaction involves interaction between players in different roles. To complicate the matter further, in these games, players belonging to different roles may also have different strategy sets. If there is any interaction among players of the same role, the interaction is termed intra-specific interaction; whereas the interaction between players of different roles is aptly termed inter-specific. In this context, {in this paper we follow the terminology that an asymmetric game is called a bimatrix game (sometimes also called truly asymmetric game) if only inter-specific interaction is present}. The purpose of this paper is to critically discuss various extant definitions of ESS in bimatrix and asymmetric games. Note that \emph{henceforth, the phrase `asymmetric game' will not include `bimatrix game' which will be used as a separate phrase}. Moreover, unless otherwise specified, we shall work assuming only two roles present in a population for simplicity and without much loss of generality.

{\color{black}The literature on bimatrix and asymmetric games has evolved over the years through several complementary directions that have collectively shaped our understanding of equilibrium structure, stability, and computability. Early studies established the existence of equilibrium points in bimatrix games and showed that, under general conditions, their number is finite~\cite{Lemke1964,Haigh1975,Vickers1988,Jansen1981}. These insights helped bridge static game-theoretic ideas with their dynamic counterparts. In asymmetric settings, symmetrization methods were introduced to connect bimatrix games with symmetric games~\cite{Gaunersdorfer1991,hofbauer_book,Sigmund2001,Cressman2003}, allowing evolutionary dynamics to be analyzed within a unified framework. In such models Nash equilibria (NE) was connected to dynamically attracting outcomes, linking game-theoretic stability to evolutionary dynamical stability~\cite{Samuelson1992,Cabrales1992,Bjornerstedt994,Cressman_Tao_2014,hofbauer_book}. More recent analytical developments have extended the concept of zero-determinant strategies~\cite{Press2012,Hilbe2013} to asymmetric games, showing that one player can unilaterally enforce a fixed relation between their own payoff and that of the opponent~\cite{Taha2020,Kang2024}.
 Alongside these conceptual developments, researchers have also explored the computational complexity of the problem—estimating the number of NEs~\cite{Savani2006,Conitzer2008} and analyzing the efficiency of algorithms used to find them~\cite{vonStengel1999,Kontogiannis2009}. Despite major progress, computing equilibria in general bimatrix games remains challenging, prompting interest in approximate and efficient numerical methods.

Applications of bimatrix and asymmetric games span a wide range of domains where interactions between distinct populations or roles are explicitly modeled. Recent studies have analyzed signaling games~\cite{skyrms2010,Banks1987,Huttegger2010} using bimatrix replicator dynamics exhibit convergence toward partial communication rather than perfect communication~\cite{Hofbauer2008,Huttegger2009,Huttegger2010}. Asymmetric game formulations have also been used to describe ecological and genotypic heterogeneity, highlighting differences between genetic and cultural evolution~\cite{McAvoy2015,Hauert2019}. Network-based extensions in asymmetric games~\cite{Du2024,Correia2022} show network topology can drive homogeneous behaviour within communities but heterogeneous outcomes across them~\cite{Broere2017,Zimmaro2024,Boodaghians2020}. Decomposition methods allow complex asymmetric games to be reduced to two symmetric counterparts, simplifying evolutionary analysis~\cite{Tuyls2018}. Other applications include replicator models of competition in electricity markets~\cite{Yin2020,Wang2023}, wireless communication~\cite{Firouzbakht2016,Garnaev2019} and auction and bidding mechanisms~\cite{Maskin2000a,Maskin2000b,Bichler2023,Kaplan2010,Bichler2023} etc. Studies of random bimatrix games~\cite{Fearnley2016} demonstrate that uniform randomization across strategies yields, with high probability, a strategy profile very close to NE~\cite{Panagopoulou2013,Panagopoulou2018}. 
In eco-evolutionary context, asymmetric interactions naturally arise in heterogeneous environments, where ecological feedbacks between strategies and local conditions can generate various complex outcomes~\cite{Hauert2019,Qin2020,Farahbakhsh2022,Shao2019,Hu2019}. These eco-evolutionary feedback models have also demonstrated how environment-dependent payoff switching can maintain oscillatory coexistence in social dilemmas~\cite{Shu2022,Betz2024}. In uncertain environments, fuzzy bimatrix games have been developed to handle ambiguous payoffs, with applications such as disaster management~\cite{Karmakar2022,Brikaa2025}. Other asymmetric formulations have been applied to volunteer dilemmas~\cite{Kawagoe2023}, studies of learning mechanisms~\cite{Lensberg2021,Bichler2023}, evolution of fairness modeled via ultimatum games~\cite{Nowak2000,Sekiguchi2015}.
Together, these developments illustrate the versatility of bimatrix and asymmetric games in capturing diverse real-world interactions across biology, social, ecological, economic and engineered systems.
	
}

 There have been multiple attempts to extend the concept of ESS to asymmetric and bimatrix games. Taylor~\cite{Taylor_1979} defined asymmetric ESS as the resilient strategy that fetches more total average payoff---summation of average payoffs of every subpopulation of individuals with distinct role---than what a mutant strategy would do. Around the same time, Selten~\cite{Selten1980} and Schuster et al.~\cite{Schuster_1981} proposed that ESS for bimatrix games could be defined either in terms of total average payoff or more strictly,
 	by requiring that the average payoffs in every role exceed those of potential mutants. Interestingly, unlike in the case of symmetric games, the ESS in bimatrix games cannot be a mixed
 	strategy~\cite{Pohley1979,Schuster_1981,Schuster_1981_3}. Consequently, alternative definitions of ESS in asymmetric and bimatrix games were proposed to allow for the mixed ESS concept. For instance, Cressman~\cite{cressman_1992,Cressman1996,Cressman_Tao_2014} weakens the aforementioned ESS condition by requiring that the average payoff in at least one role must exceed what a
 	corresponding mutant could achieve. This concept is termed two-species ESS---henceforth abbreviated as 2ESS. For bimatrix games, Hofbauer and Sigmund~\cite{hofbauer_book} provided an alternative definition of ESS to allow for the emergence of mixed ESS. The concept is termed Nash--Pareto pair.
The search for mixed ESS in asymmetric and bimatrix games is motivated by the fact that the replicator equation---an evolutionary dynamics model aligned with Darwinian principles---can exhibit asymptotically stable interior fixed points~\cite{taylorjonker1978, Cressman_Tao_2014}. It is naturally desirable that the game-theoretic concept of ESS correspond to such fixed points.

{\color{black} In parallel with these game-theoretic and dynamical developments, information theory has emerged as a powerful framework for understanding evolutionary processes. In particular, the Kullback–Leibler (KL) divergence (also called relative entropy)~\cite{kullback1951,Cover2005} plays a central role: It has been established that the KL divergence acts as a Lyapunov function for replicator dynamics in symmetric games~\cite{baez_entropy,harper2009information}, thereby ensuring that the evolutionary flow moves toward an ESS whenever the KL divergence decreases. This principle of KL-divergence reduction has also been exploited to extend the notion of ESS to non–fixed-point outcomes~\cite{Bhatacharjee_etal_2023,Dubey2024}. Beyond KL divergence, Fisher information has also been used to describe evolutionary processes. The Fisher information metric coincides with the Shahshahani metric~\cite{harper2009information}, under which the replicator dynamics of two-player partnership games can be written as the gradient flow of a potential~\cite{hofbauer_book,Weibull1997}. A variety of other works demonstrate the breadth of these information-theoretic connections: Geometrical formulations of population genetics~\cite{Akin1979}, information-theoretic models of communication and coordinated behavior~\cite{Berry2009}, interpretations of natural selection through Fisher information~\cite{FRANK2009,Frank2012}, and the role of information theory in understanding genome structure and evolvability~\cite{Wagner2017}. Furthermore, information theory has been used to connect evolutionary game dynamics with ideas from statistical physics~\cite{Wolpert2006,Kwon2021,Demetrius2022}.
These developments collectively highlight that information measures such as KL divergence and Fisher information offer powerful tools for characterizing evolutionary stability which makes information-theoretic understanding of ESS in asymmetric games an important development in evolutionary game theory.
	
Motivated by all the aforesaid observations, the present work presents a dynamical and information-theoretical framework for ESS in asymmetric (and bimatrix) games. Owing to the length and scope of the paper, let us first summarize below the main contributions and the structure of the paper. }
{\color{black}\subsection{Outline of the paper}
	
This paper has a dual nature: It serves both as a mini-review and as a research article. On one hand, it consolidates and systematizes the existing body of knowledge on evolutionary stability in bimatrix and asymmetric games, providing a comprehensive and coherent framework that integrates various definitions of ESS and shows their logical relationships. On the other hand, it advances several new theoretical results that extend the concept of evolutionary stability beyond existing frameworks—both introducing novel generalizations grounded in information theory and multi-player (hypermatrix) game settings.	
	
In the review component of this work, we systematically collect, compare, and clarify various existing definitions of ESS in bimatrix and asymmetric games. We identify three principal classes of definitions—namely, \emph{and}-, \emph{sum}-, and \emph{or}- definitions—each reflecting a distinct structure in how mutant payoffs are compared across the two interacting populations to determine whether a mutant can invade. Within these, further refinements are possible: for instance, the and-definition admits two forms, which we denoted as the \emph{primed} and \emph{unprimed} versions. We also elaborate on the two interpretations of the or-definition in bimatrix games—one is \emph{simultaneous deviation} and the other is \emph{face-value} perspective—and explain why the former formulation is essential for the existence of mixed ESS in such games. Our contribution in this review component lies precisely in providing this classification and logical organization of ESS definitions across asymmetric and bimatrix games.  Furthermore, we establish the logical relationships among these definitions and clarify their dynamical implications. These are summarized in the schematic Fig.~\ref{fig:ESS_summary}. {In particular, we discuss that strict NEs are always evolutionarily stable in both asymmetric and bimatrix games, while non-strict NE can qualify as ESS only under the unprimed and-definition and simultaneous deviation or-definition, but not under the sum- or face-value or-definition.} Importantly, we address the dynamical interpretation of mixed ESS in bimatrix games, arguing that although such equilibria are not asymptotically stable under standard replicator dynamics, they can still be stable under other evolutionary dynamics, such as the adjusted replicator dynamics.

Finally, we clarify that in asymmetric games the logical hierarchy among the definitions becomes less convoluted, following the implication chain: {and-definition$\implies$sum-definition$\implies$or-definition (see Fig.~\ref{fig:ESS_summary}).}
We also show that, unlike in bimatrix games, mixed ESS can exist in asymmetric games under both the and- and sum-definitions; however, certain mixed equilibria that are asymptotically stable are missed by the stricter and- and sum-definitions but are captured by the or-definition successfully.

In the other component of the paper, we present our new theoretical contributions that extend and unify the concept of evolutionary stability across different levels of generality. We first develop an information-theoretic interpretation of all three definitions of ESS in asymmetric games by expressing each as a particular adaptation of the Kullback–Leibler (KL) divergence~\cite{kullback1951,Cover2005}, thereby revealing a common entropic foundation underlying the stability conditions. Furthermore, we identify a link between the Fisher information~\cite{fisher, harper2009information,Frank2012} and the Shahshahani metric~\cite{shahshahani, hofbauer_book, harper2009information} in asymmetric games, and prove that the replicator dynamics can be interpreted as the Shahshahani gradient of an underlying potential function, thereby establishing an information-geometric connection with ESS.

We subsequently move beyond two-player interactions and extend the ESS framework to multiplayer asymmetric (hypermatrix) games, deriving generalized stability conditions and analyzing their dynamical implications. We prove that in bihypermatrix games, only pure ESS can exist under the and- and sum-definitions. We further show that strict NE satisfy the ESS condition under all three definitions.
This result unifies the concept of evolutionary stability across the distinct formulations of ESS, confirming that strict NE must be stable irrespective of the chosen definition. We then show that mixed ESS, when they exist in such bihypermatrix systems, correspond only to centers or saddles under the replicator dynamics. Nevertheless, we conjecture—and demonstrate through a three-player example—that the or-definition continues to hold for mixed ESS in asymmetric hypermatrix games. {Finally, we generalize the information-theoretic formulation to multiplayer settings and show that the potential function associated with the Shahshahani gradient generates the replicator dynamics in a generalization of partnership games~\cite{hofbauer_book,Weibull1997} in symmetric hypermatrix case but not so in asymmetric hypermatrix games.}

\begin{figure}[h!]
	\centering
	\includegraphics[scale=0.8]{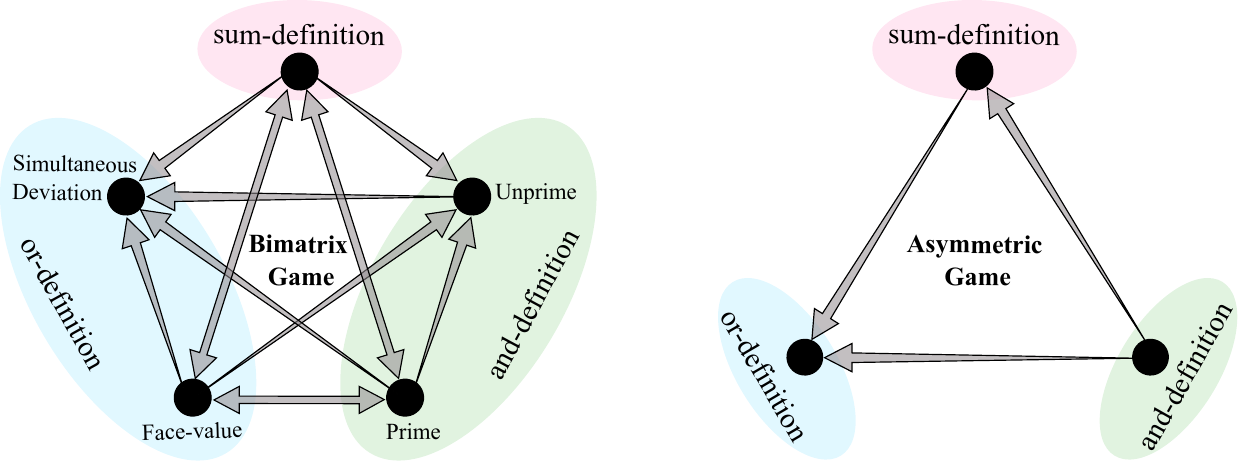}
	\caption{\color{black} Schematic summary of logical relations among ESS definitions in bimatrix and asymmetric games as discussed in this paper. Here single-headed arrows indicate implication, i.e., if the arrow points from A to B, then it means that A implies B. Similarly, the double-headed arrows indicate equivalence between definitions, i.e., if there is a double-headed arrow between A and B, then it means A is equivalent to B.}
	\label{fig:ESS_summary}
\end{figure}
}
{\color{black}In summary,} the outline of the paper is as follows: We commence with Sec.~\ref{sectionII} by outlining the existing definitions of ESS in bimatrix games. Rather than following a chronological description, we adopt a logical and coherent narrative suited to the goal of this paper. We demonstrate the equivalences between some of the definitions. Next, in Sec.~\ref{sectionII}, we highlight caveats and issues associated with these definitions in bimatrix games. We scrutinize the concept of ESS further from the angle of dynamical stability under replicator dynamics.  Sec.~\ref{sec:asymmetric} presents the implications on bimatrix games due to intra-specific interactions which essentially turns them into general asymmetric games. In Sec.~\ref{SectionVII}, we discuss some interesting information-theoretic connections of ESS. In the next section, Sec.~\ref{SectionVI}, we generalize hitherto obtained results for multiplayer hypermatrix games.  Finally, we conclude in Sec.~\ref{sec:DnC}.


\section{Equivalence between various definitions of ESS in bimatrix Games}\label{sectionII}
We begin our discussion of ESS within a population game where multiple phenotypes are present. Accordingly, it is natural to examine the frequency distribution of these phenotypes and to identify conditions under which a population state is evolutionarily stable. In other words, we seek criteria that define evolutionarily stable (population)-states. Importantly, there is a direct correspondence between an evolutionarily stable state and the mean population strategy being evolutionarily stable, that is, if a particular population state is evolutionarily stable, then the mean strategy of that population is also an evolutionarily stable strategy~\cite{Cressman1990,cressman_1992}. We explore this connection in detail in Sections~\ref{sec:strong_stability} and \ref{sec:mixed_bim} after introducing the various definitions of ESS. Henceforth, unless otherwise specified, the abbreviation ESS will refer exclusively to an evolutionarily stable \emph{state (ESS)}.

Consider a population consisting of two subpopulations denoted as X and Y. Let $\mathbf{x}=(x_1,x_2,\cdots,x_m)$ represent a state of subpopulation X, where each $x_i$ represents the frequency of the $i${th} phenotype in the subpopulation. Similarly, let $\mathbf{y}=(y_1,y_2,\cdots,y_n)$ represent a state of subpopulation Y, where each $y_j$ represents the frequency of the $j${th} type in the subpopulation. The state $(\mathbf{x},\mathbf{y})$ of the entire population is an element of simplotope, $\Sigma_m \times \Sigma_n$ [$\Sigma_m\equiv\{(x_1,\cdots,x_m)\in\mathbb{R}^m:\Sigma_{i=1}^m x_i=1~\text{and}~x_i\geq0\}$], which is a Cartesian product of two simplices of dimensions $m-1$ and $n-1$. In this context, $b_{ij}$ and $c_{ij}$, respectively, represent the payoff of type $i$ in X against type $j$ in Y and the payoff of type $i$ in Y against type $j$ in X. All such payoffs are collectively expressed through the $m\times n$ matrix ${\sf B}$ with elements $b_{ij}$'s and and $n\times m$ matrix ${\sf C}$ with elements $c_{ij}$'s.

Let $(\mathbf{x},\mathbf{y})$ denote any arbitrary state of the population. Suppose that $(\mathbf{\hat{x}},\mathbf{\hat{y}})$ represents an evolutionarily stable state (ESS). Following the idea of ESS~\cite{Maynard_Smith1982}, a state $(\mathbf{\hat{x}},\mathbf{\hat{y}})$ qualifies as an ESS if in the case of an emergence of a mutant state, say, $(\bf{x},\bf{y})$, on average an individual of host population is comparatively fitter than a mutant. Since practically mutations are rather rare, it makes sense for this criterion to hold true for infinitesimal mutant fraction in the context of infinitely large host population. Because the population under consideration has more than one subpopulation, unlike what happens in the standard case of symmetric games, the idea of ESS can be expressed in various precise mathematical definitions. 

\subsection{Mutants in both subpopulations}\label{sec:bimatrix_mbs}
Let us begin with following set of three definitions (aptly termed \emph{and}-definition) which can be shown to be equivalent:
\begin{enumerate}
\item[]\textbf{Definition 1a}: A state $(\mathbf{\hat{x}},\mathbf{\hat{y}})$ is called ESS if for any mutant state $(\mathbf{{x}},\mathbf{{y}}) \in\Sigma_m \times \Sigma_n$, such that $\mathbf{{x}}\neq\mathbf{\hat{x}}$ and $\mathbf{{y}}\neq\mathbf{\hat{y}}$, there exists an invasion barrier, $\epsilon_{(\mathbf{x},\mathbf{y})} >0$ such that, $\forall(\epsilon_x,\epsilon_y)$,  with $0 < \epsilon_x <\epsilon_{(\mathbf{x},\mathbf{y})} $ and $0 < \epsilon_y <\epsilon_{(\mathbf{x},\mathbf{y})}$,
	\begin{equation}\label{eq:ess1_fp}
		 \mathbf{\hat{x}}\cdot{\sf B}(\epsilon_y \mathbf{y}+(1-\epsilon_y) \mathbf{\hat{y}}) > \mathbf{x}\cdot {\sf B}(\epsilon_y \mathbf{y}+(1-\epsilon_y) \mathbf{\hat{y}})~\text{\emph{and}}~~
	\mathbf{\hat{y}}\cdot {\sf C}(\epsilon_x \mathbf{x}+(1-\epsilon_x) \mathbf{\hat{x}}) >\mathbf{y}\cdot {\sf C}(\epsilon_x \mathbf{x}+(1-\epsilon_x) \mathbf{\hat{x}}).
	\end{equation}
Here, $\epsilon_x$ and $\epsilon_y$ are the sufficiently small fractions of mutants inside the subpopulations X and Y respectively. Note that the constraints $\mathbf{x} \neq \mathbf{\hat{x}}$ and $\mathbf{y} \neq \mathbf{\hat{y}}$ on the mutant state are necessary---otherwise, no ESS can exist under Definition 1a. This follows trivially, as allowing all possible mutants would also include those of the form $(\mathbf{\hat{x}},\mathbf{y})$, which would always invalidate the first inequality in~(\ref{eq:ess1_fp}). This restriction on mutant states can also be imposed in a slightly different manner, leading to an alternative and-definition, as discussed in Appendix~\ref{sec:alternate_defs_and}.
\item[] \textbf{Definition 1b}: A state $(\mathbf{\hat{x}},\mathbf{\hat{y}})$ is an ESS if, for any mutant  $(\mathbf{{x}},\mathbf{{y}})$, such that $\mathbf{{x}}\neq\mathbf{\hat{x}}$ and $\mathbf{{y}}\neq\mathbf{\hat{y}}$, the following conditions hold,
\begin{equation}\label{eqn:equlibrium_1b}
\text{(i)}~\mathbf{\hat{x}}\cdot{\sf B}\mathbf{\hat{y}}\geq \mathbf{x}\cdot{\sf B}\mathbf{\hat{y}}~\text{and}~~\mathbf{\hat{y}}\cdot{\sf C}\mathbf{\hat{x}}\geq\mathbf{y}\cdot{\sf C}\mathbf{\hat{x}};
\end{equation}
(ii) and if $\mathbf{\hat{x}}\cdot{\sf B}\mathbf{\hat{y}}= \mathbf{x}\cdot{\sf B}\mathbf{\hat{y}}$, then,
\begin{equation}\label{eqn:and_definiton1b2}
\mathbf{\hat{x}}\cdot\sf{B}\mathbf{y}> \mathbf{x}\cdot{\sf B}\mathbf{y};
\end{equation}
(iii) and if $\mathbf{\hat{y}}\cdot{\sf C}\mathbf{\hat{x}}=\mathbf{y}\cdot{\sf C}\mathbf{\hat{x}}$, then,
\begin{equation}
	\mathbf{\hat{y}}\cdot{\sf C}\mathbf{x}>\mathbf{y}\cdot{\sf C}\mathbf{x}.
\end{equation}

This reminds one of what Hofbauer and Sigmund termed the ``weak ESS''~\cite{hofbauer_book}; it also includes the definitions Selten~\cite{Selten1980} and Schuster et al.~\cite{Schuster_1981,hofbauer_sigmund_1988} proposed as ESS for bimatrix games.
\item[] \textbf{Definition 1c}: A state $(\mathbf{\hat{x}},\mathbf{\hat{y}})$ is an ESS if for every population state $(\mathbf{x},\mathbf{y})$ that are sufficiently close to  $(\mathbf{\hat{x}},\mathbf{\hat{y}})$ such that $\mathbf{{x}}\neq\mathbf{\hat{x}}$ and $\mathbf{{y}}\neq\mathbf{\hat{y}}$
\begin{equation}\label{eqn:ess1_neighbourhood}
	\mathbf{\hat{x}}\cdot{\sf B}\mathbf{y}> 	\mathbf{x}\cdot{\sf B}\mathbf{y}~\text{\emph{and}}~~~\mathbf{\hat{y}}\cdot{\sf C}\mathbf{x}> \mathbf{y}\cdot{\sf C}\mathbf{x}.
\end{equation} 
This way of defining ESS  proves useful when linking the ESS definition with its dynamical stability. We emphasize that this definition is different from the previous two because it compares the fitnesses of resident states with that of only their nearby states.
\end{enumerate}

It is of use to recall that a state $(\mathbf{\hat{x}},\mathbf{\hat{y}})$ is a Nash equilibrium~\cite{Nash1950,Nash1951,hofbauer_book} (NE) in bimatrix games if the following is true for all $(\mathbf{x},\mathbf{y})\neq(\mathbf{\hat{x}},\mathbf{\hat{y}})$:
\begin{equation}\label{eqn:Bimatrix_Nash_cdn}
\mathbf{\hat{x}}\cdot{\sf B}\mathbf{\hat{y}}\geq	\mathbf{x}\cdot{\sf B}\mathbf{\hat{y}}~~\text{and}~~\mathbf{\hat{y}}\cdot{\sf C}\mathbf{\hat{x}}\geq\mathbf{y}\cdot{\sf C}\mathbf{\hat{x}}.
\end{equation}
Mere inspection reveals that NE is directly implied by and-definition; and, of course, when NE is strict, ESS under and-definition is implied. {\color{black} In fact one show using similar arguments presented in Section~\ref{sec:bihyper_pure} that ESS under this definition can only be pure.}

Rather than comparing fitnesses of host individuals against the mutants in each subpopulation separately, one could compare combined fitness of host individuals in both the subpopulations with combined fitness of mutants in both the subpopulations~\cite{Schuster_1981,Selten1980,Pohley1979,Taylor_1979}. Accordingly, we write the following set of three definitions (aptly termed \emph{sum}-definition):
\begin{enumerate}
\item[]\textbf{Definition 2a}: A state $(\mathbf{\hat{x}},\mathbf{\hat{y}})$ is called an ESS if for any mutant state $(\mathbf{{x}},\mathbf{{y}})\neq(\mathbf{\hat{x}},\mathbf{\hat{y}})$, there exists an invasion barrier, $\epsilon_{(\mathbf{x},\mathbf{y})} >0$ such that, $\forall(\epsilon_x,\epsilon_y)$,  with $0 < \epsilon_x <\epsilon_{(\mathbf{x},\mathbf{y})} $ and $0 < \epsilon_y <\epsilon_{(\mathbf{x},\mathbf{y})}$,
\begin{equation}\label{eqn:ESS_2_fp}
	\mathbf{\hat{x}}\cdot{\sf B}(\epsilon_y \mathbf{y}+(1-\epsilon_y)\mathbf{\hat{y}})+\mathbf{\hat{y}}\cdot{\sf C}(\epsilon_x \mathbf{x}+(1-\epsilon_x)\mathbf{\hat{x}})>	\mathbf{x}\cdot{\sf B}(\epsilon_y \mathbf{y}+(1-\epsilon_y)\mathbf{\hat{y}})+\mathbf{y}\cdot{\sf C}(\epsilon_x \mathbf{x}+(1-\epsilon_x)\mathbf{\hat{x}}).
\end{equation}
 Note that in this definition, one could relax the restriction---$\mathbf{x} \neq \mathbf{\hat{x}}$ \emph{and} $\mathbf{y} \neq \mathbf{\hat{y}}$---on the mutant state, as explicitly incorporated in Definition 2a itself through adoption of the condition--- $(\mathbf{x},\mathbf{y}) \neq (\mathbf{\hat{x}},\mathbf{\hat{y}})$. In this case, there is no issue with allowing mutants of the form $(\mathbf{\hat{x}},\mathbf{y})$ or $(\mathbf{x},\mathbf{\hat{y}})$. For instance, for a mutant of the form $(\mathbf{\hat{x}},\mathbf{y})$, inequality~(\ref{eqn:ESS_2_fp}) simplifies to $\mathbf{\hat{y}}\cdot{\sf C}\mathbf{\hat{x}} > \mathbf{y}\cdot{\sf C}\mathbf{\hat{x}}$. This definition was originally introduced in Refs.~\cite{Taylor_1979,Selten1980,Schuster_1981}. 

\item[]\textbf{Definition 2b}: 
A state $(\mathbf{\hat{x}},\mathbf{\hat{y}})$ is an ESS if, for any mutant $(\mathbf{{x}},\mathbf{{y}})\neq(\mathbf{\hat{x}},\mathbf{\hat{y}})$, the following conditions hold,
\begin{equation}\label{eqn:Nash_schushter}
	\text{(i)}~\mathbf{\hat{x}}\cdot{\sf B}\mathbf{\hat{y}}+\mathbf{\hat{y}}\cdot{\sf C}\mathbf{\hat{x}}\geq	\mathbf{x}\cdot{\sf B}\mathbf{\hat{y}}+\mathbf{y}\cdot{\sf C}\mathbf{\hat{x}};
\end{equation}
(ii) and $\forall(\mathbf{x},\mathbf{y})\in\Sigma_m\times\Sigma_n$ for which inequality in (i) holds as equality, then
\begin{equation}
	\mathbf{\hat{x}}\cdot{\sf B}\mathbf{y}+\mathbf{\hat{y}}\cdot{\sf C}\mathbf{x}>\mathbf{x}\cdot{\sf B}\mathbf{y}+\mathbf{y}\cdot{\sf C}\mathbf{x}.
\end{equation}
\item[]\textbf{Definition 2c}: A state $(\mathbf{\hat{x}},\mathbf{\hat{y}})$ is an ESS if for every population state $(\mathbf{x},\mathbf{y})$ that are sufficiently close but not equal to $(\mathbf{\hat{x}},\mathbf{\hat{y}})$,
\begin{equation}\label{eqn:ess2_neighbourhood}
	\mathbf{\hat{x}}\cdot{\sf B}\mathbf{y}+\mathbf{\hat{y}}\cdot{\sf C}\mathbf{x}> 	\mathbf{x}\cdot{\sf B}\mathbf{y}+ \mathbf{y}\cdot{\sf C}\mathbf{x}.
\end{equation} 
\end{enumerate}
A few remarks are in order. We first mention that Definitions 2a, 2b, and 2c are equivalent if we assume $\epsilon_x=\epsilon_y=\epsilon$.
Next, we would like to remark that in line with analogous definitions in the case of symmetric games~\cite{hofbauer_book}, versions (a), (b), and (c) in the above two set of definitions may be called first principle definition, Maynard-Smith--Price definition and neighborhood definition, respectively. 

The equilibrium condition of Definition 2b (Eq.~(\ref{eqn:Nash_schushter})) appears different from Eq.~(\ref{eqn:Bimatrix_Nash_cdn}). So while strict NE obviously implies ESS under sum-definition, it is not obvious that ESS always implies NE. To check this out, let us write inequality~(\ref{eqn:Nash_schushter}) in the following form: $(\mathbf{\hat{x}}-\mathbf{x})\cdot{\sf B}\mathbf{\hat{y}}+(\mathbf{\hat{y}}-\mathbf{y})\cdot{\sf C}\mathbf{\hat{x}}\geq0$. Since this has to be true for all $(\mathbf{x},\mathbf{y})\neq(\mathbf{\hat{x}},\mathbf{\hat{y}})$, if $(\mathbf{\hat{x}}-\mathbf{x})\cdot{\sf B}\mathbf{\hat{y}}<0$ for some $\mathbf{x}$, then one can always find a $\mathbf{y}$ such that $(\mathbf{\hat{x}}-\mathbf{x})\cdot{\sf B}\mathbf{\hat{y}}+(\mathbf{\hat{y}}-\mathbf{y})\cdot{\sf C}\mathbf{\hat{x}}<0$ which is a contradiction. This shows that both the terms individually has to satisfy $(\mathbf{\hat{x}}-\mathbf{x})\cdot{\sf B}\mathbf{\hat{y}}\geq0$ and $(\mathbf{\hat{y}}-\mathbf{y})\cdot{\sf C}\mathbf{\hat{x}}\geq0$ for all $({\bf x},{\bf y})$ so that inequality~(\ref{eqn:Nash_schushter}) holds.  {\color{black} Therefore, an ESS under sum-definition implies NE. However, as we will see in Section~\ref{sec:bimatrix:basket} that only strict NE can be ESS under sum-definition.}

We emphasize that the definitions in this section require simultaneous appearance of mutants in both the subpopulations, i.e., $\epsilon_x\neq0$ and $\epsilon_y\neq0$ as mentioned in Definitions 1a and 2a explicitly. This restriction is relaxed in the next definition as introduced in the next section.

\subsection{Mutants in both or either subpopulations}\label{sec:equv_B}
Few remarks are in order. ESSes defined through Definition 1b (and equivalently, 1a and 1c) and Definition 2b (and equivalently, 2a and 2c) can only be pure~\cite{Selten1980,Schuster_1981}. Hence, this fact prompted a need to revise the definitions so as to accommodate scenarios involving mixed ESS. The revised definitions, primarily suggested by Cressman~\cite{cressman_1992,Cressman1996,Cressman_Tao_2014}, Hofbauer and Sigmund~\cite{hofbauer_book}, relax the need for residents of both subpopulations to simultaneously outperform the respective mutants and demanding only that residents outperform the mutants in at least one subpopulation. In and- and sum-definitions, there's no provision for this scenario, as they either compare the fitnesses of both residents against their respective mutants' or compare the sum of fitnesses of both residents and both mutants. Taylor~\cite{Taylor_1979} discussed this point, highlighting that if mutants only arrive in subpopulation Y, one should compare the fitness of residents of Y versus mutants of Y while interacting with residents of population X and vice versa. He further states that if mutants arrive in both subpopulations, the comparison should involve combined fitness, as in sum-definition. In the light of above remarks, we now present the following set of equivalent definitions (aptly termed \emph{or}-definition) where one of  $\epsilon_x$ or $\epsilon_y$ can be set to  zero independently (but they cannot be zero simultaneously because then there is no mutant in any subpopulation for one to compare fitness advantages):
\begin{enumerate}
\item[]\textbf{Definition 3a}: A state $(\mathbf{\hat{x}},\mathbf{\hat{y}})$ is called ESS if for any mutant state $(\mathbf{{x}},\mathbf{{y}})\neq(\mathbf{\hat{x}},\mathbf{\hat{y}})$, there exists an invasion barrier, $\epsilon_{(\mathbf{x},\mathbf{y})} >0$ such that, $\forall(\epsilon_x,\epsilon_y)$  with $0 \leq \epsilon_x <\epsilon_{(\mathbf{x},\mathbf{y})} $ and $0 \leq \epsilon_y <\epsilon_{(\mathbf{x},\mathbf{y})}$,~$(\epsilon_x,\epsilon_y)\neq(0,0)$,
\begin{equation}\label{eqn:ess3_fp}
	\epsilon_x \mathbf{\hat{x}}\cdot {\sf B}(\epsilon_y \mathbf{y}+(1-\epsilon_y) \mathbf{\hat{y}}) >\epsilon_x \mathbf{x}\cdot {\sf B}(\epsilon_y \mathbf{y}+(1-\epsilon_y) \mathbf{\hat{y}})~~~\text{\emph{or}}~~~
	 \epsilon_y\mathbf{\hat{y}}\cdot {\sf C}(\epsilon_x \mathbf{x}+(1-\epsilon_x) \mathbf{\hat{x}}) >\epsilon_y\mathbf{y}\cdot {\sf C}(\epsilon_x \mathbf{x}+(1-\epsilon_x) \mathbf{\hat{x}}).
\end{equation}
This definition is provided in Ref.~\cite{Cressman1996} in the context of asymmetric games.

\item[]\textbf{Definition 3b}: A state $(\hat{\mathbf{x}},\hat{\mathbf{y}})$ is an ESS if, for any mutant $(\mathbf{{x}},\mathbf{{y}})\neq(\mathbf{\hat{x}},\mathbf{\hat{y}})$, the following conditions hold,
\begin{equation}\label{eq:3bi}
	\text{(i)}~~	\mathbf{\hat{x}}\cdot{\sf B}\mathbf{\hat{y}}\geq	\mathbf{x}\cdot{\sf B}\mathbf{\hat{y}}~~\text{and}~~\mathbf{\hat{y}}\cdot{\sf C}\mathbf{\hat{x}}\geq\mathbf{y}\cdot{\sf C}\mathbf{\hat{x}};
\end{equation}
(ii) and  $\forall(\mathbf{x},\mathbf{y})\in\Sigma_m\times\Sigma_n$ for which both inequalities in (i) hold as equalities, following condition holds
\begin{equation}\label{eq:3bii}
	\text{either}~	\mathbf{\hat{x}}\cdot\sf{B}\mathbf{y}>\mathbf{x}\cdot{\sf B}\mathbf{y} ~\text{or}~\mathbf{\hat{y}}\cdot{\sf C}\mathbf{x}>\mathbf{y}\cdot{\sf C}\mathbf{x}.
\end{equation}
(It basically captures the fact that both the mutants cannot simultaneously take advantage of the resident; at least one of them gets penalized.)
\item[]\textbf{Definition 3c}:  A state $(\mathbf{\hat{x}},\mathbf{\hat{y}})$ is an ESS if for every population state $(\mathbf{x},\mathbf{y})$ that are sufficiently close but not equal to $(\mathbf{\hat{x}},\mathbf{\hat{y}})$,
	\begin{equation}\label{eqn:ess3_neighbourhood}
		\text{either}~~~\mathbf{\hat{x}}\cdot{\sf B}\mathbf{y}> 	\mathbf{x}\cdot{\sf B}\mathbf{y}~~~\text{or}~~~\mathbf{\hat{y}}\cdot{\sf C}\mathbf{x}> \mathbf{y}\cdot{\sf C}\mathbf{x},
	\end{equation} 
\end{enumerate}
While Definition 3a is first principle definition when mutants in both subpopulations need not be present simultaneously, ESS in Definition 3c is known as two-species ESS (2ESS)~\cite{cressman_1992,Cressman1996,Cressman_Tao_2014} in literature. Some readers may be reminded of the Nash--Pareto pair while reading Definition 3b; however, there is a subtle difference between the two (see Appendix~\ref{sec:Nash_Pareto}).

We notice that~(\ref{eqn:Bimatrix_Nash_cdn}) directly matches with equilibrium conditions in Definition 3b. So obviously, ESS under or-definition implies NE; and strict NE trivially implies ESS. However, because of the presence of conjunction `or' in Definition 3a, there arises an interesting point: It deceptively appears that a state that is not NE, e.g., $(\mathbf{\hat{x}},\mathbf{\hat{y}})$ such that $\forall$ $(\mathbf{{x}},\mathbf{{y}})\ne(\mathbf{\hat{x}},\mathbf{\hat{y}})$, $\mathbf{\hat{x}}\cdot{\sf B}\mathbf{\hat{y}}>	\mathbf{x}\cdot{\sf B}\mathbf{\hat{y}}$ and $\mathbf{\hat{y}}\cdot{\sf C}\mathbf{\hat{x}}<\mathbf{y}\cdot{\sf C}\mathbf{\hat{x}}$ can in fact qualify as ESS---as an inspection of Definition 3a would suggest---because of the presence of conjunction `or'. However, this apparent anomaly is prevented since the definition allows all the mutant state $(\mathbf{x},\mathbf{y})\neq(\mathbf{\hat{x}},\mathbf{\hat{y}})$. For instance, if we consider a mutant state of the form $(\mathbf{\hat{x}},\mathbf{y})$,  which is distinct from $(\mathbf{\hat{x}},\mathbf{\hat{y}})$, the first inequality in~(\ref{eqn:ess3_fp}) cannot hold. Moreover, in this case the second inequality simplifies to $\epsilon_y\mathbf{\hat{y}}\cdot {\sf C}\mathbf{\hat{x}} >\epsilon_y\mathbf{y}\cdot {\sf C}\mathbf{\hat{x}}$  preventing the erroneous conclusion. In conclusion, allowing all $(\mathbf{x},\mathbf{y})\neq(\mathbf{\hat{x}},\mathbf{\hat{y}})$ prevents states that are not NEs from qualifying as ESSes. Unfortunately, this condition brings forth another issue: There cannot be any mixed ESS.

Consider Definition 3a version of or-definition. If the mutant state satisfies $\mathbf{\hat{x}} \cdot \sf{B} \mathbf{\hat{y}} = \mathbf{x} \cdot \sf{B} \mathbf{\hat{y}}$ and $\mathbf{\hat{y}} \cdot \sf{C} \mathbf{\hat{x}} = \mathbf{y} \cdot \sf{C} \mathbf{\hat{x}}$, then (\ref{eqn:ess3_fp}) reduces to $\epsilon_x\epsilon_y\mathbf{\hat{x}} \cdot {\sf B} \mathbf{y} > \epsilon_x\epsilon_y\mathbf{x} \cdot {\sf B} \mathbf{y}$ or $\epsilon_x\epsilon_y\mathbf{\hat{y}} \cdot {\sf C} \mathbf{x} > \epsilon_x\epsilon_y\mathbf{y} \cdot {\sf C} \mathbf{x}$. Consequently, setting the mutant state $(\mathbf{\hat{x}},\mathbf{y})$ or $(\mathbf{x},\mathbf{\hat{y}})$ should not be permissible, as it renders both conditions meaningless. For instance, if we take $(\mathbf{x},\mathbf{\hat{y}})$ as mutant, then both the inequalities $\epsilon_x\epsilon_y\mathbf{\hat{x}} \cdot {\sf B} \mathbf{y} > \epsilon_x\epsilon_y\mathbf{x} \cdot {\sf B} \mathbf{y}$ and $\epsilon_x\epsilon_y\mathbf{\hat{y}} \cdot {\sf C} \mathbf{x} > \epsilon_x\epsilon_y\mathbf{y} \cdot {\sf C} \mathbf{x}$ become invalid (as the first one contradicts  $\mathbf{\hat{x}} \cdot \sf{B} \mathbf{\hat{y}} = \mathbf{x} \cdot \sf{B} \mathbf{\hat{y}}$ and the second one trivially becomes an equality). Thus, if one would want to allow for a mixed ESS in bimatrix games under or-definition, then we must adopt the convention that while looking for mixed ESS  one cannot allow for mutants of the form $(\mathbf{\hat{x}},\mathbf{y})$ or $(\mathbf{x},\mathbf{\hat{y}})$. One can also argue that allowing for mixed 2ESS needs the restriction $\epsilon_x\neq0$ and $\epsilon_y\neq0$. This is because the simplified condition after putting the mixed Nash equilibrium criteria in (\ref{eqn:ess3_fp}) becomes: $\epsilon_x\epsilon_y\mathbf{\hat{x}} \cdot {\sf B} \mathbf{y} > \epsilon_x\epsilon_y\mathbf{x} \cdot {\sf B} \mathbf{y}$ or $\epsilon_x\epsilon_y\mathbf{\hat{y}} \cdot {\sf C} \mathbf{x} > \epsilon_x\epsilon_y\mathbf{y} \cdot {\sf C} \mathbf{x}$, which can never be satisfied if one allows for either of the $\epsilon_x$ or $\epsilon_y$ to go to zero. This perspective (ad hoc simultaneous imposition of $\mathbf{x}\neq\mathbf{\hat{x}},~\mathbf{y}\neq\mathbf{\hat{y}},~\epsilon_x\neq0$ and $\epsilon_y\neq0$) may appositely be called the \emph{simultaneous deviation perspective}. This view is adopted while defining the Nash--Pareto pair in Hofbauer and Sigmund~\cite{hofbauer_book}.

 An alternative perspective---which we term \emph{face-value perspective} is to permit all mutants (including $(\mathbf{\hat{x}},\mathbf{y})$ or $(\mathbf{x},\mathbf{\hat{y}})$) as outlined explicitly by the or-definition [as specified in Definition 3a, with  $(\mathbf{x},\mathbf{y})\neq (\mathbf{\hat{x}},\mathbf{\hat{y}})$,~$0 \leq \epsilon_x <\epsilon_{(\mathbf{x},\mathbf{y})}$,~$0 \leq \epsilon_y <\epsilon_{(\mathbf{x},\mathbf{y})}$~and~$(\epsilon_x,\epsilon_y)\neq(0,0)$], regardless of the equalities: $\mathbf{\hat{x}} \cdot {\sf B} \mathbf{\hat{y}} = \mathbf{x} \cdot {\sf B} \mathbf{\hat{y}}$ and $\mathbf{\hat{y}} \cdot {\sf C} \mathbf{\hat{x}} = \mathbf{y} \cdot {\sf C} \mathbf{\hat{x}}$. Under this perspective, there cannot be any mixed ESS in bimatrix games, as the ESS conditions would only be satisfied if $(\mathbf{\hat{x}}, \mathbf{\hat{y}})$ is a strict NE. This view was adopted by Cressman~\cite{Cressman_Tao_2014, cressman_1992} while defining 2ESS.
It is obvious that also for weak pure NE [e.g., $(\mathbf{\hat{x}},\mathbf{\hat{y}})$ such that $\forall$ $(\mathbf{{x}},\mathbf{{y}})\ne(\mathbf{\hat{x}},\mathbf{\hat{y}})$, $\mathbf{\hat{x}}\cdot{\sf B}\mathbf{\hat{y}}>	\mathbf{x}\cdot{\sf B}\mathbf{\hat{y}}$ and $\mathbf{\hat{y}}\cdot{\sf C}\mathbf{\hat{x}}=\mathbf{y}\cdot{\sf C}\mathbf{\hat{x}}$] the issue detailed above applies: Under  simultaneous deviation perspective, it will be 2ESS but not so under face-value perspective.

We would like to re-emphasize that by `face-value perspective', we mean strictly following the definitions without imposing any additional constraints (whether they involve allowing mutant states like $(\mathbf{\hat{x}},\mathbf{y})$ or letting $\epsilon_y$ go to zero). On the other hand, adoption of `simultaneous deviations perspective', where one puts additional constraint on the allowed mutant states and their fractions, allows for the possibility of existence of mixed ESS in bimatrix games. At this point, one might question, why we are emphasizing so much on the existence of a mixed ESS in bimatrix games, given that it can never be asymptotically stable under replicator dynamics (see Section~\ref{sec:mixed_bim}). The reason is that stability depends on the choice of evolutionary dynamics. As discussed in Section~\ref{sec:mixed_bim}, while a mixed ESS is not stable under replicator dynamics, it can be stable under other evolutionary dynamics. Since replicator dynamics is not the only possible evolutionary process, this opens the possibility that certain mixed states---though unstable under replicator dynamics---can still qualify as an ESS and exhibit dynamical stability under alternative dynamics.

\subsection{Equivalent definitions}\label{sec:bimatrix_equv}
As mentioned earlier each of aforementioned three sets (viz., and-, sum- and or-definitions) has three equivalent definitions---versions (a), (b) and (c). Let us see the line of reasoning that establishes these equivalences. For illustrative purpose, let us focus on establishing the equivalence between Definitions 3a, 3b, and 3c; equivalence between Definitions 1a, 1b, and 1c, as well as the equivalence between Definitions 2a, 2b, and 2c, follows analogously. Now, in what follows, we prove the assertion by demonstrating that Definitions 3a and 3b imply each other and Definitions 3a and 3c imply each other.

Now, let $(\mathbf{\hat{x}},\mathbf{\hat{y}})$ be an ESS according to Definition 3a, so that for all mutants $(\mathbf{x},\mathbf{y})\ne(\mathbf{\hat{x}},\mathbf{\hat{y}})$, inequalities in~(\ref{eqn:ess3_fp}) are valid.  First, we can realize that if either of the conditions $\mathbf{\hat{x}} \cdot {\sf B} \mathbf{\hat{y}} < \mathbf{x} \cdot {\sf B} \mathbf{\hat{y}}$ or $\mathbf{\hat{y}} \cdot {\sf C} \mathbf{\hat{x}} < \mathbf{y} \cdot \sf{C} \mathbf{\hat{x}}$ holds, then there is always some mutant $(\mathbf{x},\mathbf{y})$ for which the inequalities in~(\ref{eqn:ess3_fp}) do not hold. For instance, if $\mathbf{\hat{x}} \cdot \sf{B} \mathbf{\hat{y}} < \mathbf{x} \cdot \sf{B} \mathbf{\hat{y}}$, then for $(\mathbf{x},\mathbf{\hat{y}})$, the first condition in~(\ref{eqn:ess3_fp}) can clearly not be satisfied and the second condition becomes invalid. Hence, we can conclude that the plausible criteria---for resident state $(\mathbf{\hat{x}}, \mathbf{\hat{y}})$---for all allowable mutants $(\mathbf{x},\mathbf{y})$ to satisfy the inequalities in~(\ref{eqn:ess3_fp}) is:
    \begin{equation}\label{eqn:mixed_nash_3c}
	\mathbf{\hat{x}}\cdot \sf{B}\mathbf{\hat{y}} \ge \mathbf{x}\cdot \sf{B} \mathbf{\hat{y}}~\text{and}~ \mathbf{\hat{y}}\cdot \sf{C} \mathbf{\hat{x}} \ge \mathbf{y}\cdot \sf{C} \mathbf{\hat{x}}.
	\end{equation}
Thus, four explicit possibilities arise from (\ref{eqn:mixed_nash_3c}) as detailed below. We now show that if Definition 3a is satisfied under simultaneous deviation perspective, then one of the four cases (all corresponding to condition (\ref{eq:3bi})) must be satisfied with the natural requirement of condition (\ref{eq:3bii}) in the relevant case. One can trivially see that following is true when when~(\ref{eqn:ess3_fp}) holds good:
\begin{enumerate}
\item either condition---$\mathbf{\hat{x}}\cdot \sf{B}\mathbf{\hat{y}} > \mathbf{x}\cdot \sf{B} \mathbf{\hat{y}}~\text{and}~ \mathbf{\hat{y}}\cdot \sf{C} \mathbf{\hat{x}} > \mathbf{y}\cdot \sf{C} \mathbf{\hat{x}}$ is true, or 

\item $\mathbf{\hat{x}}\cdot \sf{B}\mathbf{\hat{y}} = \mathbf{x}\cdot \sf{B} \mathbf{\hat{y}}~\text{and}~ \mathbf{\hat{y}}\cdot \sf{C} \mathbf{\hat{x}} > \mathbf{y}\cdot \sf{C} \mathbf{\hat{x}}$ is true, or

\item $\mathbf{\hat{x}}\cdot \sf{B}\mathbf{\hat{y}} > \mathbf{x}\cdot \sf{B} \mathbf{\hat{y}}~\text{and}~ \mathbf{\hat{y}}\cdot \sf{C} \mathbf{\hat{x}} = \mathbf{y}\cdot \sf{C} \mathbf{\hat{x}}$ is true, or
	
\item $\mathbf{\hat{x}}\cdot{\sf B}\mathbf{\hat{y}}=	\mathbf{x}\cdot{\sf B}\mathbf{\hat{y}}$ and $\mathbf{\hat{y}}\cdot{\sf C}\mathbf{\hat{x}}=\mathbf{y}\cdot{\sf C}\mathbf{\hat{x}}$ is true but in addition, since $\mathbf{x} \ne \mathbf{\hat{x}}$ and $\mathbf{y} \ne \mathbf{\hat{y}}$, $\epsilon_x\neq0$ and $\epsilon_y\neq0$ straightforward comparison of the coefficients of $\epsilon_x$ and $\epsilon_y$ in~(\ref{eqn:ess3_fp}) tells us that the condition (\ref{eq:3bii}) must follow.

	\end{enumerate}
Above argument establishes that Definition 3a implies Definition 3b. Moreover, the conditions in Definition 3b when put in Definition 3a, under the assumption of sufficiently small $\epsilon_x$ and $\epsilon_y$, implies that whenever there is an ESS as per Definition 3b, it must be an ESS as per Definition 3a as well. In conclusion, Definitions 3a and 3b are equivalent. We remark, in passing, that this equivalence is intact under face-value perspective as well but then the only ESSes possible are the ones that are strict NE.

Next let us show that Definitions 3a and 3c imply each other. We start by assuming that $(\mathbf{\hat{x}},\mathbf{\hat{y}})$ satisfies Definition 3a. We note every $(\mathbf{x}_1,\mathbf{y}_1)$ close to  $(\mathbf{\hat{x}},\mathbf{\hat{y}})$ can written as $(\epsilon_x \mathbf{x} + (1-\epsilon_x) \mathbf{\hat{x}},\epsilon_y \mathbf{y} + (1-\epsilon_y) \mathbf{\hat{y}})$ where $\epsilon_x$ and $\epsilon_y$ are sufficiently small. Let us first show that $\epsilon_x$ and $\epsilon_y$ can be chosen independent of $(\mathbf{x},\mathbf{y})$ in a continuous manner. Consider a set, say, $\Gamma$, which is the union of all faces of $\Sigma_m\times\Sigma_n$ that does not contain $(\mathbf{\hat{x}},\mathbf{\hat{y}})$. We note that $\Gamma\equiv\Gamma_x\times\Gamma_y$, where $\Gamma_x\equiv\{\mathbf{x}\in\Sigma_m:x_i=0~\text{for some}~i\in {\rm supp}(\mathbf{\hat{x}})\}$ and $\Gamma_y\equiv\{\mathbf{y}\in\Sigma_n:y_i=0~\text{for some}~i\in {\rm supp}(\mathbf{\hat{y}})\}$, is compact~\cite{Munkres2000, hofbauer_book}. By definition, for every $(\mathbf{x},\mathbf{y})\in \Gamma$,~(\ref{eqn:ess3_fp}) holds for all $(\epsilon_x,\epsilon_y)$ such that $\epsilon_x<\epsilon_{(\mathbf{x},\mathbf{y})}$ and $\epsilon_y<\epsilon_{(\mathbf{x},\mathbf{y})}$. We can construct a continuous function~\cite{Cressman1996}, $\epsilon_{(\mathbf{x},\mathbf{y})}\in(0,1]$
and, since the set $\Gamma$  is compact, $\epsilon_{\min} \equiv\min\{\epsilon_{(\mathbf{x},\mathbf{y})}:(\mathbf{x},\mathbf{y})\in\Gamma \} $ is strictly positive. Therefore~(\ref{eqn:ess3_fp}) holds for all $\epsilon_x,\epsilon_y<\epsilon_{\min}$ but $(\epsilon_x,\epsilon_y)\ne (0,0)$.
We can now do the following steps to prove the equivalence: Starting with~(\ref{eqn:ess3_fp}), we add  $(1-\epsilon_x) \mathbf{\hat{x}}\cdot{\sf B}(\epsilon_y \mathbf{y}+(1-\epsilon_y) \mathbf{\hat{y}})$ to both sides of the first inequality of~(\ref{eqn:ess3_fp}). Similarly, we add $(1-\epsilon_y) \mathbf{\hat{y}}\cdot{\sf C}(\epsilon_x \mathbf{x}+(1-\epsilon_x) \mathbf{\hat{x}})$ to both sides of the second inequality of~(\ref{eqn:ess3_fp}). The two resultant inequality expressions can be rearranged to read:
\begin{subequations}
	\begin{eqnarray}
		\mathbf{\hat{x}}\cdot {\sf B}(\epsilon_y \mathbf{y}+(1-\epsilon_y) \mathbf{\hat{y}}) > (\epsilon_x\mathbf{x}+(1-\epsilon_x) \mathbf{\hat{x}})\cdot {\sf B}(\epsilon_y \mathbf{y}+(1-\epsilon_y) \mathbf{\hat{y}})\\
		\text{or,}~
		\mathbf{\hat{y}}\cdot {\sf C}(\epsilon_x \mathbf{x}+(1-\epsilon_x) \mathbf{\hat{x}}) >(\epsilon_y\mathbf{y}+(1-\epsilon_y) \mathbf{\hat{y}})\cdot {\sf C}(\epsilon_x \mathbf{x}+(1-\epsilon_x) \mathbf{\hat{x}}).
	\end{eqnarray}
\end{subequations}
Relabelling $(\mathbf{x}_1,\mathbf{y}_1)$ as $(\mathbf{x},\mathbf{y})$, we arrive at~(\ref{eqn:ess3_neighbourhood}) exactly, thereby demonstrating that Definition 3a implies Definition 3c. Since these steps done to reach at this implication can be reversed, Definition 3c implies Definition 3a. Thus, we have established the equivalence between both these definitions. Moreover, in the light of the already established equivalence of Definitions 3a and Definition 3b, the equivalence of Definition 3b and Definition 3c is also automatically established.

\section{Bimatrix games}\label{sec:bimatrix}
It is an interesting fact that and-, sum-, and or-definitions are not equivalent and need not imply each other in general.
So, among the three sets of definitions which one is the most useful one?  Or each of the definitions works equally well?
While there might be various arguments in favour of each one, one desirable aspect of ESS is probably undeniable: ESS is a static game-theoretic concept---it basically is about existence of an equilibrium solution; how one reaches it starting from a scenario that is not in equilibrium requires invoking some sort of dynamics. In evolutionary game theory, replicator equation~\cite{taylorjonker1978,Cressman_Tao_2014} is the most accepted paradigm in this respect. Thus, one expects the fixed points of the replicator equation that correspond to ESS to be asymptotically stable, that is, we expect the following ~proposition (cf. Thereom 4 in Ref.~\cite{Cressman_Tao_2014}, Theorem 4.5 and 4.6 in Ref.~\cite{Cressman1996}) to be satisfied an ESS:\\
\textbf{Proposition 1:} An ESS is a locally asymptotically stable fixed point of the following set of equations called the replicator equations,
\begin{subequations}\label{eqn:bimatrix_equation}
	\begin{eqnarray}
		&&\dot{x}_i=x_i\left(({\sf B}\mathbf{y})_i-\mathbf{x}\cdot{\sf B}\mathbf{y}\right),\\ 
		&&\dot{y}_i=y_i\left(({\sf C}\mathbf{x})_i-\mathbf{y}\cdot{\sf C}\mathbf{x}\right).
	\end{eqnarray}
\end{subequations}
 It is well-known that NE corresponds to a fixed point of the replicator equation~(\ref{eqn:bimatrix_equation}). Since we have already seen that under and-, sum-, and or-definitions ESS is NE, we effectively are sure that any \emph{ESS is a fixed point of the replicator equation}. 

 We can categorize ESS into two types: pure ESS and mixed ESS. Pure ESS can represent either a strict or weak NE, while mixed ESS can only correspond to a weak NE. The nuances of stability arise specifically in cases where ESS is a weak pure NE. We will examine each category separately for bimatrix and asymmetric games. However, before proceeding, let us first identify which definition aligns with the defining concept of \emph{strong stability}~\cite{hofbauer_book,hofbauer_2003}.
\subsection{Strong Stability}\label{sec:strong_stability}
We note that while $x_i$ is the frequency of $i$th type in population X and $y_i$ is the frequency of $i$th type in population Y, we never explicitly specified what the corresponding strategies are for the two types. So let us bring it into consideration. Let there be \( m \)-(pheno) types of players in population X with frequencies \(\{x_1, x_2, \dots, x_m\} (\equiv \mathbf{x})\), which are mapped to strategies \(\{\mathbf{p}_1, \mathbf{p}_2, \dots, \mathbf{p}_m\}\), respectively. Similarly, let there be \( n \)-(pheno) types in population Y with frequencies \(\{y_1, y_2, \dots, y_n\} (\equiv \mathbf{y})\), mapped to strategies \(\{\mathbf{q}_1, \mathbf{q}_2, \dots, \mathbf{q}_n\}\), respectively.
In general, $\mathbf{p}_i$ and $\mathbf{q}_i$ are, respectively, elements of an $N_1$ dimensional simplex and $N_2$ dimensional simplex; there are two corresponding payoff matrices denoted by ${\sf U}_{1}$ and ${\sf U}_{2}$,  respectively, representing the payoffs for role X and role Y players, should they adopt pure strategies in the respective simplices. The size of ${\sf U}_{1}$ is $N_{1} \times N_{2}$, while ${\sf U}_{2}$ has dimensions $N_{2} \times N_{1}$. In these notations, it is obvious that the matrices, ${\sf B}$ and ${\sf C}$, defined in the previous section are the ones with elements $b_{ij}=\mathbf{p}_i\cdot {\sf U}_{1}\mathbf{q}_j$ and $c_{ij}=\mathbf{q}_i\cdot {\sf U}_{2}\mathbf{p}_j$, respectively.

What we want to establish can be most clearly done in the simplest non-trivial setting where $m=n=2$. In any case, we say (cf.~\cite{hofbauer_book}), a strategy profile, $(\mathbf{\hat{p}},\mathbf{\hat{q}})$, is \emph{strongly stable} if, whenever $\mathbf{\hat{p}}$ is a convex combination of $\mathbf{p}_1,\cdots,\mathbf{p}_m$ and $\mathbf{\hat{q}}$ is a convex combination of $\mathbf{q}_1,\cdots,\mathbf{q}_n$, the mean population strategy profile, $(\sum_ix_i{\mathbf p}_i,\sum_iy_i{\mathbf q}_i)$, close to $(\mathbf{\hat{p}},\mathbf{\hat{q}})$, converges under the replicator equation to $(\mathbf{\hat{p}},\mathbf{\hat{q}})$. What is desired is a strongly stable strategy to be equivalent to evolutionarily stable strategy. Since the replicator equation is written using states and not strategies, we must first establish that the ESS conditions obeyed by strategies and states are equivalent.

To this end, consider general $2\times 2$ payoff matrices: 
\begin{equation}
	{\sf B} = \begin{pmatrix}
		b_{11} & b_{12} \\
		b_{21} & b_{22} 
	\end{pmatrix}~\textrm{and}
	~{\sf C} = \begin{pmatrix}
		c_{11} & c_{12} \\
		c_{21} & c_{22} 
	\end{pmatrix}.
\end{equation}
Suppose, without any loss of generality, $(\mathbf{\hat{x}},\mathbf{\hat{y}})=\left(\begin{pmatrix} 1 ,0  \end{pmatrix},\begin{pmatrix} 1, 0  \end{pmatrix}\right)$ is an ESS as per Definition 1b. It is trivial to see that $\mathbf{\hat{x}}\cdot{\sf B}\mathbf{\hat{y}}\ge\mathbf{x}\cdot{\sf B}\mathbf{\hat{y}}\implies b_{11}\ge b_{21}\implies \mathbf{p}_1\cdot {\sf U}_{1}\mathbf{q}_1\ge \mathbf{p}_2\cdot {\sf U}_{1}\mathbf{q}_1$ and $\mathbf{\hat{y}}\cdot{\sf C}\mathbf{\hat{x}}\ge\mathbf{y}\cdot{\sf C}\mathbf{\hat{x}}\implies c_{11}\ge c_{21}\implies\mathbf{q}_1\cdot {\sf U}_{2}\mathbf{p}_1\ge\mathbf{q}_2\cdot {\sf U}_{2}\mathbf{p}_1$. Furthermore in case the equalities hold in the relation between terms containing states, then it is easily seen that $\mathbf{\hat{x}}\cdot{\sf B}\mathbf{y}>\mathbf{x}\cdot{\sf B}\mathbf{y}\implies b_{12}>b_{22}$, which in turn means $\mathbf{p}_1\cdot {\sf U}_{1}\mathbf{q}_2>\mathbf{p}_2\cdot {\sf U}_{1}\mathbf{q}_2$ when $\mathbf{p}_1\cdot {\sf U}_{1}\mathbf{q}_1=\mathbf{p}_2\cdot {\sf U}_{1}\mathbf{q}_1$; similarly, we can show $\mathbf{\hat{y}}\cdot{\sf C}\mathbf{x}>\mathbf{y}\cdot{\sf C}\mathbf{x}\implies \mathbf{q}_1\cdot {\sf U}_{2}\mathbf{p}_2>\mathbf{q}_2\cdot {\sf U}_{2}\mathbf{p}_2$ when $\mathbf{q}_1\cdot {\sf U}_{2}\mathbf{p}_1=\mathbf{q}_2\cdot {\sf U}_{2}\mathbf{p}_1$. In conclusion, if Definition 1b is satisfied by the state $(\mathbf{\hat{x}},\mathbf{\hat{y}})$, then the conditions of definition are also satisfied by the corresponding strategy $(\mathbf{p}_1,\mathbf{q}_1)$. It is rather repetitive to show that the conclusion remain unchanged if $(\mathbf{\hat{x}},\mathbf{\hat{y}})=\left(\begin{pmatrix} 0 ,1  \end{pmatrix},\begin{pmatrix} 0, 1  \end{pmatrix}\right)$---corresponding strategy profile being $(\mathbf{p}_2,\mathbf{q}_2)$---is chosen to be an ESS as per Definition 1b. Furthermore, employing these arguments, one can observe that if we had chosen to work with Definitions 2b and 3b instead of Definition 1b, the state-strategy equivalence in the definition of respective ESSes remains intact.

Now, to track the evolution of the mean population strategy, the bimatrix replicator equation for the case in hand is best presented  as
\begin{subequations}\label{eqn:bimatrix_equation2}
	\begin{eqnarray}
		&&\dot{x}=x(1-x)[y(\mathbf{p}_1\cdot {\sf U}_1\mathbf{q}_1-\mathbf{p}_2\cdot {\sf U}_1\mathbf{q}_1)-(1-y)(\mathbf{p}_2\cdot {\sf U}_1\mathbf{q}_2-\mathbf{p}_1\cdot {\sf U}_1\mathbf{q}_2)], \\ 
		&&\dot{y}=y(1-y)[x(\mathbf{q}_1\cdot {\sf U}_2\mathbf{p}_1-\mathbf{q}_2\cdot {\sf U}_2\mathbf{p}_1)-(1-x)(\mathbf{q}_2\cdot {\sf U}_2\mathbf{p}_2-\mathbf{q}_1\cdot {\sf U}_2\mathbf{p}_2)]. 
	\end{eqnarray}
\end{subequations}
We now have to show that if an $(\mathbf{\hat{x}},\mathbf{\hat{y}})$, say, $\left(\begin{pmatrix} 0 ,1  \end{pmatrix},\begin{pmatrix} 0, 1  \end{pmatrix}\right)$---equivalently, $(\mathbf{p}_2,\mathbf{q}_2)$---is ESS, then it is approached by neighbourhood initial conditions---equivalently, $(\sum_ix_i{\mathbf p}_i,\sum_iy_i{\mathbf q}_i)$ reaches $(\mathbf{p}_2,\mathbf{q}_2)$. This can be argued as follows: Here $(x,y)=(0,0)$ denotes the ESS. We observe that near $(0,0)$, the coefficients of $(1-x)$ and $(1-y)$ terms inside the square bracket dominates and if those terms vanish, the coefficients of $x$ and $y$ terms inside the square bracket are the deciding  factor for the stability of $(0,0)$. Consequently, comparing the dominating terms, the local asymptotic stability criteria for $(x,y)=(0,0)$ explicitly comes out to be $(a)$ $\mathbf{p}_2\cdot {\sf U}_1\mathbf{q}_2>\mathbf{p}_1\cdot {\sf U}_1\mathbf{q}_2$ and $\mathbf{q}_2\cdot {\sf U}_2\mathbf{p}_2>\mathbf{q}_1\cdot {\sf U}_2\mathbf{p}_2$, which is the case of strict NE. However, (b) if $\mathbf{p}_2\cdot {\sf U}_1\mathbf{q}_2=\mathbf{p}_1\cdot {\sf U}_1\mathbf{q}_2$ then $\mathbf{p}_2\cdot {\sf U}_1\mathbf{q}_1>\mathbf{p}_1\cdot {\sf U}_1\mathbf{q}_1$ does not ensure that all the trajectories starting in the neighbourhood of $(0,0)$ would flow towards $(0,0)$---some flows to other phase points on the line $x=0$. Hence, local asymptotic stability of $(0,0)$ is not guaranteed; the fixed point is at best Lyapunov stable, a weaker form of stability. Similar `weak' stability can be seen for the remaining two cases: (c) if $\mathbf{q}_2\cdot {\sf U}_2\mathbf{p}_2=\mathbf{q}_1\cdot {\sf U}_2\mathbf{p}_2$, then $\mathbf{q}_2\cdot {\sf U}_2\mathbf{p}_1>\mathbf{q}_1\cdot {\sf U}_2\mathbf{p}_1$; and (d) if $\mathbf{p}_2\cdot {\sf U}_1\mathbf{q}_2=\mathbf{p}_1\cdot {\sf U}_1\mathbf{q}_2$ and $\mathbf{q}_2\cdot {\sf U}_2\mathbf{p}_2=\mathbf{q}_1\cdot {\sf U}_2\mathbf{p}_2$, then $\mathbf{p}_2\cdot {\sf U}_1\mathbf{q}_1>\mathbf{p}_1\cdot {\sf U}_1\mathbf{q}_1$ and  $\mathbf{q}_2\cdot {\sf U}_2\mathbf{p}_1>\mathbf{q}_1\cdot {\sf U}_2\mathbf{p}_1$. This weak stability criteria is equivalent to---as one can easily see from state-strategy correspondence discussed above---condition (ii) and (iii) in Definition 1b (and-definition) which is, thus, also known as weak ESS~\cite{hofbauer_book}.
	
In summary, ESS under any of the three definitions is strongly stable only if it is a strict NE, since the mean population strategy converges towards $(\mathbf{p}_2,\mathbf{q}_2)$. However, aforementioned weak stability can only be guaranteed by the and-definition.

\subsection{Pure ESS}\label{sec:pure_bimatrix}
The pure ESS can be either strict NE or non-strict NE. Let's first discuss what happens when it is a strict NE.
	
	\subsubsection{Strict NE as ESS}\label{sec:pure_ESS_bimatrix}
	One can actually find that and-, sum-, and or- definitions are equivalent in the case ESS is strict NE; and linear stability analysis shows that being a strict NE guarantees the local asymptotic stability of the corresponding fixed point. However, since this exercise will be extended to more general asymmetric games---where both inter- and intra-specific interactions are considered---in Section~\ref{sec:asym_pure_strict}, we refrain from providing the proof right here to avoid repetition. The argument presented in Section~\ref{sec:asym_pure_strict} can be easily adapted to the case of bimatrix games.
	
	Here, because it is analytically doable for bimatrix games, let us go beyond linear stability analysis for a pure ESS $(\mathbf{\hat{x}},\mathbf{\hat{y}})$ that happens to be an isolated fixed point as well. Consider the following function:
	\begin{equation}\label{eqn:DKL_1}
		V_1(\mathbf{x},\mathbf{y})\equiv\sum_{i=1}^m\hat{x}_i\log\left(\frac{\hat{x}_i}{x_i}\right)+\sum_{i=1}^n\hat{y}_i\log\left(\frac{\hat{y}_i}{y_i}\right).
	\end{equation}
Realizing that the two sums in $V_1$ represent the Kullback--Leibler (KL) divergence~\cite{kullback1951, Cover2005} of  $\mathbf{\hat{x}}$ from $\mathbf{x}$ and the KL-divergence of  $\mathbf{\hat{y}}$ from $\mathbf{y}$, evidently $V_1$ must be positive except at $(\mathbf{\hat{x}},\mathbf{\hat{y}})$ where $V_1=0$. If we can additionally show that $\dot{V}_1<0$ $\forall(\mathbf{x},\mathbf{y})$ in some neighbourhood of $(\mathbf{\hat{x}},\mathbf{\hat{y}})$, then $V_1$ qualifies as Lyapunov function~\cite{jordan-smith,wiggins_2003} and the ESS must be locally asymptotically stable fixed point. To this end, on taking the time derivative, we get,
	\begin{equation}\label{eqn:r_term1}
		\dot{V_1}=-(\mathbf{\hat{x}}\cdot{\sf B}\mathbf{y}-\mathbf{x}\cdot{\sf B}\mathbf{y})-(\mathbf{\hat{y}}\cdot{\sf C}\mathbf{x}-\mathbf{y}\cdot{\sf C}\mathbf{x}),
	\end{equation}
	where we have used Eq.~(\ref{eqn:bimatrix_equation}). Of course, $\dot{V_1}<0$ when either inequalities in~(\ref{eqn:ess1_neighbourhood}) or inequality in~(\ref{eqn:ess2_neighbourhood}) is satisfied. Thus, any pure ESS that is an isolated fixed point, satisfying either Definition 1c or Definition 2c has $V_1$ as a Lyapunov function, ensuring that the corresponding fixed point is locally asymptotically stable. We mention that while a Lyapunov function for 2ESS (or-definition) is not known, the fixed point corresponding to the pure 2ESS can be shown to be locally asymptotically stable through alternative means (see Ref.~\cite{Cressman1996}).
	
\subsubsection{Non-strict pure NE as ESS}\label{sec:bimatrix_nuances}
\begin{figure}[h!]
	\centering
	\includegraphics[scale=0.8]{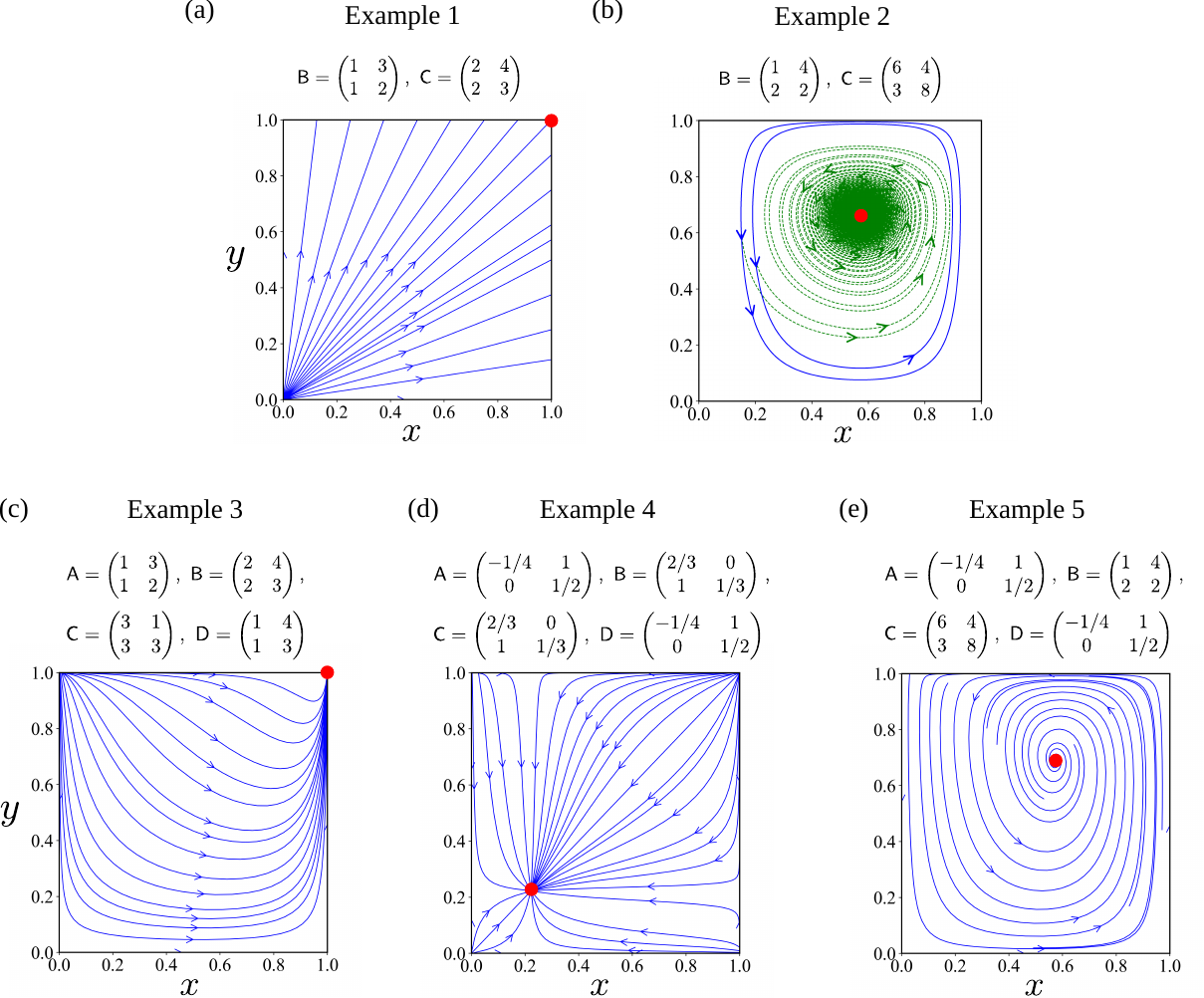}
	\caption{Phase portraits supplementing Examples 1--5 of the main text which should be referred for further details. The red circular marker indicates the ESS (fixed point) under focus. The blue lines are trajectories under replicator equation whereas the green dashed lines are trajectories under adjusted replicator equation.}
	\label{fig:battle_of_sexes}
\end{figure}
The case of pure ESS that is non-strict (or weak) NE is laden with subtleties, mainly because the corresponding fixed point happens not to be isolated. For instance, the equivalence between the two definitions (and- and sum-definitions) in the context of linear stability and Lyapunov function, discussed above, may be broken; after all, the linear stability analysis would fail as one eigenvalue should be zero and the Lyapunov function argument cannot be applicable since the fixed point is not isolated. Below we highlight such points through a simple example.
	\\
	\\
	\indent
	 \textbf{Example 1 (see Fig.~\ref{fig:battle_of_sexes}a):} \emph{ESS under and-definition need not be an ESS under or-definition if one adopts face-value perspective, but it is an ESS under or-definition with simultaneous deviation perspective. It is not an ESS} under sum-definition as well.

	Consider the following example:
	\begin{eqnarray}\label{eqn:weakESS_mat2}
		{\sf B} = \begin{pmatrix}
			b_{1} & b_{2} \\
			b_1 & b_{3} 
		\end{pmatrix}~\textrm{and}
		~{\sf C} = \begin{pmatrix}
			c_1 & c_2 \\
			c_1 & c_3 
		\end{pmatrix},
	\end{eqnarray}
	where $b_2>b_3$ and $c_2>c_3$. We can see that $(\hat{x},\hat{y})=(1,1)$ is a non-strict NE and we can again take neighbourhood state as $(\mathbf{x}, \mathbf{y})\equiv \left(\begin{pmatrix}
		1-\epsilon , \epsilon  \end{pmatrix},\begin{pmatrix} 1-\delta ,\delta  \end{pmatrix}\right)$. We find that $\mathbf{\hat{x}}\cdot{\sf B}\mathbf{{y}}-\mathbf{x}\cdot{\sf B}\mathbf{{y}}=(b_2-b_3)\epsilon\delta$ and $\mathbf{\hat{y}}\cdot{\sf C}\mathbf{x}-\mathbf{y}\cdot{\sf C}\mathbf{x}=(c_2-c_3)\epsilon\delta$ which are positive for infinitesimally small $\epsilon$ and $\delta$. Therefore, the NE \emph{satisfies and-definition} but it \emph{does not satisfy or-definition in face-value perspective} (since for $\epsilon=0$ both the terms become zero). However, \emph{adapting simultaneous deviation perspective} puts restrictions that  $\epsilon\neq0$ and $\delta\neq0$, thereby making it an \emph{ESS under or-definition}. Finally, to justify the non-equivalence of and- and sum-definitions, we calculate $\mathbf{\hat{x}}\cdot{\sf B}\mathbf{{y}}+\mathbf{\hat{y}}\cdot{\sf C}\mathbf{x}-\mathbf{x}\cdot{\sf B}\mathbf{{y}}-\mathbf{y}\cdot{\sf C}\mathbf{x}=(b_2-b_3)\epsilon\delta+(c_2-c_3)\epsilon\delta$ as one can see that the fixed point $(1,1)$ does not satisfy the sum-definition (for $\epsilon=0$ and $\delta\neq0$ the term becomes zero).   
		
To analyze the dynamical stability of the ESS, we cast the corresponding replicator equations as:  
\begin{align}
	\dot{x} &= (b_2 - b_3)x(1 - x)(1 - y), \\
	\dot{y} &= (c_2 - c_3)y(1 - y)(1 - x).
\end{align}
It is evident that both $\dot{x}$ and $\dot{y}$ are positive, indicating that the dynamics is driven away from $x=0$ and $y=0$ by any infinitesimal perturbation. Note that since $x=1$ and $y=1$ are line of fixed points, once a trajectory reaches at either boundary, $x=1$ or $y=1$, it terminates. Furthermore, since $\frac{dy}{dx} = \frac{k y}{x}$, where $k \equiv \frac{c_2 - c_3}{b_2 - b_3}$, one arrives at $y = K x^k$ ($K$ being an integration constant) as equation of trajectories. This result shows that \emph{only a single trajectory can terminate at $(1,1)$}; others terminate at other points on the lines $x = 1$ and $y = 1$. This implies the absence of an open neighbourhood of $(1,1)$ such that all the phase points in the neighbourhood be asymptotically attracted towards $(1,1)$. Consequently, the \emph{non-strict NE as ESS need not satisfy Proposition 1}.

\subsection{Mixed ESS}\label{sec:mixed_bim}
In the case of mixed states, the state-strategy correspondence---discussed in Section IIIA---shows that evolutionarily stable state $(\mathbf{\hat{x}},\mathbf{\hat{y}})$ corresponds to evolutionarily stable strategy $(\mathbf{\hat{p}},\mathbf{\hat{q}})$ such that $\mathbf{\hat{p}}=\hat{x}\mathbf{p}_1+(1-\hat{x})\mathbf{p}_2$ and $\mathbf{\hat{q}}=\hat{y}\mathbf{q}_1+(1-\hat{y})\mathbf{q}_2$. In this case, as well, a strongly stable $(\mathbf{\hat{p}},\mathbf{\hat{q}})$ would attract $(\sum_ix_i{\mathbf p}_i,\sum_iy_i{\mathbf q}_i)$, close enough to $(\hat{\mathbf p},\hat{\mathbf q})$. For mixed state case, $\mathbf{\hat{x}}\cdot{\sf B}\mathbf{\hat{y}}=\mathbf{x}\cdot{\sf B}\mathbf{\hat{y}}$ and $\mathbf{\hat{y}}\cdot{\sf C}\mathbf{\hat{x}}=\mathbf{y}\cdot{\sf C}\mathbf{\hat{x}}$. The former implies
\begin{equation}	
	\hat{x}(\hat{y}b_{11}+(1-\hat{y})b_{12})+(1-\hat{x})(\hat{y}b_{21}+(1-\hat{y})b_{22})=x(\hat{y}b_{11}+(1-\hat{y})b_{12})+(1-x)(\hat{y}b_{21}+(1-\hat{y})b_{22}).
\end{equation}
Substituting the matrix elements in terms of ${\sf U}_1$ and rearranging the resulting expression we arrive at $\mathbf{\hat{p}}\cdot {\sf U}_1\mathbf{\hat{q}}=\mathbf{p}\cdot {\sf U}_1\mathbf{\hat{q}}.$ Similarly, $\mathbf{\hat{y}}\cdot{\sf C}\mathbf{\hat{x}}=\mathbf{y}\cdot{\sf C}\mathbf{\hat{x}}$ implies	$\mathbf{\hat{q}}\cdot {\sf U}_2\mathbf{\hat{p}}=\mathbf{q}\cdot {\sf U}_2\mathbf{\hat{p}}$. In fact, similar manipulation of terms yield that $\mathbf{\hat{x}}\cdot{\sf B}\mathbf{y}>\mathbf{x}\cdot{\sf B}\mathbf{y}$ implies $\mathbf{\hat{p}}\cdot {\sf U}_1\mathbf{q}>\mathbf{p}\cdot {\sf U}_1\mathbf{q}$ and $\mathbf{\hat{y}}\cdot{\sf C}\mathbf{x}>\mathbf{y}\cdot{\sf C}\mathbf{x}$ implies $\mathbf{\hat{q}}\cdot {\sf U}_2\mathbf{p}>\mathbf{q}\cdot {\sf U}_2\mathbf{p}$. Summarizing, if Definition 1b is satisfied by the mixed state profile $(\mathbf{\hat{x}},\mathbf{\hat{y}})$, then the conditions of definition are also satisfied by the corresponding strategies $(\mathbf{\hat{p}},\mathbf{\hat{q}})$. Similarly, there is such correspondence for Definitions 2b and 3b as well.

Now we recall the well-known result~\cite{Schuster_1981,Selten1980} that ESS of Definitions 1c or 2c can never be mixed. There may, however, be mixed 2ESS---the ESS of Definition 3c (as shown below) under simultaneous deviation perspective. Since 2ESS must be NE which must be a fixed point of the replicator equation, let us assume that it corresponds to a fixed point $(\hat{x},\hat{y})$. A glance at replicator equation~(\ref{eqn:bimatrix_equation2}) reveals that with
$\alpha_1\equiv \mathbf{p}_2\cdot {\sf U}_1\mathbf{q}_2-\mathbf{p}_1\cdot {\sf U}_1\mathbf{q}_2$,~~$\beta_1\equiv \mathbf{p}_1\cdot {\sf U}_1\mathbf{q}_1-\mathbf{p}_2\cdot {\sf U}_1\mathbf{q}_1$,~~$\alpha_2\equiv \mathbf{q}_2\cdot {\sf U}_2\mathbf{p}_2-\mathbf{q}_1\cdot {\sf U}_2\mathbf{p}_2$~and~$\beta_2\equiv \mathbf{q}_1\cdot {\sf U}_2\mathbf{p}_1-\mathbf{q}_2\cdot {\sf U}_2\mathbf{p}_1$, the internal fixed point must be $(\hat{x},\hat{y})=\left({\alpha_2}/({\alpha_2+\beta_2}),{\alpha_1}/({\alpha_1+\beta_1})\right)$. The eigenvalues of Jacobian in linearization about this fixed points can be calculated to be $\pm\sqrt{({\alpha_1\beta_1\alpha_2\beta_2})/[{(\alpha_1+\beta_1)(\alpha_2+\beta_2)}]}$. Evidently, the fixed point can either be a center or a saddle. We know that (linear) center found using the linear stability analysis may not actually be a robust nonlinear center. However, it is known that the system is Hamiltonian~\cite{hofbauer_book} in the interior of the phase space, and hence any center in this case is, in fact, nonlinear center. Thus, ESS of any definition is not strongly stable. 

In conclusion, we find that Proposition 1 is invalid if the 2ESS is mixed. The proposition, however, can be saved by either adopting face-value perspective which allows for no mixed ESS or, even more interestingly, using another version of the replicator equation, viz., adjusted replicator equation~\cite{Mukhopadhyay2021, Maynard_Smith1982, hofbauer_book}:
\begin{subequations}\label{eqn:adjusted_replicator_equation}
	\begin{eqnarray}
		&&\dot{x}_i=x_i\frac{({\sf B}\mathbf{y})_i-\mathbf{x}\cdot{\sf B}\mathbf{y}}{\mathbf{x}\cdot{\sf B}\mathbf{y}},\\ 
		&&\dot{y}_i=y_i\frac{({\sf C}\mathbf{x})_i-\mathbf{y}\cdot{\sf C}\mathbf{x}}{\mathbf{y}\cdot{\sf C}\mathbf{x}}.
	\end{eqnarray}
\end{subequations}
Under this dynamics, the mixed state $(\mathbf{\hat{x}},\mathbf{\hat{y}})$ can be asymptotically stable (see Fig.~\ref{fig:battle_of_sexes}b).
\\
\\
	\textbf{Example 2 (see Fig.~\ref{fig:battle_of_sexes}b):} \emph{Mixed 2ESS under simultaneous deviation perspective is center-type fixed point in $2\times 2$ bimatrix games.}
	
	To simplify the calculation while keeping generality, we use two key facts: first, one can always modify the payoff matrix by adding a matrix with identical rows and by scaling the payoff matrices, which preserves both the dynamics and the ESS. Specifically, ${\sf B}$ and $c{\sf B}+{\sf \Omega}$ are equivalent, where $c$ is a positive scalar and ${\sf\Omega}_{ij}=\omega_j$. Second, using this transformation, one can show that any $2\times 2$ bimatrix game can be reduced to either a rescaled zero-sum game or a rescaled partnership game~\cite{hofbauer_book}. Therefore, we can work with the following payoff matrix:
\begin{equation}\label{eqn:payoff_matrices_example2}
	{\sf B} = \begin{pmatrix}
		\beta_1 & 0 \\
		0 & \alpha_1 
	\end{pmatrix}~\text{and}~
	{\sf C}=\begin{pmatrix}
	\beta_2 & 0 \\
		0 & \alpha_2 
	\end{pmatrix},
\end{equation}
where for rescaled partnership game, $\alpha_1=c\alpha_2$ and $\beta_1=c\beta_2$, on the other hand for rescaled zero-sum game $\alpha_1=-c\alpha_2$ and $\beta_1=-c\beta_2$~\cite{hofbauer_book}. Now, the internal fixed point---if it exist in the simplotope---is $(\mathbf{\hat{x}},\mathbf{\hat{y}})=\left(\begin{pmatrix}
	\alpha_2/(\alpha_2+\beta_2) , {\beta_2}/({\alpha_2+\beta_2})  \end{pmatrix},\begin{pmatrix} {\alpha_1}/({\alpha_1+\beta_1}) , {\beta_1}/({\alpha_1+\beta_1})  \end{pmatrix}\right)$ and the neighbourhood vector is $(\mathbf{x},\mathbf{y})=\left(\begin{pmatrix}
	\alpha_2/(\alpha_2+\beta_2)+\delta_x , {\beta_2}/({\alpha_2+\beta_2})-\delta_x  \end{pmatrix},\begin{pmatrix} {\alpha_1}/({\alpha_1+\beta_1})+\delta_y , {\beta_1}/({\alpha_1+\beta_1})-\delta_y \end{pmatrix}\right)$, where $\delta_x$ and $\delta_y$ are infinitesimally small which can have positive or negative sign but neither of them are allowed to be zero as per simultaneous deviation perspective. We find that $\mathbf{\hat{x}}\cdot{\sf B}\mathbf{{y}}-\mathbf{x}\cdot{\sf B}\mathbf{y}=-\delta_x\delta_y(\alpha_1 +\beta_1)$ and $\mathbf{\hat{y}}\cdot{\sf C}\mathbf{x}-\mathbf{y}\cdot{\sf C}\mathbf{x}=-\delta_x\delta_y(\alpha_2 +\beta_2)$. We can see that the fixed point cannot be an ESS under or-definition if $\alpha_1+\beta_1$ and $\alpha_2+\beta_2$ have the same sign, which is the case of rescaled partnership game. The \emph{fixed point is always an ESS under or-definition if they have different sign, this is the case of rescaled zero-sum game}. Next we note that the eigenvalues (of Jacobian during linearization about the fixed point) simplify to $\pm\sqrt{c}({\alpha_2\beta_2/(\alpha_2+\beta_2)})$ for rescaled partnership game and $\pm i\sqrt{c}({\alpha_2\beta_2/(\alpha_2+\beta_2)})$ for \emph{rescaled zero-sum game, which verifies the claim of ESS being a center}. 
	
Therefore, we conclude that mixed ESS under or-definition is possible, however it does not tell anything about asymptotic stability---an ESS can at best be Lyapunov stable as a center. This also tells us that unlike the case of strict NE, and-, sum- and or-definitions are not equivalent when it comes to mixed NE; in fact there is no scope of mixed ESS under and- and sum-definitions. Finally, of course, \emph{as per the face-value perspective the interior fixed point is not an ESS}, since $\delta_x=0$ makes $\mathbf{\hat{x}}\cdot{\sf B}\mathbf{{y}}-\mathbf{x}\cdot{\sf B}\mathbf{y}=\mathbf{\hat{y}}\cdot{\sf C}\mathbf{x}-\mathbf{y}\cdot{\sf C}\mathbf{x}=0$.

{\color{black}\subsection{Comparison of all the ESSes in bimatrix games}\label{sec:bimatrix:basket}
Let us now compare all the ESSes that may appear under different definitions and find how much one definition is stricter compared to another. We begin by recalling that, given any bimatrix game, its ESSes under any of the definitions must be NEs. One can partition the set of NEs into three sets:
$\mathbb{S},~
\mathbb{W}$~and~$
\mathbb{M}$ which are, respectively, the set of strict NEs, the set of weak pure NEs and the set of mixed NEs. If the set of all ESSes of the bimatrix game under a particular definition is denoted by $\mathbb{E}$, then the relationship between $\mathbb{S}$, $\mathbb{W}$, $\mathbb{M}$, and $\mathbb{E}$ (as shown through Venn diagrams in Fig.~\ref{fig:bimatrix_venn}) will clarify and justify all the arrows shown in Fig.~\ref{fig:ESS_summary}. \\

\begin{figure}[h!]
	\centering
	\includegraphics[scale=0.80]{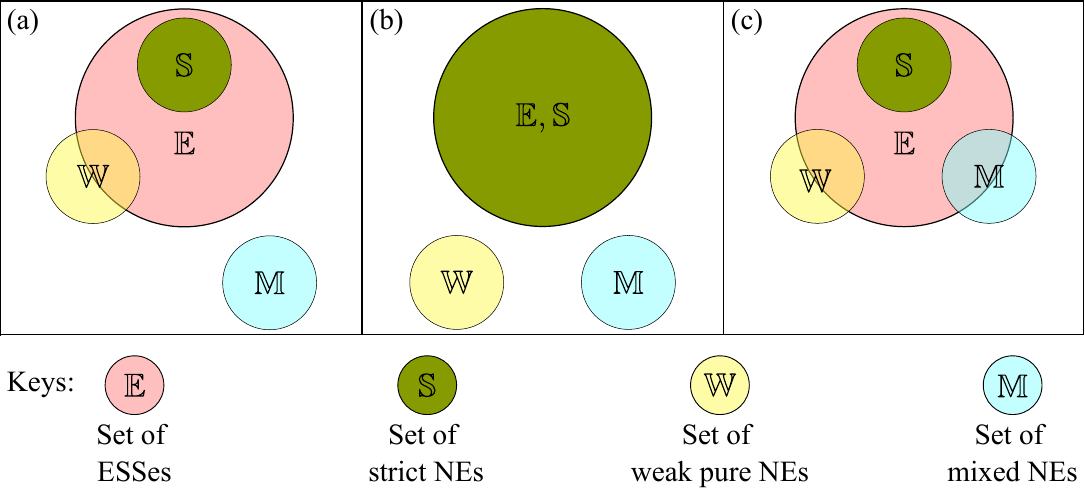}
	\caption{\color{black}Venn diagrams showing how the set of ESSes ($\mathbb{E}$) is related to the set of strict NEs ($\mathbb{S}$), the set of weak pure NEs ($\mathbb{W}$), and the set of mixed NEs ($\mathbb{M}$): Subfigure (a) corresponds to the unprimed and-definition of ESS. Subfigure (b) shows the Venn diagram for the primed and-definition, the sum-definition, and the face-value or-definition. Finally, subfigure (c) depicts the Venn diagram for the simultaneous-deviation or-definition of ESS.}
\label{fig:bimatrix_venn}
\end{figure}

\noindent\textbf{Unprimed and-definition} (Definition 1c in Section~\ref{sec:bimatrix_mbs}):
\begin{itemize}
	\item  \textbf{Strict NE:} Without loss of any generality, we consider the vector $(\mathbf{\hat{x}},\mathbf{\hat{y}})=((0,\cdots,0,1),(0,\cdots,0,1))$ which we test for its qualification as ESS.
	We write the arbitrary vector in the  neighborhood of this vector as $({\mathbf{x}},{\mathbf{y}})=((\delta_{x_1},\cdots,\delta_{x_{m-1}}, 1-\delta_{x_1}-\cdots-\delta_{x_{m-1}}),(\delta_{y_1},\cdots,\delta_{y_{n-1}}, 1-\delta_{y_1}-\cdots-\delta_{y_{n-1}}))$, where the $\delta$'s are infinitesimal and non-negative. Consider the following expressions: 
	\begin{subequations}\label{eqn:comparision}
		\begin{eqnarray}
			\mathbf{\hat{x}}\cdot{\sf B}\mathbf{y}-	\mathbf{x}\cdot{\sf B}\mathbf{y}&&=-\sum_{k=1}^{m-1}(b_{kn}-b_{mn})\delta_{x_k}-\sum_{k=1}^{m-1}\sum_{l=1}^{n-1}(b_{kl}-b_{ml}-b_{kn}+b_{mn})\delta_{x_k}\delta_{y_l},\\
			\mathbf{\hat{y}}\cdot{\sf C}\mathbf{x}-	\mathbf{y}\cdot{\sf C}\mathbf{x}&&=-\sum_{l=1}^{n-1}(c_{lm}-c_{nm})\delta_{y_l}-\sum_{l=1}^{n-1}\sum_{k=1}^{m-1}(c_{lk}-c_{nk}-c_{lm}+c_{nm})\delta_{x_k}\delta_{y_l}.
		\end{eqnarray}	
	\end{subequations}
For $(\mathbf{\hat{x}},\mathbf{\hat{y}})$ to be ESS, the unprimed and-definition requires $\mathbf{\hat{x}}\cdot{\sf B}\mathbf{y}-	\mathbf{x}\cdot{\sf B}\mathbf{y}>0$ and  $	\mathbf{\hat{y}}\cdot{\sf C}\mathbf{x}-	\mathbf{y}\cdot{\sf C}\mathbf{x}>0$ for neighboring ${\mathbf{x}} \neq \hat{\mathbf{x}}$ and $\mathbf{y} \neq \hat{\mathbf{y}}$. If $(\hat{\mathbf{x}},\hat{\mathbf{y}})=((0,\ldots,0,1),(0,\ldots,0,1))$ 
is a strict NE, then $b_{kn}-b_{mn}$ and $c_{lm}-c_{nm}$ are negative and the overall expression becomes positive to the leading order. Thus, $\mathbb{S} \subset \mathbb{E}$.

\item \textbf{Weak pure NE:} To argue this suppose that some of the differences $b_{kn}-b_{mn}$ are zero—for instance,
$b_{1n}-b_{mn}=0$—while the remaining $b_{kn}-b_{mn}$ terms are negative.  
Then $(\hat{\mathbf{x}},\hat{\mathbf{y}})=((0,\ldots,0,1),(0,\ldots,0,1))$ is a weak pure NE.  
In this situation, the coefficient of $\delta_{x_1}$ vanishes, but the coefficients of the 
other $\delta_{x_k}$'s remain positive. Thus, depending on the choice of neighbourhood, either the first order term is positive, or it vanishes in which case the exact value of $b_{1l}-b_{ml}-b_{1n}+b_{mn}$ decides whether the entire R.H.S. is positive. In other words, $\mathbb{E}\cap\mathbb{W}$ need not be empty.

\item \textbf{Mixed NE:} As discussed in Section~\ref{sec:bimatrix_mbs}, ESS under this definition can only be pure hence, $\mathbb{M} \not\subset \mathbb{E}$.
\end{itemize}
\noindent\textbf{Primed and-definition} (Definition 1b$^{\prime\prime}$ in Appendix~\ref{sec:alternate_defs_and}):
\begin{itemize}
	\item \textbf{Strict NE:} The next case is straightforward: The primed and-definition is exactly the definition of a 
	strict NE; therefore, $\mathbb{E}=\mathbb{S}$.
	\item \textbf{Weak pure NE:} From the above fact it trivially follows that $\mathbb{W} \not\subset \mathbb{E}$.
	\item \textbf{Mixed NE:} Similarly, $\mathbb{M} \not\subset \mathbb{E}$.
\end{itemize}

\noindent\textbf{Sum-definition} (Definition 2c in Section~\ref{sec:bimatrix_mbs}):
\begin{itemize}
	\item \textbf{Strict NE:} Following the notations used in Eq.~(\ref{eqn:comparision}), for $(\mathbf{\hat{x}},\mathbf{\hat{y}})$ to be ESS under the sum-definition
	$\mathbf{\hat{x}}\cdot{\sf B}\mathbf{y}+\mathbf{\hat{y}}\cdot{\sf C}\mathbf{x}-	\mathbf{x}\cdot{\sf B}\mathbf{y}-	\mathbf{y}\cdot{\sf C}\mathbf{x}>0$. Using arguments similar to given for the case of unprimed and-definition, one can conclude: $\mathbb{S} \subset \mathbb{E}$.
	
	\item \textbf{Weak pure NE:} Consider the following explicit expression:	\begin{eqnarray}\label{eqn:comparision_sum}
		\mathbf{\hat{x}}\cdot{\sf B}\mathbf{y}+\mathbf{\hat{y}}\cdot{\sf C}\mathbf{x}-	\mathbf{x}\cdot{\sf B}\mathbf{y}-	\mathbf{y}\cdot{\sf C}\mathbf{x}=-\sum_{k=1}^{m-1}(b_{kn}-b_{mn})\delta_{x_k}-\sum_{k=1}^{m-1}\sum_{l=1}^{n-1}(b_{kl}-b_{ml}-b_{kn}+b_{mn})\delta_{x_k}\delta_{y_l}\nonumber\\
		-\sum_{l=1}^{n-1}(c_{lm}-c_{nm})\delta_{y_l}-\sum_{l=1}^{n-1}\sum_{k=1}^{m-1}(c_{lk}-c_{nk}-c_{lm}+c_{nm})\delta_{x_k}\delta_{y_l}.
	\end{eqnarray}	
	By definition, at least one of the coefficients of $\delta$'s (in the leading order) must be zero. Therefore, a choice of neighbourhood in which the $\delta$'s corresponding to the vanishing coefficients, are non-zero renders the entire R.H.S. of  Eq.~(\ref{eqn:comparision_sum}) zero. Meaning, $\mathbb{W}\not\subset\mathbb{E}$.
	
	\item \textbf{Mixed NE:} As already discussed in Section~\ref{sec:bimatrix_mbs}, any ESS under sum-definition can only be pure NE; hence, there cannot be any mixed ESS under sum-definition. Consequently, $\mathbb{M}\not\subset\mathbb{E}$.

In view of the above relations--- $\mathbb{S}\subset \mathbb{E}$, $\mathbb{W}\not\subset\mathbb{E}$ and $\mathbb{M}\not\subset\mathbb{E}$, and the fact that ESS must be NE, we conclude that $\mathbb{E}=\mathbb{S}$. 
\end{itemize}

\noindent\textbf{Face-value or-definition} (Definition 3c in Section~\ref{sec:bimatrix_mbs}): With a view to avoid trivial repetition of analogous arguments given for the case of sum-definition, here we tersely mention the logical relations:
\begin{itemize}
	\item \textbf{Strict NE:} $\mathbb{S}\subset\mathbb{E}$.
	\item \textbf{Weak pure NE:} $\mathbb{W}\not\subset\mathbb{E}$.
	\item \textbf{Mixed NE:}  From Section~\ref{sec:equv_B}, mixed NE cannot be ESS under the face-value or-definition (which, we recall, motivated the requirement of the simultaneous-deviation perspective). So, $\mathbb{M}\not\subset\mathbb{E}$. \\
	Here also one can finally arrive at $\mathbb{E}=  \mathbb{S}$.
\end{itemize}

\noindent\textbf{Simultaneous deviation or-definition} (Definition 3c in Section~\ref{sec:bimatrix_mbs}):
\begin{itemize}
	\item \textbf{Strict NE:} Since for the strict NE, face-value and simultaneous perspectives coincide, $\mathbb{S}\subset\mathbb{E}$.

	\item \textbf{Weak pure NE:} For weak pure NE, the arguments follow along the line already seen in the case of the unprimed and-definition. Hence, $\mathbb{E}\cap\mathbb{W}$ need not be empty.
	\item \textbf{Mixed NE:} Recall from Section~\ref{sec:equv_B} that mixed NE may satisfy this definition: $\mathbb{M}\cap\mathbb{E}$ need not be empty.
	
\end{itemize}
}

\section{Asymmetric games}\label{sec:asymmetric}
It is now natural to ask whether this situation changes introducing some intra-specific interactions, i.e., whether Proposition 1 renders---with replicator equation modified to represent the resultant asymmetric game---true for a mixed ESS and makes it possible for a mixed NE to be an ESS under and- and sum-definitions (modified appropriately for asymmetric games). In the present section, we will discuss how adding intra-specific interactions addresses these questions and makes the concept of 2ESS more useful in the context of asymmetric games.

Due to the presence of intra-specific interaction, we must introduce two additional square payoff matrices---denoted as $\sf{A}$ and $\sf{D}$, respectively of sizes $m$ and $n$---accounting for the payoff gained by players of population X and players of population Y, respectively. The new payoff components are just the ones that one uses very commonly in symmetric games~\cite{Maynard_Smith1982}. Furthermore, rather than working with Definitions 1b, 2b, and 3b, it turns out to be more convenient to work with Definitions 1c, 2c, and 3c, which, however, must be modified now to include the intra-specific interactions. Consequently, we write the extensions---Definitions 1d, 2d, and 3d---of Definitions 1c, 2c, and 3c, respectively, as follows:
\begin{enumerate}
	\item[] \textbf{Definition 1d (and-definition)}: A state $(\mathbf{\hat{x}},\mathbf{\hat{y}})$ is an ESS if for every population state $(\mathbf{x},\mathbf{y})$ that are sufficiently close to  $(\mathbf{\hat{x}},\mathbf{\hat{y}})$ such that $\mathbf{{x}}\neq\mathbf{\hat{x}}$ and $\mathbf{{y}}\neq\mathbf{\hat{y}}$,
	\begin{equation}\label{eqn:essA1_neighbourhood}
		\mathbf{\hat{x}}\cdot{\sf A}\mathbf{x}+\mathbf{\hat{x}}\cdot{\sf B}\mathbf{y}>\mathbf{x}\cdot{\sf A}\mathbf{x}+\mathbf{x}\cdot{\sf B}\mathbf{y}~\text{\emph{and}}~\mathbf{\hat{y}}\cdot{\sf C}\mathbf{x}+\mathbf{\hat{y}}\cdot{\sf D}\mathbf{y}>\mathbf{y}\cdot{\sf C}\mathbf{x}+\mathbf{y}\cdot{\sf D}\mathbf{y}.
	\end{equation} 
	Taylor~\cite{Taylor_1979}, however, proposes that for $(\mathbf{\hat{x}}, \mathbf{\hat{y}})$ to be an ESS the total payoff of the resident against the mutant must be higher than that of the mutant. Mathematically, we write,
	\item[] \textbf{Definition 2d (sum-definition):} A state $(\mathbf{\hat{x}},\mathbf{\hat{y}})$ is an ESS if for every population state $(\mathbf{x},\mathbf{y})$ that are sufficiently close but not equal to $(\mathbf{\hat{x}},\mathbf{\hat{y}})$,
	\begin{equation}\label{eq:d2d}
		\mathbf{\hat{x}}\cdot{\sf A}\mathbf{x}+\mathbf{\hat{x}}\cdot{\sf B}\mathbf{y}+\mathbf{\hat{y}}\cdot{\sf C}\mathbf{x}+\mathbf{\hat{y}}\cdot{\sf D}\mathbf{y}>\mathbf{x}\cdot{\sf A}\mathbf{x}+\mathbf{x}\cdot{\sf B}\mathbf{y}+\mathbf{y}\cdot{\sf C}\mathbf{x}+\mathbf{y}\cdot{\sf D}\mathbf{y}.
	\end{equation}
	As discussed in Section~\ref{sectionII}, since the above definitions lead to only pure ESS, Cressman~\cite{cressman_1992,Cressman1996} gave a weaker definition of ESS for asymmetric games. According to this definition, an ESS is a state-pair ($\mathbf{\hat{x}}$,$\mathbf{\hat{y}}$) where, for any small perturbation, at least one of the states is more advantageous. Mathematically,
	\item[] \textbf{Definition 3d (or-definition):}  A state $(\mathbf{\hat{x}},\mathbf{\hat{y}})$ is an ESS if for every population state $(\mathbf{x},\mathbf{y})$ that are sufficiently close but not equal to $(\mathbf{\hat{x}},\mathbf{\hat{y}})$,
	\begin{equation}\label{eqn:ess_dfinition_neighbourhood}
		\text{either}~~\mathbf{\hat{x}}\cdot{\sf A}\mathbf{x}+\mathbf{\hat{x}}\cdot{\sf B}\mathbf{y}> \mathbf{x}\cdot{\sf A}\mathbf{x}+\mathbf{x}\cdot{\sf B}\mathbf{y}~~\text{\emph{or}}~~\mathbf{\hat{y}}\cdot{\sf C}\mathbf{x}+\mathbf{\hat{y}}\cdot{\sf D}\mathbf{y}> \mathbf{y}\cdot{\sf C}\mathbf{x}+\mathbf{y}\cdot{\sf D}\mathbf{y}.
	\end{equation} 
ESS under this definition is also called 2ESS as before. When both inter- and intra-specific interactions are simultaneously present, mixed 2ESS may appear even without the imposing of additional simultaneous deviation perspective, which thus is no longer adopted henceforth.
\end{enumerate}

{\color{black} The logical implications among the three definitions in asymmetric games (as shown in Fig.~\ref{fig:ESS_summary}) is rather clean. As we shall later illustrate in this section that each definition can admit strict NE, weak pure NE, and mixed NE as ESSes. It is obvious, by inspection, that the sum-definition (Definition~2d) implies the or-definition. Furthermore, using similar kind of arguments used in the the bimatrix case (Section~\ref{sec:bimatrix:basket}), it is not hard to establish that the and-definition implies the sum-definition; consequently, the and-definition implies the or-definition.}

Let us also, for clarity, reiterate Proposition 1 refereed as  Proposition 1$^\prime$ (cf. Thereom 4 in Ref.~\cite{Cressman_Tao_2014}, Theorem 4.5 and 4.6 in Ref.~\cite{Cressman1996} and Theorem 3.4.2 in Ref.~\cite{cressman_1992}) in the context of asymmetric games.
\begin{enumerate}
	\item[]\textbf{Proposition ${\bm 1^{\prime}}$:} An ESS is a locally asymptotically stable rest point of the following replicator equations,
	\begin{subequations}\label{eqn:asymmetric_rep1}
		\begin{eqnarray}
			&&\dot{x}_i=x_i\left(({\sf A}\mathbf{x})_i+({\sf B}\mathbf{y})_i-\mathbf{x}\cdot{\sf A}\mathbf{x}-\mathbf{x}\cdot{\sf B}\mathbf{y}\right), \label{eq:t1`a}\\ 
			&&\dot{y}_j=y_j\left(({\sf C}\mathbf{x})_j+({\sf D}\mathbf{y})_j-\mathbf{y}\cdot{\sf C}\mathbf{x}-\mathbf{y}\cdot{\sf D}\mathbf{y}\right).\label{eq:t1`b}
		\end{eqnarray}
	\end{subequations}
\end{enumerate}
We now turn to discussing pure and mixed ESSes separately in the context of asymmetric games.
\subsection{Pure ESS}\label{sec:asym_pure_strict}
In case of a strict NE being ESS, all the eigenvalues of the Jacobian about the corresponding fixed point of the replicator dynamics are  negative making the fixed point locally asymptotically stable under all the three definitions (Definitions 1d, 2d, and 3d). To demonstrate this, we go beyond the restriction of two strategies per subpopulation and address the matter in its full generality. We remind ourselves that these definitions are valid for any finite number of strategies, i.e., for arbitrary $m$ and $n$ values. In fact---as mentioned in Section~\ref{sec:pure_bimatrix}---even though our focus here is on asymmetric games, it formally extends our conclusion on bimatrix games for arbitrary $m$ and $n$ values---all one has to do is set ${\sf A}={\sf D}={\sf 0}$ in what follows.
\subsubsection{Strict NE as ESS}\label{sec:asymmetric_pure_ESS}
Let $a_{ij}$'s, $b_{ij}$'s, $c_{ij}$'s and $d_{ij}$'s denote the elements of the payoff matrices ${\sf A}$, ${\sf B}$, ${\sf C}$ and ${\sf D}$, respectively. Without any loss of generality, consider the state $(\mathbf{\hat{x}},\mathbf{\hat{y}})=((0,\cdots,0,1),(0,\cdots,0,1))$ which is a fixed point of the set of equations given by Eqs.~(\ref{eq:t1`a})~and~(\ref{eq:t1`b}). On doing linear stability analysis about it yields a \emph{diagonal} Jacobian matrix, and hence the diagonal elements can be easily read as the eigenvalues---say, $\lambda_1, \lambda_2,\cdots,\lambda_{m-1}, \lambda'_1, \lambda'_2,\cdots,\lambda'_{n-1}$---which respectively are 
$a_{1m}+b_{1n}-a_{mm}-b_{mn},a_{2m}+b_{2n}-a_{mm}-b_{mn},\cdots,a_{m-1,m}+b_{m-1,n}-a_{mm}-b_{mn},c_{1m}+d_{1n}-c_{nm}-d_{nn},c_{2m}+d_{2n}-c_{nm}-d_{nn},\cdots,c_{n-1,m}+d_{n-1,n}-c_{nm}-d_{nn}$. Here, comma has been put between subscripts wherever it is needed for clarity. It is clear that all the eigenvalues must be negative if the fixed point  $(\mathbf{\hat{x}},\mathbf{\hat{y}})=((0,\cdots,0,1),(0,\cdots,0,1))$ is a strict NE; thus, linear stability is automatically implied in such cases.

Now we would like to figure out whether the strict NE is an ESS. For this purpose, we write the arbitrary vector in the  neighbourhood of $(\mathbf{\hat{x}},\mathbf{\hat{y}})=((0,\cdots,0,1),(0,\cdots,0,1))$ as $({\mathbf{x}},{\mathbf{y}})=((\delta_{x_1},\cdots,\delta_{x_{m-1}}, 1-\delta_{x_1}-\cdots-\delta_{x_{m-1}}),(\delta_{y_1},\cdots,\delta_{y_{n-1}}, 1-\delta_{y_1}-\cdots-\delta_{y_{n-1}}))$, where the $\delta$'s are infinitesimal and non-negative. A glance at Definition 1d directs us to calculate the following terms, which we do up to first order in $\delta$'s: 
\begin{subequations}
	\begin{eqnarray}
		\mathbf{\hat{x}}\cdot{\sf A}\mathbf{x}&&=a_{mm}+(a_{m1}-a_{mm})\delta_{x_1}+(a_{m2}-a_{mm})\delta_{x_2}+\cdots+(a_{m,m-1}-a_{mm})\delta_{x_{m-1}},\\
		\mathbf{\hat{x}}\cdot{\sf B}\mathbf{y}&&=b_{mn}+(b_{m1}-b_{mn})\delta_{y_1}+(b_{m2}-b_{mn})\delta_{y_2}+\cdots+(b_{m,n-1}-b_{mn})\delta_{y_{n-1}},\\
		\mathbf{x}\cdot{\sf A}\mathbf{x}&&=a_{mm}+(a_{1m}-a_{mm})\delta_{x_1}+(a_{2m}-a_{mm})\delta_{x_2}+\cdots+(a_{m-1,m}-a_{mm})\delta_{x_{m-1}}\nonumber\\&&~~~~~+(a_{m1}-a_{mm})\delta_{x_1}+(a_{m2}-a_{mm})\delta_{x_2}+\cdots+(a_{m,m-1}-a_{mm})\delta_{x_{m-1}}\text{+ h.o.t},\\
		\mathbf{x}\cdot{\sf B}\mathbf{y}&&=b_{mn}+(b_{m1}-b_{mn})\delta_{y_1}+(b_{m2}-b_{mn})\delta_{y_2}+\cdots+(b_{m,n-1}-b_{mn})\delta_{y_{n-1}}\nonumber\\&&~~~~~+(b_{1n}-b_{mn})\delta_{x_1}+(b_{2n}-b_{mn})\delta_{x_2}+\cdots+(b_{m-1,n}-b_{mn})\delta_{x_{m-1}}\text{+ h.o.t}.
	\end{eqnarray}	
\end{subequations}
Therefore, recalling the forms of the eigenvalues, we find that 
\begin{eqnarray}
	\mathbf{\hat{x}}\cdot{\sf A}\mathbf{x}+\mathbf{\hat{x}}\cdot{\sf B}\mathbf{y}-\mathbf{x}\cdot{\sf A}\mathbf{x}-\mathbf{x}\cdot{\sf B}\mathbf{y}&=-\delta_{x_1}\lambda_1-\delta_{x_2}\lambda_2-\cdots-\delta_{x_{m-1}}\lambda_{m-1}\text{+ h.o.t}.\label{eqn:ess_ms1}
\end{eqnarray}
Similarly, we can calculate
\begin{eqnarray}
	\mathbf{\hat{y}}\cdot{\sf C}\mathbf{x}+\mathbf{\hat{y}}\cdot{\sf D}\mathbf{y}-\mathbf{y}\cdot{\sf C}\mathbf{x}-\mathbf{y}\cdot{\sf D}\mathbf{y}&=-\delta_{y_1}\lambda'_1-\delta_{y_2}\lambda'_2-\cdots-\delta_{y_{n-1}}\lambda'_{n-1}\text{+ h.o.t}.\label{eqn:ess_ms2}
\end{eqnarray}
We can see that the first $m-1$ eigenvalues, as the coefficient of the $\delta_{x_i}$'s, come with negative signs in Eq.~(\ref{eqn:ess_ms1}). Similarly, the last $n-1$ eigenvalues, as the coefficient of $\delta_{y_j}$'s, come with negative signs in Eq.~(\ref{eqn:ess_ms2}). If the fixed point is an ESS under and-definition then for both equations, the RHS's have to be positive in the entire neighbourhood. This requirement equivalently demands that all the eigenvalues  be negative because all the $\delta$'s are independent of each other and all are not allowed to be simultaneously zero.  This demand is automatically satisfied if the fixed point is strict NE.

We furthermore find that if we add Eq.~(\ref{eqn:ess_ms1}) and Eq.~(\ref{eqn:ess_ms2}) we get,
\begin{eqnarray}
	\mathbf{\hat{x}}\cdot{\sf A}\mathbf{x}+\mathbf{\hat{x}}\cdot{\sf B}\mathbf{y}+	\mathbf{\hat{y}}\cdot{\sf C}\mathbf{x}+\mathbf{\hat{y}}\cdot{\sf D}\mathbf{y}-\mathbf{x}\cdot{\sf A}\mathbf{x}-\mathbf{x}\cdot{\sf B}\mathbf{y}-\mathbf{y}\cdot{\sf C}\mathbf{x}-\mathbf{y}\cdot{\sf D}\mathbf{y}=-\sum_{i=1}^{m-1}\delta_{x_i}\lambda_{i}-\sum_{i=1}^{n-1}\delta_{y_i}\lambda'_i\text{+ h.o.t}.
\end{eqnarray}
The LHS has to be positive for the fixed point to be ESS under sum-definition. It is easy to see that exactly same argument as given above for the ESS under and-definition applies to this case. Hence, in the context of linear stability, and- and sum-definitions appear equivalent. 

Lastly, in the light of Eq.~(\ref{eqn:ess_ms1}) and Eq.~(\ref{eqn:ess_ms2}), or-definition implies 
\begin{eqnarray}\label{eqn:or_eig_val}
	\text{either }-\sum_{i=1}^{m-1}\delta_{x_i}\lambda_{i}>0~~\text{or}~~-\sum_{i=1}^{n-1}\delta_{y_i}\lambda'_i>0,\label{eq:col}
\end{eqnarray}
for the fixed point to be ESS. The validity of condition~(\ref{eqn:or_eig_val}) in the entire neighbourhood requires the negativity of all the eigenvalues, or equivalently, requires the fixed point to be strict NE. It might be tempting to assume that not all eigenvalues need to be negative for condition~(\ref{eqn:or_eig_val}) to hold, given the presence of conjunction or. However, it is important to recall that in Definition 3d, mutants in either or both subpopulations are allowed, as discussed in Section~\ref{sec:equv_B}. Therefore it is allowed to have either all $\delta_x$'s or all $\delta_y$'s to be zero rendering either the former or the latter inequality in condition~(\ref{eqn:or_eig_val}) invalid. As a result, the inequalities in~(\ref{eqn:or_eig_val}) imply that all eigenvalues must be negative. In conclusion, strict NE is 2ESS as well.

\subsubsection{Non-strict NE as ESS}
~\label{sec:weak_pure_asym}
The equivalence between ESSes under and-, sum- and or-definitions goes away if the ESS is non-strict NE. Following example highlights this.
\\
\\
\textbf{Example 3 (see Fig.~\ref{fig:battle_of_sexes}c):} \emph{Non-strict pure NE as ESS under or-definition need not be ESS under and- and sum-definitions.}

Consider
	\begin{equation}\label{eqn:payoff_matrices}
		{\sf A} = \begin{pmatrix}
			a_1 & a_2 \\
			a_1 & a_3 
		\end{pmatrix},
		~~{\sf B} = \begin{pmatrix}
			b_1 & b_2 \\
			b_1 & b_3
		\end{pmatrix},~~
		{\sf C}=\begin{pmatrix}
			c_1 & c_2 \\
			c_1 & c_3
		\end{pmatrix}~\text{and}
		~~{\sf D}=\begin{pmatrix}
			d_1 & d_2 \\
			d_1 & d_3 
		\end{pmatrix},
	\end{equation}
	where $a_2>a_3$, $b_2>b_3$, $c_2<c_3$ and $d_2>d_3$. We can see that $(\mathbf{\hat{x}},\mathbf{\hat{y}})=\left(\begin{pmatrix}
		1 ,0  \end{pmatrix},\begin{pmatrix} 1 ,0 \end{pmatrix}\right)$ is a non-strict NE whose neighbourhood state is given as $(\mathbf{x}, \mathbf{y})\equiv \left(\begin{pmatrix}
		1-\epsilon , \epsilon  \end{pmatrix},\begin{pmatrix} 1-\delta ,\delta  \end{pmatrix}\right)$. To check whether it is an ESS we calculate, $\mathbf{\hat{x}}\cdot{\sf A}\mathbf{x}+\mathbf{\hat{x}}\cdot{\sf B}\mathbf{y}-\mathbf{x}\cdot{\sf A}\mathbf{x}-\mathbf{x}\cdot{\sf B}\mathbf{y}=\epsilon\delta(b_2-b_3)+\epsilon^2(a_2-a_3)$ and $\mathbf{\hat{y}}\cdot{\sf C}\mathbf{x}+\mathbf{\hat{y}}\cdot{\sf D}\mathbf{y}-\mathbf{y}\cdot{\sf C}\mathbf{x}-\mathbf{y}\cdot{\sf D}\mathbf{y}=\epsilon\delta(c_2-c_3)+\delta^2(d_2-d_3)$. It is an \emph{ESS under or-definition} since at least one of them is positive for all allowed values of $\epsilon$ and $\delta$. However, it is \emph{not an ESS under and-definition} since $c_2-c_3<0$; furthermore, if $|c_2-c_3|>|b_2-b_3|$, then it \emph{may not satisfy the ESS condition under sum-definition}. The fixed point remains asymptotically stable as may be checked numerically, thus, \emph{it satisfies Proposition 1$^\prime$}.
\\
\\
	Before going to the case of mixed ESS, we would like to bring it to the fore that, for Definitions 1d and 2d, there are further arguments based on Lyapunov function for them to obey Proposition 1$^\prime$ in case of pure ESS corresponding to an isolated fixed point. One can show using the same Lyapunov function given in Eq.~(\ref{eqn:DKL_1}), that satisfying Definition 1d or 2d imply asymptotic stability for the pure ESS (corner fixed point of the simplotope). However, the general proof that Definition 3d imply asymptotic stability for pure ESS is not so straightforward: A proof using center manifold theory is available in the literature~\cite{Cressman1996}.

\subsection{Mixed ESS}\label{sec:mixed_nash_asym}
 The introduction of intra-specific interactions leads to circumventing certain restrictions present in bimatrix games. Specifically now and- and sum-definitions can lead to mixed ESS (see Example 4) and mixed ESS under or-definition can be locally asymptotically stable (see Example 5). While the former is illustrated in Example 4 below, the latter is easily seen on choosing $m=n=2$, in which case the form of internal fixed point, if it exists, is
\begin{equation}\label{eqn:asym_fixed_point}
	\left(\hat{x},\hat{y}\right)=\left(\frac{-a' d +b c'+b d' -b'd }{a d-bc}, \frac{-ac' -a d' +a'c+b' c}{a d-bc }\right).
\end{equation}
Here, $a\equiv a_{11}-a_{12}-a_{21}+a_{22}$,~$a'\equiv a_{12}-a_{22}$,~$b\equiv b_{11}-b_{12}-b_{21}+b_{22}$,~$b'\equiv b_{12}-b_{22}$,~$c\equiv c_{11}-c_{12}-c_{21}+c_{22}$,~$c'\equiv c_{12}-c_{22}$,~ $d\equiv d_{11}-d_{12}-d_{21}+d_{22}$ and $d'=d_{12}-d_{22}$. On linearization, the eigenvalues, say, $\lambda_-$ and $\lambda_+$, of the Jacobian matrix comes out as
\begin{equation}\label{eqn:eigenvalues}
	\lambda_{\pm}=\frac{(a\hat{x}(1-\hat{x})+ d\hat{y}(1-\hat{y}))\pm i\sqrt{4\hat{x}(1-\hat{x})\hat{y}(1-\hat{y})( ad-cb)-(a\hat{x}(1-\hat{x})+ d\hat{y}(1-\hat{y}))^2}}{2}.
\end{equation}
It is evident that now there is a possibility of focus or node types of fixed points under suitable conditions. For clear visualization, consider the example of asymmetric game given in (\ref{eqn:payoff_matrices_asym_2ess}), where change in the stability of the fixed point---due to addition of intra-specific interaction---is demonstrated through phase portraits as plotted in Fig.~\ref{fig:battle_of_sexes}(b) and Fig.~\ref{fig:battle_of_sexes}(e). Furthermore, Example 5 (see below) makes an important point clear that 2ESS may help find new evolutionarily important asymptotically stable states which may escape and- and sum-definitions.
\\
\\
\textbf{Example 4 (see Fig.~\ref{fig:battle_of_sexes}d):} \emph{Mixed ESS is possible under and- and sum-definitions in asymmetric games.}
	
Suppose that
\begin{equation}
	{\sf A} = \begin{pmatrix}
		-\frac{1}{4} & 1 \\
		0 & \frac{1}{2} 
	\end{pmatrix},
	~~{\sf B} = \begin{pmatrix}
		\frac{2}{3} & 0 \\
		1 & \frac{1}{3} 
	\end{pmatrix},~~
	{\sf C}=\begin{pmatrix}
		\frac{2}{3} & 0 \\
		1 & \frac{1}{3} 
	\end{pmatrix}~\text{and}
	~~{\sf D}=\begin{pmatrix}
		-\frac{1}{4} & 1 \\
		0 & \frac{1}{2} 
	\end{pmatrix}.
\end{equation}
One can easily check that the internal fixed point (mixed NE) in this case is $(\mathbf{\hat{x}},\mathbf{\hat{y}})=\left(\begin{pmatrix}
	2/9 , 7/9  \end{pmatrix},\begin{pmatrix} 2/9 ,7/9 \end{pmatrix}\right)$. Therefore the neighbourhood vector has the form---$(\mathbf{x},\mathbf{y})=\left(\begin{pmatrix}
	{2}/{9}+\delta_x , {7}/{9}-\delta_x  \end{pmatrix},\begin{pmatrix} {2}/{9}+\delta_y ,{7}/{9}-\delta_y \end{pmatrix}\right)$, where $\delta_x$ and $\delta_y$ are infinitesimally small and can be positive or negative but they can't be zero simultaneously. We calculate, $\mathbf{\hat{x}}\cdot{\sf A}\mathbf{x}+\mathbf{\hat{x}}\cdot{\sf B}\mathbf{y}-\mathbf{x}\cdot{\sf A}\mathbf{x}-\mathbf{x}\cdot{\sf B}\mathbf{y}={3 \delta_x^2}/{4}$ and $\mathbf{\hat{y}}\cdot{\sf C}\mathbf{x}+\mathbf{\hat{y}}\cdot{\sf D}\mathbf{y}-\mathbf{y}\cdot{\sf C}\mathbf{x}-\mathbf{y}\cdot{\sf D}\mathbf{y}={3 \delta_y^2}/{4}$. Both terms are positive for all infinitesimal values of $\delta_x$ and $\delta_y$, therefore it \emph{satisfies and-definition}. For sum-definition one just need to add both the terms and get, $\mathbf{\hat{x}}\cdot{\sf A}\mathbf{x}+\mathbf{\hat{x}}\cdot{\sf B}\mathbf{y}+\mathbf{\hat{y}}\cdot{\sf C}\mathbf{x}+\mathbf{\hat{y}}\cdot{\sf D}\mathbf{y}-\mathbf{x}\cdot{\sf A}\mathbf{x}+\mathbf{x}\cdot{\sf B}\mathbf{y}-\mathbf{y}\cdot{\sf C}\mathbf{x}+\mathbf{y}\cdot{\sf D}\mathbf{y}={3 (\delta_x^2+\delta_y^2)}/{4}$, \emph{which means sum-definition is also satisfied}. Fig.~\ref{fig:battle_of_sexes}(d) shows that the fixed point $(\mathbf{\hat{x}},\mathbf{\hat{y}})=\left(\begin{pmatrix} 2/9 , 7/9  \end{pmatrix},\begin{pmatrix} 2/9 ,7/9 \end{pmatrix}\right)$ is asymptotically stable, \emph{confirming Proposition 1$^\prime$ for and- and sum-definitions}.
\\
\\
\textbf{Example 5 (see Fig.~\ref{fig:battle_of_sexes}e):} \emph{Mixed 2ESS need not be ESS under and- and sum-definitions.}
	
Consider the payoff matrices,
\begin{equation}\label{eqn:payoff_matrices_asym_2ess}
	{\sf A} = \begin{pmatrix}
		-\frac{1}{4} & 1 \\
		0 & \frac{1}{2} 
	\end{pmatrix},
	~~{\sf B} = \begin{pmatrix}
		1 & 4 \\
		2 & 2 
	\end{pmatrix},~~
	{\sf C}=\begin{pmatrix}
		6 & 4 \\
		3 & 8 
	\end{pmatrix}~\text{and}
	~~{\sf D}=\begin{pmatrix}
		-\frac{1}{4} & 1 \\
		0 & \frac{1}{2} 
	\end{pmatrix}.
\end{equation}
Now, let see whether the internal fixed point is an ESS. One can calculate the internal fixed point  $(\mathbf{\hat{x}},\mathbf{\hat{y}})=\left(\begin{pmatrix}
	66/115 , 49/115  \end{pmatrix},\begin{pmatrix} 238/345 ,107/345 \end{pmatrix}\right)$, the neighbourhood vector therefore is,  $(\mathbf{x},\mathbf{y})=\left(\begin{pmatrix}
	\frac{66}{115}+\delta_x , \frac{49}{115}-\delta_x  \end{pmatrix},\begin{pmatrix} \frac{238}{345}+\delta_y ,\frac{238}{345}-\delta_y \end{pmatrix}\right)$, where $\delta_x$ and $\delta_y$ are infinitesimally small. We calculate, $\mathbf{\hat{x}}\cdot{\sf A}\mathbf{x}+\mathbf{\hat{x}}\cdot{\sf B}\mathbf{y}-\mathbf{x}\cdot{\sf A}\mathbf{x}-\mathbf{x}\cdot{\sf B}\mathbf{y}=3\delta_x\delta_y+\frac{3 \delta_x^2}{4}$ and $\mathbf{\hat{y}}\cdot{\sf C}\mathbf{x}+\mathbf{\hat{y}}\cdot{\sf D}\mathbf{y}-\mathbf{y}\cdot{\sf C}\mathbf{x}-\mathbf{y}\cdot{\sf D}\mathbf{y}=-7\delta_x\delta_y+\frac{3 \delta_y^2}{4}$. It is not very hard observe that at least one of the terms in the RHS of the two expressions is positive for all possible infinitesimal values of $\delta_x$ and  $\delta_y$, which means the internal fixed point is 2ESS---\emph{an ESS under or-definition}. Since both the terms cannot remain simultaneously positive in the entire neighbourhood, it is definitely \emph{not an ESS under and-definition}. To check for sum-definition we add the two terms and get, $\mathbf{\hat{x}}\cdot{\sf A}\mathbf{x}+\mathbf{\hat{x}}\cdot{\sf B}\mathbf{y}+\mathbf{\hat{y}}\cdot{\sf C}\mathbf{x}+\mathbf{\hat{y}}\cdot{\sf D}\mathbf{y}-\mathbf{x}\cdot{\sf A}\mathbf{x}-\mathbf{x}\cdot{\sf B}\mathbf{y}-\mathbf{y}\cdot{\sf C}\mathbf{x}-\mathbf{y}\cdot{\sf D}\mathbf{y}=-4\delta_x\delta_y+\frac{3 (\delta_x^2+\delta_y^2)}{4}$, which is not guaranteed to be always positive in the neighbourhood; hence it is \emph{not an ESS under sum-definition} as well. Fig.~\ref{fig:battle_of_sexes}(e) shows that the fixed point, a 2ESS, is asymptotically stable, \emph{confirming Proposition 1$^\prime$ for or-definition}.
	\\
	\\
The stability of the mixed ESS can be understood through Lyapunov function. Let us consider the general case of mixed ESS, $(\mathbf{\hat{x}},\mathbf{\hat{y}})$ when $m$ and $n$ are arbitrary, using the function given in Eq.~(\ref{eqn:DKL_1}). We calculate the time derivative using Eq.~(\ref{eqn:asymmetric_rep1}) and get,
\begin{equation}\label{eqn:r_term1}
	\dot{V_1}=-(\mathbf{\hat{x}}\cdot{\sf A}\mathbf{x}+\mathbf{\hat{x}}\cdot{\sf B}\mathbf{y}-\mathbf{x}\cdot{\sf A}\mathbf{x}-\mathbf{x}\cdot{\sf B}\mathbf{y})-(\mathbf{\hat{y}}\cdot{\sf C}\mathbf{x}+\mathbf{\hat{y}}\cdot{\sf D}\mathbf{y}-\mathbf{y}\cdot{\sf C}\mathbf{x}-\mathbf{y}\cdot{\sf D}\mathbf{y}),
\end{equation}
We can see that $\dot{V_1}<0$ $\forall ({\mathbf{x}},{\mathbf{y}})$ in some neighbourhood of $({\mathbf{\hat{x}}},{\mathbf{\hat{y}}})$ when either inequalities~(\ref{eqn:essA1_neighbourhood}) or inequalities~(\ref{eq:d2d}) are satisfied. Hence, any {mixed} ESS under either and-definition or sum-definition has $V_1$ as a Lyapunov function and the corresponding fixed point is locally asymptotically stable. Note that, this conclusion for mixed ESS does not carry over to bimatrix games (${\sf A}={\sf D}={\sf 0}$) since we know only pure ESS is possible there. As far as the or-definition is concerned  Theorem 1 (see the lemma in Appendix A of Ref.~\cite{Cressman1996} and the supplementary material of Ref.~\cite{Cressman_Tao_2014}) below comes to our help:
\begin{enumerate}
	\item[]\textbf{Theorem 1:} $(\mathbf{\hat{x}},\mathbf{\hat{y}})$ is a \emph{mixed} 2ESS if and only if there exists a positive $r$ (termed `weight factor') such that $\forall(\mathbf{x},\mathbf{y})\ne(\mathbf{\hat{x}},\mathbf{\hat{y}})$,
	\begin{equation}\label{r_definition}
		\mathbf{\hat{x}}\cdot{\sf A}\mathbf{x}+\mathbf{\hat{x}}\cdot{\sf B}\mathbf{y}-\mathbf{x}\cdot{\sf A}\mathbf{x}-\mathbf{x}\cdot{\sf B}\mathbf{y}+r(\mathbf{\hat{y}}\cdot{\sf D}\mathbf{y}+\mathbf{\hat{y}}\cdot{\sf C}\mathbf{x}-\mathbf{y}\cdot{\sf D}\mathbf{y}-\mathbf{y}\cdot{\sf C}\mathbf{x})>0.
	\end{equation}
\end{enumerate}
In other words, this condition is an equivalent qualifying condition for 2ESS defined in Definition 3d. Consequently, following similar steps given earlier, it is now straightforward to see that 
\begin{equation}\label{eqn:lyapunov_asym_DKL}
	V_r(\mathbf{x},\mathbf{y})\equiv\sum_{i=1}^m\hat{x}_i\log\left(\frac{\hat{x}_i}{x_i}\right)+r\sum_{i=1}^n\hat{y}_i\log\left(\frac{\hat{y}_i}{y_i}\right),
\end{equation}
qualifies as the Lyapunov function for a mixed 2ESS which, thus, is locally asymptotically stable fixed point. There is a subtlety to be noted: This existence of this Lyapunov function does not extend to bimatrix games (see Appendix~\ref{appendix_cressman_proof}) formed by setting ${\sf A}={\sf 0}$ and ${\sf D}={\sf 0}$.
\section{Information theoretical aspects}\label{section_information}\label{SectionVII}
The mention of KL-divergence (also, known as relative entropy) in the preceding sections is quite intriguing. The fact that a core concept of information theory is tied to the stability notion of ESS, reminds us of a known result in symmetric games in line with the general principle of minimum relative entropy~\cite{bomze1991cross,Karev2010,qian1991relative,floerchinger2020thermodynamics}: The KL-divergence of ESS from another time-evolving state monotonically decreases in symmetric games~\cite{baez_entropy}. In asymmetric games, however, the presence of multiple definitions of ESS (the and-, sum-, and or-definitions) necessitates different adaptations of the KL-divergence to establish the relationship between KL-divergence and ESS. Furthermore, since Theorem 1 applies only to mixed states, the information-theoretic connection of the or-definition is also valid solely for mixed states.

For completeness, let us recall that mathematically KL-divergence of a probability distribution with probability mass function (\textcolor{black}{PMF}) $p$ from another probability distribution ($q$ being the \textcolor{black}{PMF}) defined over the same random variable is expressed as
\begin{equation}\label{eqn:KL_divergence}
	D_{\rm KL}({p}||{q})=\sum_{i}p_i\log\left[ \frac{p_i}{q_i}\right],
\end{equation}
where $i$ belongs to the support of the \textcolor{black}{PMF}'s $p$ and $q$. The KL-divergence is a useful tool for quantifying the dissimilarity (or distance) between two probability distributions. In the context of population dynamics, it may be employed to measure the difference between two population states as they are nothing but frequencies which qualify as probabilities~\cite{chakraborty2024}.

What we have noted in the preceding section is that $D_{\rm KL}(\mathbf{\hat{x}}||\mathbf{{x}})+D_{\rm KL}(\mathbf{\hat{y}}||\mathbf{{y}})$ decreases monotonically as the system races towards ESS of and- and sum-definitions. However, what decreases in case of  \emph{mixed} 2ESS (or-definition) is $D_{\rm KL}(\mathbf{\hat{x}}||\mathbf{{x}})+rD_{\rm KL}(\mathbf{\hat{y}}||\mathbf{{y}})$ (for some $r>0$). In passing, we may note that an even stricter condition on KL-divergence is compatible with and-definition: It is obvious that the fixed point corresponding to ESS of and-definition, leads to monotonic decrease of $D_{\rm KL}(\mathbf{\hat{x}}||\mathbf{{x}})$ and $D_{\rm KL}(\mathbf{\hat{y}}||\mathbf{{y}})$ separately. 

We find the weight factor, $r$, interesting and would like to find some game-theoretic connection for it. Below, we establish such a connection at least for a specific category of games, which we call asymmetric partnership game. In an asymmetric partnership game, the symmetric component mirrors a partnership game (i.e., ${\sf A}^T={\sf A}$ and ${\sf D}^T={\sf D}$) and the bimatrix component of the asymmetric game mirrors a rescaled partnership game~\cite{hofbauer_book}, i.e.,  a game with ${\sf B}={\sf G} +{\sf \Omega}$ and ${\sf C}=c{\sf G}^T+{\sf\Omega'}$ for some ${\sf G}$ and $c>0$, where ${\sf \Omega}$ and ${\sf\Omega'}$ matrices are such ${\sf \Omega}_{ij}=\omega_j$ and ${\sf \Omega}'_{ij}=\omega'_j$, with $\omega_j,\omega'_j\in\mathbb{R}$. We intend to show that $r=c^{-1}$ in asymmetric partnership game. We accomplish this in two steps: In the first step, we determine the potential function whose Shahshahani gradient corresponds to asymmetric replicator dynamics; in the second step, we construct a Lyapunov function from this potential and demonstrate that the time derivative of this Lyapunov function is a negative quantity, implying that the associated fixed point happens to be asymptotically stable.

{\color{black} Before presenting the mathematical proof, let us briefly motivate the significance of the partnership game and its generalization to asymmetric games. In evolutionary game theory, a \emph{partnership game} represents a special class of interactions where the interests of the players are perfectly aligned—each player has an incentive to work together with the other. Consequently, the evolutionary dynamics can be derived from a single potential function~\cite{hofbauer_book,Weibull1997}. Mathematically, this corresponds to a symmetric payoff matrix satisfying ${\sf A}^T={\sf A}$, which ensures that the replicator dynamics are gradient of some potential and admits a Lyapunov function~\cite{hofbauer_book}. Biologically, such games describe cooperative interactions or mutually beneficial tasks in which both participants gain or lose in equal proportion, such as joint foraging, cooperative hunting, or other mutualistic behaviours.

One can generalize this setting to two-population interactions by introducing a scaling factor $c>0$ that quantifies the degree of coupling strength between the two interacting populations. The bimatrix components of the partnership game are written as ${\sf B}={\sf G}+{\sf\Omega}$ and ${\sf C}=c{\sf G}^T+{\sf\Omega}'$, where the parameter $c$ scales how strongly the payoff of one population depends on the strategy of the other. Thus, $c$ acts as a \emph{coupling parameter} between the two species: When $c>1$, the  X-player’s gain (or loss) from interacting with the other population is amplified, meaning that both players receive different payoffs while partnering on the same task. Biologically, this represents an \emph{unequal partnership}, where one population experiences a stronger effect of the same encounter—examples include parasite–host systems or worker–queen asymmetries in eusocial colonies.
When $c=1$, the system reduces to a \emph{true partnership game}, restoring full symmetry, where both sides experience the same magnitude of influence.
When $0<c<1$, the situation is reversed; the other player now experiences a comparatively weaker coupling.
Therefore, the parameter $c$ provides a continuous measure of interaction asymmetry, smoothly interpolating between equal and unequal partnerships, and allows one to analyze how deviations from perfect symmetry influence the stability and evolution of cooperative behaviour.

The next question one can ask is: What is the significance of the parameter $r$? 
While defining evolutionary stability for a two-species system, one must compare payoff deviations in both populations on an equal footing. 
However, these deviations may naturally occur on different scales---selection intensities, payoff magnitudes, or update rates can all be asymmetric between species. 
The weight factor $r>0$ in the mixed 2ESS condition~(\ref{r_definition}) compensates for these scale differences by balancing the X-population’s payoff perturbation against that of the Y-population. 
Formally, the $r$-weighted inequality renders the stability test invariant to differences in selection intensity across the two populations. 
Biologically, one may interpret $r$ as a measure of the \emph{relative evolutionary importance} of the species---reflecting differences in effective population size, selection strength, or adaptation rate. 

In the specific case of a partnership game, the cross-effect of the $X$-population on the $Y$-population is multiplied by a coupling parameter $c>0$. 
The $r$-weighted test then depends on the \emph{product} $r c$ [put the relation ${\sf B} \sim\frac{1}{c}{\sf C}^T$ in~(\ref{r_definition})]; maintaining the same stability balance therefore requires $r \propto 1/c$. 
This provides the natural intuition behind the mapping $r = 1/c$ proved below: stronger payoff coupling (an increase in $c$) must be offset by a proportionally smaller evolutionary weight on that side (a decrease in $r$), thereby preserving the neutrality of the payoff comparison while taking about evolutionary stability.

}

We begin by putting forward the potential function whose Shahshahani gradient~\cite{shahshahani} yields the asymmetric replicator dynamics (Eq.~(\ref{eqn:asymmetric_rep1})). First, we define an inner product in the tangent space: 
\begin{equation}\label{eqn:c_gradient}
	\langle(\bm{\xi}',\bm{\eta}'),(\bm{\xi},\bm{\eta})\rangle=\sum_{i=1}^{m}\frac{\xi_i'\xi_i}{x_i}+\frac{1}{c}\sum_{i=1}^{n}\frac{\eta_i'\eta_i}{y_i},
\end{equation}
where $\bm{\xi}'$~and~$\bm{\xi}$ are tangent vectors tangent to $\Sigma_m$ whereas $\bm{\eta}'$ and $\bm{\eta}$ are tangent vectors tangent to $\Sigma_n$. To write the metric involved in the inner product explicitly, we define $\bm{\zeta}'=(\bm{\xi}',\bm{\eta}')$ ~and~ $\bm{\zeta}=(\bm{\xi},\bm{\eta})$, where the first $m$ components are $x$-components and form $m+1$ to $m+n$ are $y$-components. Under the coordinate transformation, $z_i=x_i$ for $i$ from $1$ to $m$ and $z_i=cy_{i-m}$ for $i$ from $m+1$ to $n+m$, we may express the metric as,
\begin{equation}\label{eqn:shahshahani_modified}
	g_{ij}(z)=\frac{\delta_{ij}}{z_i},
\end{equation}
which is the Shahshahani metric. Moreover, we can recast Eq.~(\ref{eqn:c_gradient}) as, 
\begin{equation}\label{eqn:c_gradient3}
	\langle\bm{\zeta}',\bm{\zeta}\rangle=\sum_{i=1}^{m+n}\frac{\zeta'_i\zeta_i}{z_i}=\sum_{i,j=1}^{m+n}g_{ij}\zeta'_i\zeta_j.
\end{equation}
We would like to make a passing remark here that a well-celebrated concept in the theory of statistical inference, viz., Fisher information~\cite{fisher_1922} metric (${\sf J}$)---which represents a metric of distance between two nearby probability distributions---is related to the Shahshahani metric: In symmetric games, by mapping the probability mass function $(p(1),p(2),\cdots,p(n))$ to $(x_1,x_2,\cdots,x_n)$, we can easily show~\cite{harper2009information} that the Fisher information metric gets converted into Shahshahani metric:
\begin{equation}\label{eqn:shahshahani_1}
{\sf J}_{ij}\equiv\mathbb{E}\left[\frac{\partial \log p}{\partial x_i}\frac{\partial \log p}{\partial x_j}\right]\to\mathbb{E}\left[\frac{\partial \log x}{\partial x_i}\frac{\partial \log x}{\partial x_j}\right]=\sum_kx_k\frac{\delta_{ik}}{x_k}\frac{\delta_{jk}}{x_j}=\frac{\delta_{ij}}{x_i}=g_{ij}.
\end{equation} 
This relation is true even in asymmetric games, where under a similar mapping, $1/(1+c)$-scaled Fisher information metric becomes Shahshahani metric.

Our aim is to search for a potential $U$ such that the following is true (a potential whose Shahshahani gradient is the replicator equation):
\begin{equation}\label{eqn:gradient}
	\langle(\dot{\mathbf{x}},\dot{\mathbf{y}}),(\bm{\xi},\bm{\eta})\rangle=\sum_{i=1}^m\frac{\partial U}{\partial x_i}\xi_i+\sum_{i=1}^n\frac{\partial U}{\partial y_i}\eta_i.
\end{equation}
We calculate the L.H.S. of this equation, using Eq.~(\ref{eqn:c_gradient}) and replicator equation, Eq.~(\ref{eqn:asymmetric_rep1}):
\begin{equation}\label{eqn:shahshahani1}
	\langle(\dot{\mathbf{x}},\dot{\mathbf{y}}),(\bm{\xi},\bm{\eta})\rangle=\sum_{i=1}^m\xi_i[({\sf A}\mathbf{x})_i+({\sf B}\mathbf{y})_i]+\frac{1}{c}\sum_{i=1}^n\eta_i[({\sf C}\mathbf{x})_i+({\sf D}\mathbf{y})_i],
\end{equation}
since $\bm{\xi}$ and $\bm{\eta}$, being tangent vectors, satisfy $\sum_i\xi_i=0$ and $\sum_i\eta_i=0$, respectively. 
As far as the R.H.S. of Eq.~(\ref{eqn:shahshahani1}) is concerned, one can easily verify that it is the R.H.S of Eq.~(\ref{eqn:gradient}) with the function:
\begin{equation}
	U\equiv\frac{1}{2}\left[\mathbf{x}\cdot{\sf A}\mathbf{x}+\mathbf{x}\cdot{\sf B}\mathbf{y}-\mathbf{x}\cdot{\sf\Omega}\mathbf{y}+\frac{1}{c}\mathbf{y}\cdot{\sf D}\mathbf{y}+\frac{1}{c}\mathbf{y}\cdot{\sf C}\mathbf{x}-\frac{1}{c}\mathbf{y}\cdot{\sf\Omega'}\mathbf{x} \right].
\end{equation}
So $U$ is the potential whose Shahshahani gradient is the replicator dynamics.
To arrive at this conclusion, we must invoke the properties of ${\sf\Omega}$ and ${\sf\Omega}'$ that $\sum_i\xi_i({\sf\Omega} \mathbf{y})_i=0$ and $\sum_i\eta_i({\sf\Omega}' \mathbf{x})_i=0$ owing to the respective facts that $({\sf\Omega} \mathbf{y})_i$ and $({\sf\Omega}' \mathbf{x})_i$ are independent of the index $i$.

In the next step, we construct a Lyapunov function $V$ from the potential $U$. Let $V(\mathbf{x},\mathbf{y})=U(\mathbf{\hat{x}},\mathbf{\hat{y}})-U(\mathbf{x},\mathbf{y})$. If $(\mathbf{\hat{x}},\mathbf{\hat{y}})$ is the point where $U$ takes the maximum value, then $V(\mathbf{x},\mathbf{y})>0~ \forall(\mathbf{x},\mathbf{y})\ne(\mathbf{\hat{x}},\mathbf{\hat{y}})$ and at $(\mathbf{\hat{x}},\mathbf{\hat{y}})$, $V$ takes the minimum value which is $0$. Now an extremum of $V$ is attained at a point where
\begin{subequations}
\begin{eqnarray}
\frac{\partial V}{\partial x_i}=0\implies({\sf A}\mathbf{x})_i+({\sf B}\mathbf{y})_i=\mathbf{x}\cdot{\sf A}\mathbf{x}+\mathbf{x}\cdot{\sf B}\mathbf{y},\\
\frac{\partial V}{\partial y_i}=0\implies({\sf C}\mathbf{x})_i+({\sf D}\mathbf{y})_i=\mathbf{y}\cdot{\sf C}\mathbf{x}+\mathbf{y}\cdot{\sf D}\mathbf{y}.
\end{eqnarray}
\end{subequations}
This means that the extremum $(\mathbf{\hat{x}},\mathbf{\hat{y}})$ is nothing but the internal fixed point of the replicator dynamics~(\ref{eqn:asymmetric_rep1}). For this fixed point to be minimum, the Hessian matrix,
\begin{eqnarray}
	\hspace{-0.3cm}{\sf H}=\begin{bmatrix}
		\left[\frac{\partial^2 V}{\partial x_i\partial x_j}\right]_{(m-1)\times (m-1)}&\left[\frac{\partial^2 V}{\partial x_i\partial y_j}\right]_{(m-1)\times (n-1)}\\
		\left[\frac{\partial^2 V}{\partial y_i\partial x_j}\right]_{(n-1)\times (m-1)}&\left[\frac{\partial^2 V}{\partial y_i\partial y_j}\right]_{(n-1)\times (n-1)}
	\end{bmatrix}=
	\begin{bmatrix}
	\left[-( a_{ij}- a_{im}- a_{mj}+ a_{mm})\right]&\left[-( b_{ij}- b_{in}- b_{mj}+ b_{mn})\right]\\
	\left[-\frac{1}{c}( c_{ij}- c_{im}- c_{nj}+ c_{nm})\right]&	\left[-\frac{1}{c}( d_{ij}- d_{in}- d_{nj}+ d_{nn})\right]
	\end{bmatrix}
\end{eqnarray}
has to be positive definite. Here, as before, we have used $a_{ij}$'s, $b_{ij}$'s, $c_{ij}$'s and $d_{ij}$'s as elements of the payoff matrices ${\sf A}$, ${\sf B}$, ${\sf C}$ and ${\sf D}$, respectively. Using the relation between ${\sf B}$ and ${\sf C}$ (i.e., ${\sf C}=c{\sf B}^T-c{\sf\Omega}^T+{\sf \Omega}'$), one can show that,
\begin{equation}
{\sf H }=\begin{bmatrix}
	\left[-( a_{ij}- a_{im}- a_{mj}+ a_{mm})\right]&\left[-( b_{ij}- b_{in}- b_{mj}+ b_{mn})\right]\\
\left[-( b_{ji}- b_{mi}- b_{jn}+ b_{mn})\right]&	\left[-\frac{1}{c}( d_{ij}- d_{in}- d_{nj}+ d_{nn})\right]
\end{bmatrix}.
\end{equation}
We can see that ${\sf H}$ is a symmetric matrix which if positive definite makes the extremum at $(\mathbf{\hat{x}},\mathbf{\hat{y}})$ a minimum~\cite{Kuttle_2008}. Supposing a minimum  exists at $(\mathbf{\hat{x}},\mathbf{\hat{y}})$, we evaluate the time derivative of $V$ to be
\begin{equation}\label{eqn:c_gradient2}
	\dot{V}=-\sum_{i=1}^m\dot{x}_i[({\sf A}\mathbf{x})_i+({\sf B}\mathbf{y})_i]-\frac{1}{c}\sum_{i=1}^n\dot{y}_i[({\sf C}\mathbf{x})_i+({\sf D}\mathbf{y})_i].
\end{equation}
Here we have used the following to get Eq.~(\ref{eqn:c_gradient2}): $\frac{d}{dt}(\mathbf{x}\cdot{\sf A}\mathbf{x})=2\sum_i\dot{x}_i({\sf A}\mathbf{x})_i$, $\frac{d}{dt}(\mathbf{x}\cdot{\sf B}\mathbf{y})=\sum_i\dot{x}_i({\sf B}\mathbf{y})_i+\frac{1}{c}\sum_i\dot{y}_i({\sf C}\mathbf{x})_i+\sum_i\dot{y}_i({\Omega}^T\mathbf{x})_i$,~ $\frac{d}{dt}(\frac{\mathbf{y}\cdot{\sf C}\mathbf{x}}{c})=\frac{1}{c}\sum_i\dot{y}_i({\sf C}\mathbf{x})_i+\sum_i\dot{x}_i({\sf B}\mathbf{y})_i+\frac{1}{c}\sum_i\dot{x}_i({\sf \Omega'}^T\mathbf{y})_i$ and $\frac{d}{dt}(\mathbf{y}\cdot{\sf D}\mathbf{y})=2\sum_i\dot{y}_i({\sf D}\mathbf{y})_i$. Furthermore, using the replicator equation Eq.~(\ref{eqn:asymmetric_rep1}), we get,
\begin{equation}
	\frac{dV}{dt}=-\sum_{i=1}^mx_i[({\sf A}\mathbf{x})_i+({\sf B}\mathbf{y})_i-\mathbf{x}\cdot{\sf A}\mathbf{x}-\mathbf{x}\cdot{\sf B}\mathbf{y}]^2-\frac{1}{c}\sum_{i=1}^ny_i[({\sf C}\mathbf{x})_i+({\sf D}\mathbf{y})_i-\mathbf{y}\cdot{\sf C}\mathbf{x}-\mathbf{y}\cdot{\sf D}\mathbf{y}]^2.
\end{equation}
This implies that $\dot{V}<0$ except at $(\mathbf{\hat{x}},\mathbf{\hat{y}})$ where $\dot{V}=0$. Consequently, $V$ serves as the Lyapunov function indicating that $(\mathbf{\hat{x}},\mathbf{\hat{y}})$ is locally asymptotically stable.

Since $V(\mathbf{\hat{x}},\mathbf{\hat{y}})<V(\mathbf{x},\mathbf{y})$, by definition, 
\begin{equation}
	\mathbf{\hat{x}}\cdot{\sf A}\mathbf{\hat{x}}+\mathbf{\hat{x}}\cdot{\sf B}\mathbf{\hat{y}}-\mathbf{\hat{x}}\cdot{\sf\Omega}\mathbf{\hat{y}}+\frac{1}{c}\hat{\mathbf{y}}\cdot{\sf D}\mathbf{\hat{y}}+\frac{1}{c}\mathbf{\hat{y}}\cdot{\sf C}\mathbf{\hat{x}}-\frac{1}{c}\mathbf{\hat{y}}\cdot{\sf\Omega'}\mathbf{\hat{x}}>\mathbf{x}\cdot{\sf A}\mathbf{x}+\mathbf{x}\cdot{\sf B}\mathbf{y}-\mathbf{x}\cdot{\sf\Omega}\mathbf{y}+\frac{1}{c}\mathbf{y}\cdot{\sf D}\mathbf{y}+\frac{1}{c}\mathbf{y}\cdot{\sf C}\mathbf{x}-\frac{1}{c}\mathbf{y}\cdot{\sf\Omega'}\mathbf{x},
\end{equation}
for every $(\mathbf{x},\mathbf{y})\neq(\mathbf{\hat{x}},\mathbf{\hat{y}})$ in an appropriate neighbourhood of $(\mathbf{\hat{x}},\mathbf{\hat{y}})$. Now replacing ($\mathbf{x}$, $\mathbf{y}$) by $(2\mathbf{x}-\mathbf{\hat{x}},2\mathbf{y}-\mathbf{\hat{y}}$), which is also in the neighbourhood of ($\mathbf{\hat{x}},\mathbf{\hat{y}}$) (see Ref.~\cite{hofbauer_book}), we get,
\begin{equation}
	\mathbf{\hat{x}}\cdot{\sf A}\mathbf{x}+\mathbf{\hat{x}}\cdot{\sf B}\mathbf{y}-\mathbf{x}\cdot{\sf A}\mathbf{x}-\mathbf{x}\cdot{\sf B}\mathbf{y}+\frac{1}{c}(\mathbf{\hat{y}}\cdot{\sf D}\mathbf{y}+\mathbf{\hat{y}}\cdot{\sf C}\mathbf{x}-\mathbf{y}\cdot{\sf D}\mathbf{y}-\mathbf{y}\cdot{\sf C}\mathbf{x})>0.
\end{equation}
On comparing this expression with Eq.~(\ref{r_definition}), one concludes that $r=1/c$---what we aimed to prove.

\section{ESS in Multiplayer hypermatrix games}\label{SectionVI}
In the preceding sections, we delved into the definitions of ESS and elucidated their respective domains of applicability, i.e., we clarified which definitions are useful for characterizing pure ESS and which are useful for mixed ESS, all within the context of two-player interactions. However, what if the strategic interaction involves more than two players? In such scenarios, we must shift our focus to the framework of multiplayer games~\cite{Broom1997,Broom2022,Gokhale2010,Palm1984,Gokhale2014,Broom2016,Omidshafiei2020,Couto2023}. Within the realm of multiplayer games, the questions might arise what is the definition of ESS? Are there multiple definitions of ESS, and what are their respective domains of applicability? In this section, we address these questions. Multiplayer games can be somewhat intricate. For the purpose of this paper, it suffices to focus---unless otherwise specified---on \emph{multiplayer hypermatrix games} where the payoffs are linear in their respective arguments~\cite{Broom2022}.

In multiplayer games, the payoff hypermatrix depends on multiple arguments. We denote by ${\sf U}[\mathbf{p}_1;\mathbf{p}_2,\cdots,\mathbf{p}_\nu]$ the payoff of a player with strategy $\mathbf{p}_1$ interacting with other $\nu-1$ players with different strategies. If $d_i$ opponents are of same strategy, $\mathbf{p}_i$ ($1<i\le\nu$), we compactly denote the payoff as ${\sf U}[\mathbf{p}_1;\mathbf{p}_2^{d_2},\cdots,\mathbf{p}_\nu^{d_\nu}]$; furthermore, if $d_i=1$, we suppress the superscript and $d_i=0$ means the absence of an opponent employing strategy $\mathbf{p}_i$.   

In order to get familiarized with the notations to be used in this section, let's consider the scenario of symmetric hypermatrix games with two strategies, $\mathbf{\hat{p}}$ and $\mathbf{p}$, both belong to some $\Sigma_N$, in the population. By definition, the game is symmetrical among all players, meaning that the order of the players is irrelevant and the set of strategies available is the same for all players.  In case all the interacting players adopt the same mixed strategy, $\mathbf{\hat{p}}$, the payoff should notationaly be ${\sf U}[\mathbf{\hat{p}};\mathbf{\hat{p}}^{\nu-1}]$. For $\mathbf{\hat{p}}$ to be an Evolutionary Stable Strategy (ESS), it should be advantageous against small mutations of type $\mathbf{p}$. Let $\epsilon$ represents the frequency of players choosing $\mathbf{p}$ and $1-\epsilon$ denotes the frequency of players choosing $\mathbf{\hat{p}}$. In other words,  a strategy $\mathbf{\hat{p}}$ is an ESS if for any mutant strategy $\mathbf{p}$,~$\mathbf{p}\neq\mathbf{\hat{p}}$, there exists an invasion barrier $\epsilon_{\mathbf{p}}>0$, such that $\forall\epsilon$, with $0<\epsilon<\epsilon_{\mathbf{p}}$,
	\begin{equation}\label{eqn:multi_ess1}
		{\sf U}\left[\mathbf{\hat{p}};(\epsilon\mathbf{p}+(1-\epsilon)\mathbf{\hat{p}})^{\nu-1}\right]>{\sf U}\left[\mathbf{p};(\epsilon\mathbf{p}+(1-\epsilon)\mathbf{\hat{p}})^{\nu-1}\right].
	\end{equation}
Since the payoff is linear in arguments, we recognize that
\begin{equation}\label{eqn:ess_expansion}
	{\sf U}\left[\mathbf{\hat{p}};(\epsilon\mathbf{p}+(1-\epsilon)\mathbf{\hat{p}})^{\nu-1}\right]=\sum_{k=0}^{\nu-1}\binom{\nu-1}{k}\epsilon^k(1-\epsilon)^{\nu-k-1}{\sf U}[\mathbf{\hat{p}};\mathbf{p}^k,\mathbf{\hat{p}}^{\nu-k-1}].
\end{equation}

Introducing a short hand notation for average population strategy as $\tilde{\mathbf{p}}=\epsilon\mathbf{p}+(1-\epsilon)\mathbf{\hat{p}}$,  inequality (\ref{eqn:multi_ess1}) can be rewritten as, ${\sf U}\left[\mathbf{\hat{p}};\tilde{\mathbf{p}}^{\nu-1}\right]>{\sf U}\left[\mathbf{p};\tilde{\mathbf{p}}^{\nu-1}\right]$. (Note that $\tilde{\mathbf{p}}$ is in the infinitesimal neighbourhood of $\mathbf{\hat{p}}$, for sufficiently small $\epsilon$.) We multiply both sides of inequality~(\ref{eqn:multi_ess1}) by $\epsilon$ and add $(1-\epsilon){\sf U}\left[\mathbf{\hat{p}};\tilde{\mathbf{p}}^{\nu-1}\right]$ to both sides to get,
\begin{equation}
	\epsilon {\sf U}\left[\mathbf{\hat{p}};\tilde{\mathbf{p}}^{\nu-1}\right]+(1-\epsilon){\sf U}\left[\mathbf{\hat{p}};\tilde{\mathbf{p}}^{\nu-1}\right]>\epsilon {\sf U}\left[\mathbf{p};\tilde{\mathbf{p}}^{\nu-1}\right]+(1-\epsilon){\sf U}\left[\mathbf{\hat{p}};\tilde{\mathbf{p}}^{\nu-1}\right].
\end{equation}
We mention that as done in two‑player symmetric games, one can choose an $\bar{\epsilon}$ independent of $\mathbf{p}$ such that inequality~(\ref{eqn:multi_ess1}) is valid for all $\epsilon<\bar{\epsilon}$, since the union of all faces of the simplex that exclude $\mathbf{\hat{p}}$ is compact (see Ref.~\cite{Broom2022,Bukowski2004}). Now, absorbing the $\epsilon$ and $(1-\epsilon)$ inside the arguments of the payoff, we get
\begin{equation}
	{\sf U}\left[\epsilon\mathbf{\hat{p}}+(1-\epsilon)\mathbf{\hat{p}};\tilde{\mathbf{p}}^{\nu-1}\right]>{\sf U}\left[\epsilon\mathbf{p}+(1-\epsilon)\mathbf{\hat{p}};\tilde{\mathbf{p}}^{\nu-1}\right].
\end{equation}
Thus, we arrive at the final conclusion that a strategy $\mathbf{\hat{p}}$ is an ESS if for every strategy $\tilde{\mathbf{p}}$ that are sufficiently close but not equal to $\mathbf{\hat{p}}$,
	\begin{equation}\label{eqn:ess_neigh1}
		{\sf U}\left[\mathbf{\hat{p}};\tilde{\mathbf{p}}^{\nu-1}\right]>{\sf U}\left[\tilde{\mathbf{p}};\tilde{\mathbf{p}}^{\nu-1}\right].
	\end{equation}
This definition is analogous to Definitions 1c, 1d, 2c, 2d, 3c, and 3d of ESS in the case of two-player games.

However, note that those definitions are in terms of the states and not strategies. When the above formalism of hypermatrix games is transferred to a population with phenotypes identified with pure strategies, say $\mathbf{p}_1,\mathbf{p}_2,\cdots,\mathbf{p}_m$, the fitness of $i$-type may be denoted as ${\sf A}[\mathbf{p}_i;\mathbf{x}^{\nu-1}]$, where the $k$-th component ($k=1,2,\cdots,m$) of the $m$-dimensional vector $\mathbf{x}$ represents the frequency of the $k$-th strategy. Here, we have adopted a notational convention: When $\mathbf{p}_i$ appears inside square brackets, as in ${\sf A}\left[\mathbf{p}_i;\mathbf{x}^{\nu-1}\right]$, it represents the unit vector in the direction of the $i$-{th} pure state on the simplex $\Sigma_m$, rather than the strategy $\mathbf{p}_i$ itself.  {\color{black} Thus, notationally, we have 
\begin{equation}
{\sf A}[\mathbf{p}_i;\mathbf{x}^{\nu-1}]\equiv\sum_{{i_2}{i_3},\cdots,{i_\nu}}x_{i_2}x_{i_3}\cdots x_{i_\nu}{\sf U}\left[\mathbf{p}_{i};\mathbf{p}_{i_2},\cdots,\mathbf{p}_{i_\nu}\right].\end{equation} 
Similarly, average fitness in the population is
\begin{equation}
{\sf A}\left[\mathbf{x};\mathbf{x}^{\nu-1}\right]\equiv \sum_{{i_1}{i_2},\cdots,{i_\nu}}x_{i_1}x_{i_2}\cdots x_{i_\nu}{\sf U}\left[\mathbf{p}_{i_1};\mathbf{p}_{i_2},\cdots,\mathbf{p}_{i_\nu}\right].\label{eq:A[xx]}
\end{equation} 
For the sake of notational convenience later in the paper, we observe the illustrative situation of two-player game: Suppose we adopt the convention ${\sf A}[\mathbf{p}_i;\mathbf{p}_j]\equiv{\sf A}_{ij}$. Then because $({\sf A}\mathbf{x})_i=\sum_j {\sf A}_{ij} x_j= \sum_j \, \mathbf{p}_i \cdot ({\sf U}\mathbf{p}_j)x_j$ which in the above notation is written as $\sum_j {\sf U}[\mathbf{p}_i;\mathbf{p}_j]x_j$, it is obvious that the symbol $\sf{U}$ is synonymous to the symbol ${\sf A}$ whenever all the arguments are only pure strategies. This synonymity can be straightforwardly argued to be extending to the case of hypermatrices as well.}

Finally, in the context of states we can say that a state $\mathbf{\hat{x}}$ is an Evolutionary Stable State (ESS) if 
	\begin{equation}\label{eqn:mESSn}
		{\sf A}\left[\mathbf{\hat{x}};\mathbf{x}^{\nu-1}\right]>{\sf A}\left[\mathbf{x};\mathbf{x}^{\nu-1}\right].
	\end{equation}
for every $\mathbf{x}\in\Sigma_m$ that are sufficiently close but not equal to $\mathbf{\hat{x}}$.

Of course, in the notation above, the replicator dynamics for the symmetric hypermatrix games can be written as,
\begin{equation}\label{eq:symmetric_multi_repl}
	\dot{x}_i=x_i\left({\sf A}\left[\mathbf{p}_i;\mathbf{x}^{\nu-1}\right]-{\sf A}\left[\mathbf{x};\mathbf{x}^{\nu-1}\right]\right).
\end{equation}
To explore asymptotic stability of its fixed points, we propose the following Lyapunov function: $V_s(\mathbf{x})=\sum_i\hat{x}_i\log({\hat{x}_i}/{x_i})$. This function is positive definite by definition of KL-divergence and $V_s(\mathbf{\hat{x}})=0$. We evaluate
\begin{equation}\label{eqn:KL_div_lyap_sym}
	\dot{V}_s=-\sum_i\hat{x}_i\frac{\dot{x}_i}{x_i}=-\sum_i\hat{x}_i\left({\sf A}\left[\mathbf{p}_i;\mathbf{x}^{\nu-1}\right]-{\sf A}\left[\mathbf{x};\mathbf{x}^{\nu-1}\right]\right)=-\left({\sf A}\left[\mathbf{\hat{x}};\mathbf{x}^{\nu-1}\right]-{\sf A}\left[\mathbf{x};\mathbf{x}^{\nu-1}\right]\right).
\end{equation}
This implies that if ${\sf A}\left[\mathbf{\hat{x}};\mathbf{x}^{\nu-1}\right]>{\sf A}\left[\mathbf{x};\mathbf{x}^{\nu-1}\right]$ (i.e., if $\mathbf{\hat{x}}$ is ESS), then the corresponding fixed point is (locally) asymptotically stable as then $\dot{V}_s<0$.
\subsection{Asymmetric hypermatrix games}\label{sec:asy_hyg}
We are now ready to delve into the definitions of ESS in asymmetric hypermatrix games in a population consisting of two subpopulations. When X-subpopulation is in state $\mathbf{x}$ ($m$-component vector) and Y-subpopulation is in state $\mathbf{y}$ ($n$-component vector), we define the average payoff of a player in X-subpopulation as ${\sf A}[\mathbf{x};\mathbf{x}^{\nu_x-1}]+{\sf B}[\mathbf{x};\mathbf{y}^{\gamma_x}]$. Note that a player in X is involved in strategic game with $\nu_x-1$ players of X and $\gamma_x$ players of Y. Similarly, the average payoff of a Y-player is given as ${\sf C}[\mathbf{y};\mathbf{x}^{\nu_y}]+{\sf D}[\mathbf{y};\mathbf{y}^{\gamma_y-1}]$---a player in Y is involved in strategic game with $\nu_y$ players of X and $\gamma_y-1$ players of Y. To remain notationaly consistent with the previous sections, we have used ${\sf A}$ and ${\sf D}$ for intra-specific interactions and ${\sf B}$ and ${\sf C}$ for inter-specific interactions. Letting boldfaced letters ${\mathbf{p}}$ and ${\mathbf{q}}$ denote the strategies in X and Y subpopulations, respectively, the replicator equation can be expressed as
\begin{subequations}\label{eqn:replicator_asym_2}
	\begin{eqnarray}
		\dot{x}_i=x_i\left({\sf A}[\mathbf{p}_i;\mathbf{x}^{\nu_x-1}]+{\sf B}[\mathbf{p}_i;\mathbf{y}^{\gamma_x}]-{\sf A}[\mathbf{x};\mathbf{x}^{\nu_x-1}]-{\sf B}[\mathbf{x};\mathbf{y}^{\gamma_x}]\right),\\
		\dot{y}_j=y_j\left({\sf C}[\mathbf{q}_j;\mathbf{x}^{\nu_y}]+{\sf D}[\mathbf{q}_j;\mathbf{y}^{\gamma_y-1}]-{\sf C}[\mathbf{y};\mathbf{x}^{\nu_y}]-{\sf D}[\mathbf{y};\mathbf{y}^{\gamma_y-1}]\right).
	\end{eqnarray}
\end{subequations}

Now, one could define three non-equivalent definitions of ESS:
\begin{enumerate}
	\item[] 
	\textbf{Definition 1e (and-definition):} A state ($\mathbf{\hat{x}},\mathbf{\hat{y}}$) is an ESS in asymmetric hypermatrix game if for every population state $(\mathbf{x},\mathbf{y})$ that are sufficiently close to  $(\mathbf{\hat{x}},\mathbf{\hat{y}})$ such that $\mathbf{{x}}\neq\mathbf{\hat{x}}$ and $\mathbf{{y}}\neq\mathbf{\hat{y}}$,
	\begin{equation}\label{eqn:mul_and_ESS}
		{\sf A}[\mathbf{\hat{x}};\mathbf{x}^{\nu_x-1}]	+{\sf B}[\mathbf{\hat{x}};\mathbf{y}^{\gamma_x}]>{\sf A}[\mathbf{x};\mathbf{x}^{\nu_x-1}]+{\sf B}\left[\mathbf{x};\mathbf{y}^{\gamma_x}\right]~\text{\emph{and}}~	{\sf C}[\mathbf{\hat{y}};\mathbf{x}^{\nu_y}]+{\sf D}[\mathbf{\hat{y}};\mathbf{y}^{\gamma_y-1}]>{\sf C}\left[\mathbf{y};\mathbf{x}^{\nu_y}\right]+{\sf D}[\mathbf{y};\mathbf{y}^{\gamma_y-1}].
	\end{equation}
	\item[]\textbf{Definition 2e (sum-definition):} A state ($\mathbf{\hat{x}},\mathbf{\hat{y}}$) is an ESS in asymmetric hypermatrix game if for every population state $(\mathbf{x},\mathbf{y})$ that are sufficiently close but not equal to $(\mathbf{\hat{x}},\mathbf{\hat{y}})$,
	\begin{equation}
		{\sf A}[\mathbf{\hat{x}};\mathbf{x}^{\nu_x-1}]	+{\sf B}[\mathbf{\hat{x}};\mathbf{y}^{\gamma_x}]+{\sf C}[\mathbf{\hat{y}};\mathbf{x}^{\nu_y}]+{\sf D}[\mathbf{\hat{y}};\mathbf{y}^{\gamma_y-1}]>{\sf A}[\mathbf{x};\mathbf{x}^{\nu_x-1}]+{\sf B}\left[\mathbf{x};\mathbf{y}^{\gamma_x}\right]+{\sf C}\left[\mathbf{y};\mathbf{x}^{\nu_y}\right]+{\sf D}[\mathbf{y};\mathbf{y}^{\gamma_y-1}].
	\end{equation}
	\item[]\textbf{Definition 3e (or-definition):} A state ($\mathbf{\hat{x}},\mathbf{\hat{y}}$) is an ESS in asymmetric hypermatrix game if for every population state $(\mathbf{x},\mathbf{y})$ that are sufficiently close but not equal to $(\mathbf{\hat{x}},\mathbf{\hat{y}})$,
	\begin{equation}\label{eqn:2ESS_mul}
		{\sf A}[\mathbf{\hat{x}};\mathbf{x}^{\nu_x-1}]	+{\sf B}[\mathbf{\hat{x}};\mathbf{y}^{\gamma_x}]>{\sf A}[\mathbf{x};\mathbf{x}^{\nu_x-1}]+{\sf B}\left[\mathbf{x};\mathbf{y}^{\gamma_x}\right]~\text{\emph{or}}~	{\sf C}[\mathbf{\hat{y}};\mathbf{x}^{\nu_y}]+{\sf D}[\mathbf{\hat{y}};\mathbf{y}^{\gamma_y-1}]>{\sf C}\left[\mathbf{y};\mathbf{x}^{\nu_y}\right]+{\sf D}[\mathbf{y};\mathbf{y}^{\gamma_y-1}].
	\end{equation}
	In this case, the ESS may be called 2ESS in keeping with the earlier discussions on two-player games.
\end{enumerate}
The corresponding definitions of ESS for bihypermatrix games can be trivially read off from above three definitions on putting ${\sf A}={\sf D}=0$ in the above definitions. 

\subsubsection{Pure ESS}\label{sec:bihyper_pure}
ESS under and-definition for bihypermatrix games (${\sf A}={\sf D}={\sf 0}$) can only be pure. To prove this assertion, we recast the first inequality in (\ref{eqn:mul_and_ESS}) as
\begin{eqnarray}
	{\sf B}\left[\mathbf{\hat{x}};\mathbf{y}^{\gamma_x}\right]-{\sf B}\left[\mathbf{x};\mathbf{y}^{\gamma_x}\right]=(\hat{x}_1-x_1)\sum_{i_1,i_2,\cdots,i_{\gamma_x}=1}^ny_{i_1}y_{i_2}\cdots y_{i_{\gamma_x}}\left({\sf B}[\mathbf{p}_1;\mathbf{q}_{i_1}\cdots\mathbf{q}_{i_{\gamma_x}}]-{\sf B}[\mathbf{p}_m;\mathbf{q}_{i_1}\cdots\mathbf{q}_{i_{\gamma_x}}]\right)+\cdots\nonumber\\+(\hat{x}_{m-1}-x_{m-1})\sum_{i_1,i_2,\cdots,i_{\gamma_x}=1}^ny_{i_1}y_{i_2}\cdots y_{i_{\gamma_x}}\left({\sf B}[\mathbf{p}_{m-1};\mathbf{q}_{i_1}\cdots\mathbf{q}_{i_{\gamma_x}}]-{\sf B}[\mathbf{p}_m;\mathbf{q}_{i_1}\cdots\mathbf{q}_{i_{\gamma_x}}]\right)>0,
\end{eqnarray}
and the second inequality in (\ref{eqn:mul_and_ESS}) as
\begin{eqnarray}
	{\sf C}\left[\mathbf{\hat{y}};\mathbf{x}^{\nu_y}\right]-{\sf C}\left[\mathbf{y};\mathbf{x}^{\nu_y}\right]=(\hat{y}_1-y_1)\sum_{i_1,i_2,\cdots,i_{\nu_y}=1}^mx_{i_1}x_{i_2}\cdots x_{i_{\nu_y}}\left({\sf C}[\mathbf{q}_1;\mathbf{p}_{i_1}\cdots\mathbf{p}_{i_{\nu_y}}]-{\sf C}[\mathbf{q}_n;\mathbf{p}_{i_1}\cdots\mathbf{p}_{i_{\nu_y}}]\right)+\cdots\nonumber\\+(\hat{y}_{n-1}-y_{n-1})\sum_{i_1,i_2,\cdots,i_{\nu_y}=1}^mx_{i_1}x_{i_2}\cdots x_{i_{\nu_y}}\left({\sf C}[\mathbf{q}_{n-1};\mathbf{p}_{i_1}\cdots\mathbf{p}_{i_{\nu_y}}]-{\sf C}[\mathbf{q}_n;\mathbf{p}_{i_1}\cdots\mathbf{p}_{i_{\nu_y}}]\right)>0.
\end{eqnarray}
{\color{black} Note that the notational convention is in accordance with the remark made just after Eq.~(\ref{eq:A[xx]})}.
This inequality has to hold for all values of $(\mathbf{x},\mathbf{y})$ in some neighbourhood of the ESS, $(\mathbf{\hat{x}},\mathbf{\hat{y}})$. If the ESS is not pure, then we can choose $x_2=\hat{x}_2$, $\cdots$, $x_{n-1}=\hat{x}_{n-1}$ and $y_2=\hat{y}_2$, $\cdots$, $y_{n-1}=\hat{y}_{n-1}$. Thus, one can always set the unfixed components $x_1$ and $y_1$ in such a manner that sign of the remaining term can be flipped; this, however, is not possible if the ESS is pure as $\hat{x}_1$ and $\hat{y}_1$ can be either zero or one. This proves the assertion. In fact, same assertion is valid for any ESS under sum-definition (see Appendix~\ref{sec:app:2eESS}).

\textcolor{black}{In general asymmetric hypermatrix games, ESSes can be either pure or mixed under all the three definitions.} The next logical question is to ask ESS (of either bihypermatrix game or asymmetric hypermatrix game) under which definition is (locally) asymptotically stable fixed point of the replicator equation (\ref{eqn:replicator_asym_2}). As far as the pure ESSes under and- and sum-definitions are concerned, the same Lyapunov function as used in the case of two-player game (see Sec.~\ref{sec:pure_bimatrix}), $V_1(\mathbf{x},\mathbf{y})=D_{\rm KL}(\mathbf{\hat{x}}||\mathbf{{x}})+D_{\rm KL}(\mathbf{\hat{y}}||\mathbf{{y}})$, exists for the corresponding fixed points of Eq.~(\ref{eqn:replicator_asym_2}). Thus, one can show that and- and sum-definitions imply the asymptotic stability of the (isolated) fixed point corresponding to a pure ESS---all one has to do is to take the time derivative of $V_1$ and use Eq.~(\ref{eqn:replicator_asym_2}) to arrive at
\begin{equation}\label{eqn:sum_and_lyp_multi}
	\frac{dV_1}{dt}=-\left({\sf A}[\mathbf{\hat{x}};\mathbf{x}^{\nu_x-1}]	+{\sf B}[\mathbf{\hat{x}};\mathbf{y}^{\gamma_x}]+{\sf C}[\mathbf{\hat{y}};\mathbf{x}^{\nu_y}]+{\sf D}[\mathbf{\hat{y}};\mathbf{y}^{\gamma_y-1}]-{\sf A}[\mathbf{x};\mathbf{x}^{\nu_x-1}]-{\sf B}\left[\mathbf{x};\mathbf{y}^{\gamma_x}\right]-{\sf C}\left[\mathbf{y};\mathbf{x}^{\nu_y}\right]-{\sf D}[\mathbf{y};\mathbf{y}^{\gamma_y-1}]\right),
\end{equation}
and subsequently demand the $\dot{V}_1<0$ to validate the implication. Moreover, following exactly similar arguments as done in the two-player case (Sec.~\ref{sec:asym_pure_strict}), linear stability analysis establishes that pure 2ESS is asymptotically stable fixed point of the replicator equation (\ref{eqn:replicator_asym_2}) even in the multiplayer case when it is a strict NE. We omit the straightforward steps as they are analogous and repetitive, as one can see from the eigenvalues $\lambda_1, \lambda_2,\cdots,\lambda_{m-1}, \lambda'_1, \lambda'_2,\cdots,\lambda'_{n-1}$---which respectively are {\color{black}the following:
\begin{eqnarray}
&&({\sf A}[\mathbf{p}_1;\mathbf{p}_m^{\nu_x-1}]+{\sf B}[\mathbf{p}_1;\mathbf{q}_n^{\gamma_x}]-{\sf A}[\mathbf{p}_m;\mathbf{p}_m^{\nu_x-1}]-{\sf B}[\mathbf{p}_m;\mathbf{q}_n^{\gamma_x}]),\nonumber\\
&&({\sf A}[\mathbf{p}_2;\mathbf{p}_m^{\nu_x-1}]+{\sf B}[\mathbf{p}_2;\mathbf{q}_n^{\gamma_x}]-{\sf A}[\mathbf{p}_m;\mathbf{p}_m^{\nu_x-1}]-{\sf B}[\mathbf{p}_m;\mathbf{q}_n^{\gamma_x}]),\nonumber\\
&&~~~~~~~~~~~~~~~~~~~~~~~~~~~~~~~~~\vdots\nonumber\\
&&({\sf A}[\mathbf{p}_{m-1};\mathbf{p}_m^{\nu_x-1}]+{\sf B}[\mathbf{p}_{m-1};\mathbf{q}_n^{\gamma_x}]-{\sf A}[\mathbf{p}_m;\mathbf{p}_m^{\nu_x-1}]-{\sf B}[\mathbf{p}_m;\mathbf{q}_n^{\gamma_x}]),\nonumber\\
&&({\sf C}[\mathbf{q}_1;\mathbf{p}_m^{\nu_y}]+{\sf D}[\mathbf{q}_1;\mathbf{q}_n^{{\gamma_y}-1}]-{\sf C}[\mathbf{q}_n;\mathbf{p}_m^{\nu_y}]-{\sf D}[\mathbf{q}_n;\mathbf{q}_n^{\gamma_y-1}]),\nonumber\\
&&({\sf C}[\mathbf{q}_2;\mathbf{p}_m^{\nu_y}]+{\sf D}[\mathbf{q}_2;\mathbf{q}_n^{\gamma_y-1}]-{\sf C}[\mathbf{q}_n;\mathbf{p}_m^{\nu_y}]-{\sf D}[\mathbf{q}_n;\mathbf{q}_n^{\gamma_y-1}])\nonumber\\
&&~~~~~~~~~~~~~~~~~~~~~~~~~~~~~~~~~\vdots\nonumber\\
&&({\sf C}[\mathbf{q}_{n-1};\mathbf{p}_m^{\nu_y}]+{\sf D}[\mathbf{q}_{n-1};\mathbf{q}_n^{\gamma_y-1}]-{\sf C}[\mathbf{q}_n;\mathbf{p}_m^{\nu_y}]-{\sf D}[\mathbf{q}_n;\mathbf{q}_n^{\gamma_y-1}]),\nonumber
\end{eqnarray}}
at the fixed point $(\hat{\mathbf{x}},\hat{\mathbf{y}})=((0,\cdots,0,1),(0,\cdots,0,1))$. Using the same argument we can also show that all three definitions are equivalent in case of strict NE.

It is not very hard to see that, for ESS that is non-strict pure NE, the situation is more nuanced than for ESS that is strict NE. In fact, similar or even more intricate subtleties may arise, as observed in two-player games.
\subsubsection{Mixed ESS}
The above arguments using Lyapunov function, $V_1$, carry over to the mixed ESSes under and- and sum-definitions as well, leading to the conclusion: Mixed ESS (except mixed 2ESS) is always locally asymptotically fixed point of the replicator equation. The case of 2ESS (or-definition) is interesting. 

First, we recall that interior fixed point can only be a center or saddle node in bimatrix games (see Sec.~\ref{sec:mixed_bim}). We now show that---extending the argument presented in Ref.~\cite{hofbauer_book,Schuster_1981} for two player setting---a similar assertion holds true even for multiplayer bihypermatrix games, i.e., an interior fixed point can neither be a source or sink.
It is straightforward to see that, by definition, the internal fixed point ($\mathbf{\hat{x}},\mathbf{\hat{y}}$) of Eq.~(\ref{eqn:replicator_asym_2}) with ${\sf A}={\sf D}=0$ satisfies the following condition:
\begin{subequations}\label{eqn:bihyper_matrix_int}
	\begin{eqnarray}
		&&{\sf B}[\mathbf{p}_1;\mathbf{\hat{y}}^{\gamma_x}]=\cdots={\sf B}[\mathbf{p}_m;\mathbf{\hat{y}}^{\gamma_x}]={\sf B}[\mathbf{\hat{x}};\mathbf{\hat{y}}^{\gamma_x}],\\
		&&{\sf C}[\mathbf{q}_1;\mathbf{\hat{x}}^{\nu_y}]=\cdots={\sf C}[\mathbf{q}_n;\mathbf{\hat{x}}^{\nu_y}]={\sf C}[\mathbf{\hat{y}};\mathbf{\hat{x}}^{\nu_y}].
	\end{eqnarray}
\end{subequations}
On linearizing the replicator equation about the fixed point, yields the Jacobian matrix
\begin{equation}
	{\sf J}=\begin{bmatrix}
		{\sf 0}&{\sf\alpha}(\hat{x},\hat{y})\\
		{\sf\beta}(\hat{x},\hat{y})&{\sf 0}
	\end{bmatrix},
\end{equation}
where diagonal elements are zero matrices of sizes $m-1\times m-1$ and $n-1\times n-1$, respectively, while the other two are finite matrices denoted by ${\sf\alpha}$ and ${\sf\beta}$ of sizes $m-1\times n-1$ and $n-1\times m-1$, respectively. The reason for the appearance of zero matrices is the fact that the corresponding elements of Jacobian matrix at $(\mathbf{\hat{x}},\mathbf{\hat{y}})$ can be calculated to be
\begin{eqnarray}
	\frac{\partial \dot{x}_i}{\partial x_k}&&=\frac{\partial}{\partial x_k}\left\{x_i\left({\sf B}[\mathbf{p}_i;\mathbf{y}^{\gamma_x}]-{\sf B}[\mathbf{x};\mathbf{y}^{\gamma_x}]\right)\right\}=-x_i\frac{\partial}{\partial x_k}{\sf B}[\mathbf{x};\mathbf{y}^{\gamma_x}]\nonumber\\
	&&=-x_i\frac{\partial}{\partial x_k}\left\{\sum_{j=1}^{m-1}x_j{\sf B}[\mathbf{p}_j;\mathbf{y}^{\gamma_x}]+\left(1-\sum_{j=1}^{m-1}x_j\right){\sf B}[\mathbf{p}_m;\mathbf{y}^{\gamma_x}]\right\}=-x_i\left({\sf B}[\mathbf{p}_k;\mathbf{\hat{y}}^{\gamma_x}]-{\sf B}[\mathbf{p}_m;\mathbf{\hat{y}}^{\gamma_x}]\right)=0.
\end{eqnarray}
Similarly, we can prove that ${\partial \dot{y}_j}/{\partial y_k}=0$. Now, the characteristic polynomial, $p(\lambda)$ (say), can be obtained as $\det({\sf J}-\lambda I)$. It is evident that $p(\lambda)=(-1)^{(m-1)+(n-1)}p(-\lambda)$---a result derived by changing the signs of the first $m-1$ columns and the last $n-1$ rows of $\det({\sf J}-\lambda I)$. Hence, if $\lambda$ is an eigenvalue of ${\sf J}$, then $-\lambda$ must also be an eigenvalue, thereby preventing the interior fixed point from being sink or source. 

Next, we come to mixed 2ESS in asymmetric hypermatrix games. One could expect that as done previously in Sec.~\ref{sec:mixed_nash_asym}, $V_r(\mathbf{x},\mathbf{y})=D_{\rm KL}(\mathbf{\hat{x}}||\mathbf{{x}})+rD_{\rm KL}(\mathbf{\hat{y}}||\mathbf{{y}})$ as a candidate for Lyapunov function. For this function to be a valid Lyapunov function one needs an analog of Theorem 1 for multiplayer games, that is, {\color{black}the following needs to be true: $(\mathbf{\hat{x}},\mathbf{\hat{y}})$ is a mixed 2ESS in asymmetric hypermatrix games if and only if there exists an $r>0$ such that $\forall(\mathbf{x},\mathbf{y})\ne(\mathbf{\hat{x}},\mathbf{\hat{y}})$,
	\begin{equation}\label{r_definition_mult}
		{\sf A}[\mathbf{\hat{x}};\mathbf{x}^{\nu_x-1}]	+{\sf B}[\mathbf{\hat{x}};\mathbf{y}^{\gamma_x}]-{\sf A}[\mathbf{x};\mathbf{x}^{\nu_x-1}]-{\sf B}\left[\mathbf{x};\mathbf{y}^{\gamma_x}\right]+r\left({\sf C}[\mathbf{\hat{y}};\mathbf{x}^{\nu_y}]+{\sf D}[\mathbf{\hat{y}};\mathbf{y}^{\gamma_y-1}]-{\sf C}\left[\mathbf{y};\mathbf{x}^{\nu_y}\right]-{\sf D}[\mathbf{y};\mathbf{y}^{\gamma_y-1}]\right)>0.
	\end{equation}}
Of course, it is trivially true that if inequality (\ref{r_definition_mult}) is satisfied, then $(\mathbf{\hat{x}},\mathbf{\hat{y}})$ is a 2ESS, meaning that inequalities (\ref{eqn:2ESS_mul}) are also satisfied. {\color{black}However, the converse may not be true, that is, if $(\mathbf{\hat{x}},\mathbf{\hat{y}})$ is a 2ESS then inequality~(\ref{r_definition_mult}) may not be satisfied}. We  present an example below to elucidate this fact. However, it is clear that if a 2ESS satisfies inequality (\ref{r_definition_mult}), then $V_r$ is a right Lyapunov function to chose to prove that 2ESS implies asymptotic stability. The failure to recognize appropriate Lyapunov function doesn't falsify the statement that 2ESS is asymptotically stable fixed point of the replicator equation. In fact, in our numerical search, we could not find a situation when 2ESS in asymmetric hypermatrix games is not locally asymptotically stable fixed point of the replicator equation. Hence, we propose a conjecture that needs to be proved in due course:
\begin{enumerate}
	\item[]\textbf{Conjecture 1:} A mixed 2ESS in asymmetric hypermatrix games is locally asymptotically stable fixed point of the replicator equation (\ref{eqn:replicator_asym_2}).
\end{enumerate}

\begin{figure}[h!]
	\hspace*{-1cm}\centering
	\includegraphics[scale=0.80]{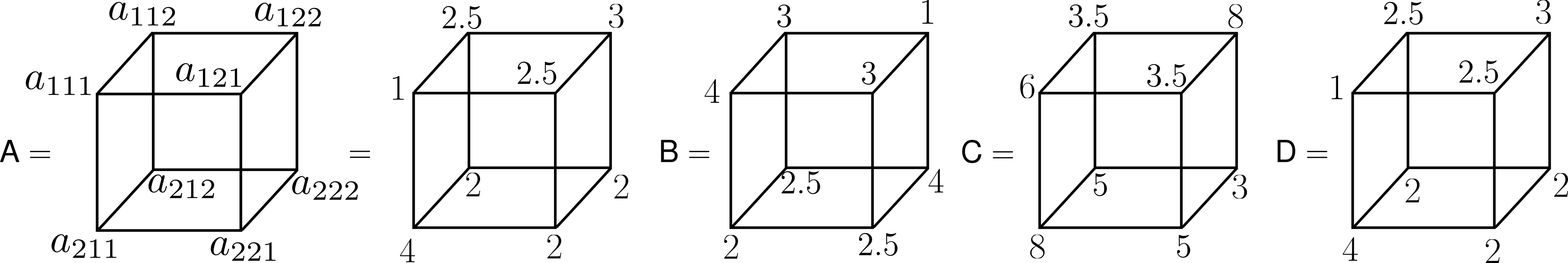}
	\caption{Figure lists the 3-player game payoff hypermatrices: The four hypermatrices represent the payoffs ${\sf A}$, ${\sf B}$, ${\sf C}$ and ${\sf D}$. The number written at the corners of each of the four hypermatrices are payoff elements. In the first hypermatrix, we have illustrated the convention with which we have assigned the indices of the $2\times2\times2$ dimensional hypermatrices.}
	\label{fig:payoff_cube}
\end{figure}

Let's consider an insightful example of a 3-player $2\times2\times2$ hypermatrix game demonstrating that {\color{black} inequality~(\ref{r_definition_mult}) is not necessarily satisfied for all payoff hypermatrices}. We consider payoff hypermatrices as shown in Fig.~\ref{fig:payoff_cube}.
In order to render calculations analytically tractable, we adopt the assumption that intra-specific interactions are infinitesimal, due to a small parameter $\mu$ that is multiplied to ${\sf A}$ and ${\sf D}$. Using the payoff hypermatrices of Fig.~\ref{fig:payoff_cube}, we can rewrite Eq.~(\ref{eqn:replicator_asym_2}) as
\begin{subequations}\label{eqn:replicator_mpa2}
	\begin{eqnarray}
		&&\dot{x}=x(1-x)[2y^2+y(1-y)-3(1-y)^2+\mu\left(-3x^2+x(1-x)+(1-x)^2\right)],\\
		&&\dot{y}=y(1-y)[-2x^2-3x(1-x)+5(1-x)^2+\mu\left(-3y^2+y(1-y)+(1-y)^2\right)].
	\end{eqnarray}
\end{subequations}

Let us define the following short hand notations: $f_1(y)\equiv 2y^2+y(1-y)-3(1-y)^2$,~$f_2(x)\equiv -3x^2+x(1-x)+(1-x)^2$,~$g_1(x)\equiv -2x^2-3x(1-x)+5(1-x)^2$ and $g_2(y)\equiv -3y^2+y(1-y)+(1-y)^2$. For finding interior fixed points, one has to solve $f_1(y)+\mu f_2(x)=0$ and $g_1(x)+\mu g_2(y)=0$ simultaneously. We adopt the ansatz that fixed point $(\hat{x},\hat{y})=(x_0,y_0)+\mu(x_1,y_1)+\mu^2(x_2,y_2)+\cdots$. We find the zeroth order solution as $(x_0,y_0)=(1/2,1/2)$ for which $f_1(y_0)=0$ and $g_1(x_0)=0$. Proceeding further, up to second order in $\mu$ we perturbatively find
\begin{subequations}\label{eqn:internal_fp1}
	\begin{eqnarray}
		\hat{x}=x_0-\mu\frac{ g_2(y_0)}{g_1'(x_0)}+\frac{\mu ^2}{g_1'(x_0)} \left[\frac{f_2(x_0)}{f_1'(y_0)} g_2'(y_0)-\frac{1}{2} \left(\frac{g_2(y_0)}{g_1'(x_0)}\right)^2g_1''(x_0)\right]+\mathcal{O}(\mu^3),\\
		\hat{y}=y_0-\mu\frac{f_2(x_0)}{f_1'(y_0)}+\frac{\mu ^2}{f_1'(y_0)} \left[\frac{g_2(y_0) }{g_1'(x_0)}f_2'(x_0)-\frac{1}{2} \left(\frac{f_2(x_0)}{f_1'(y_0)}\right)^2f_1''(y_0)\right]+\mathcal{O}(\mu^3).
	\end{eqnarray}
\end{subequations}
Here, prime on a function denotes derivative with respect to the argument of that very function. Rather explicitly, we have
\begin{equation}
	(\hat{x},\hat{y})=\left(\frac{1}{2}-\frac{\mu }{28}-\frac{377 \mu ^2}{13720},\frac{1}{2}+\frac{\mu }{20}-\frac{193 \mu ^2}{7000}\right).\label{eq:fpmu}
\end{equation}
Furthermore, on linearizing about this fixed point, the Jacobian is of the form
\begin{equation}
	J=\begin{bmatrix}
		\mu\hat{x}(1-\hat{x})f_2'(\hat{x})&\hat{x}(1-\hat{x})f_1'(\hat{y})\\
		\hat{y}(1-\hat{y})g_1'(\hat{x})&\mu\hat{y}(1-\hat{y})g_2'(\hat{y})
	\end{bmatrix},
\end{equation}
whose eigenvalues can be calculated, up to first order of $\mu$, to be $-\mu\pm ({i}/{4})\sqrt{35+{26\mu}/{35}}$. Thus, the fixed point is a (locally asymptotically) stable spiral.

The neighborhood of $(\mathbf{\hat{x}},\mathbf{\hat{y}})$ [i.e., Eq.~(\ref{eq:fpmu})] is given as $(\mathbf{x},\mathbf{y})=((x,1-x),(y,1-y))=((\hat{{x}}+\delta_x, 1-\hat{x}-\delta_x),(\hat{{y}}+\delta_y, 1-\hat{y}-\delta_y))$, where $\delta_x$ and $\delta_y$ are infinitesimally small but can have positive or negative sign, allowing $(\mathbf{x},\mathbf{y})$ to cover an entire neighbourhood around the point $(\hat{\mathbf{x}},\hat{\mathbf{y}})$. After some calculations, we can find that
\begin{subequations}\label{eqn:3plESS}
	\begin{eqnarray}
		\mu {\sf A}[\mathbf{\hat{x}};\mathbf{x},\mathbf{x}]	+{\sf B}[\mathbf{\hat{x}};\mathbf{y},\mathbf{y}]-\mu {\sf A}[\mathbf{x};\mathbf{x},\mathbf{x}]-{\sf B}\left[\mathbf{x};\mathbf{y},\mathbf{y}\right]&=&-5\delta_x \delta_y+ \frac{\mu\delta_x\delta_y}{5}+4\mu\delta_x^2+2\delta_y^2\delta_x+\text{h.o.t},\\
		{\sf C}[\mathbf{\hat{y}};\mathbf{x},\mathbf{x}]+\mu {\sf D}[\mathbf{\hat{y}};\mathbf{y},\mathbf{y}]-{\sf C}\left[\mathbf{y};\mathbf{x},\mathbf{x}\right]-\mu {\sf D}[\mathbf{y};\mathbf{y},\mathbf{y}]&=&+7\delta_y\delta_x+\frac{3}{7}\mu\delta_x\delta_y+4\mu\delta_y^2-6\delta_x^2\delta_y+\text{h.o.t}.
	\end{eqnarray}
\end{subequations}
Since signs of $\delta_x$ and $\delta_y$ can be positive or negative, at least one of the R.H.S.'s of Eq.~(\ref{eqn:3plESS}) is positive in the entire neighbourhood; thus, the criteria of or-definition is satisfied but it does not meet the criteria of ESS of and- and sum-definitions. Thus, the locally asymptotically stable fixed point given by Eq.~(\ref{eq:fpmu}) is a 2ESS. But does it satisfy inequality~(\ref{r_definition_mult})? As we see below, it does not (when $\mu$ is very small as assumed in the calculations above).

\begin{figure}[h!]
	\centering
	\includegraphics[scale=0.75]{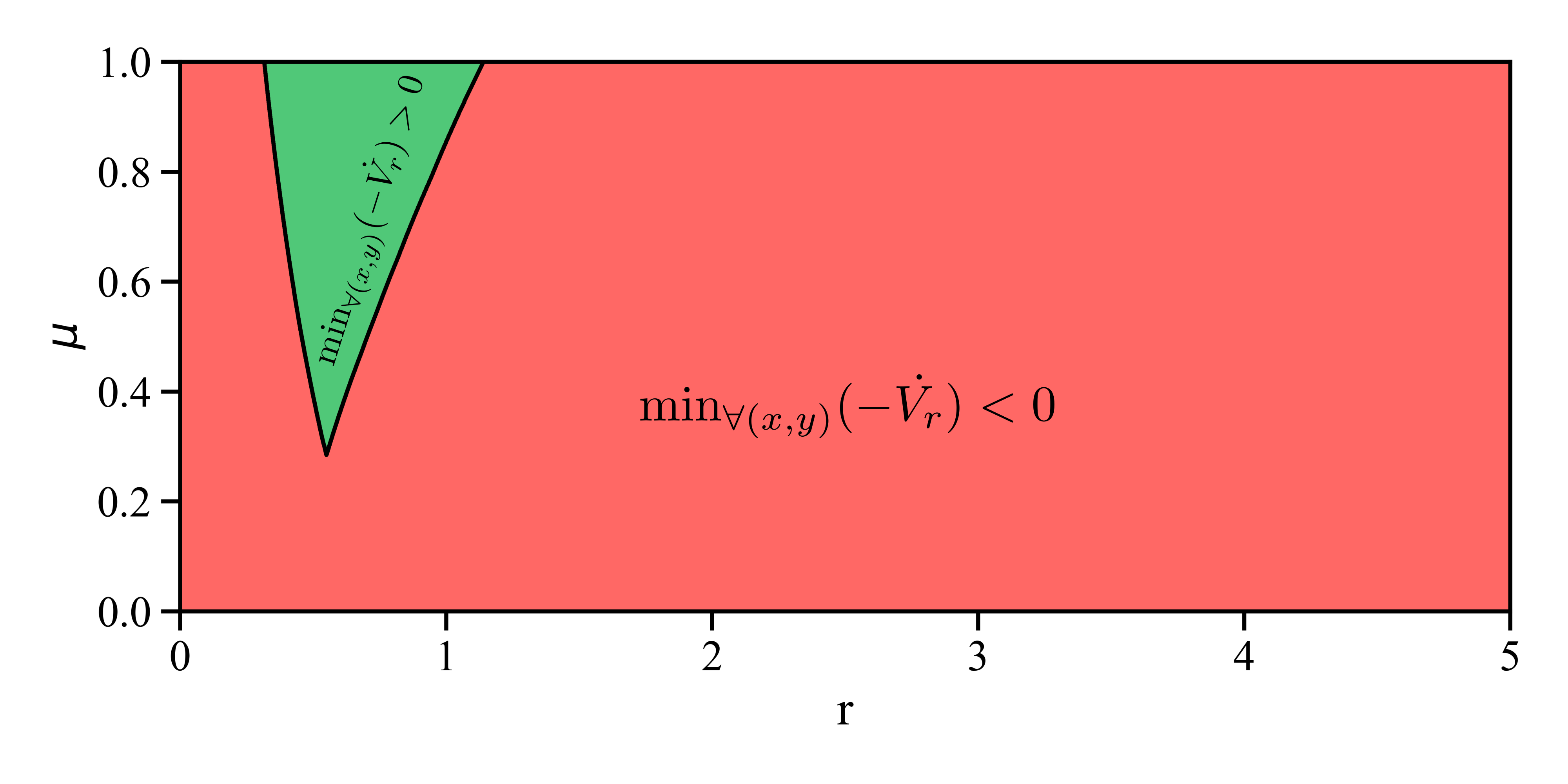}
	\caption{Numerical demonstration that showing {\color{black}inequality~(\ref{r_definition_mult}) is not always satisfied}, yet Conjecture 1 may be valid. We use payoff hypermatrices from Fig.~\ref{fig:payoff_cube} and multiply the parameter $\mu$ with matrices ${\sf A}$ and ${\sf D}$ to control the strength of intra-specific interactions. The figure illustrates two distinct regions: In the first region  (green colour),  {\color{black}inequality~(\ref{r_definition_mult}) is satisfied} for all $(x,y)$ and in second region (red colour), it doesn't. However, all the interior fixed points under consideration are checked to be locally asymptotically stable and 2ESS.}
	\label{fig:rproof}
\end{figure}

Since we forced to do numerical check, we no longer need to assume $\mu$ to be infinitesimal. In other words, we now consider Eq.~(\ref{eqn:replicator_mpa2}) with positive real $\mu$ and find its interior fixed points numerically for various values of $\mu$. We also validate that all the fixed points are 2ESS as well. For each such fixed points, we check if inequality~(\ref{r_definition_mult}) is satisfied. To investigate that, we divide the entire $(x,y)$ space into a uniform grid of size $50\times50$. We then determine the sign of minimum value of L.H.S. of inequality~(\ref{r_definition_mult})---denoted as $\min_{\forall (x,y)}(-\dot{V_r})$---for different values of $\mu$ and $r$ in entire $(x,y)$ grid. Positive $\min_{\forall (x,y)}(-\dot{V_r})$ implies that inequality~(\ref{r_definition_mult}) is satisfied. As seen in Fig.~\ref{fig:rproof}, we find a (green) region (in $r$-$\mu$ space discretized as $5000\times1000$ grid in numerics) bounded by black solid curves, where $\min_{\forall (x,y)}(-\dot{V_r})>0$; elsewhere, we find that $\min_{\forall (x,y)}(-\dot{V_r})<0$ (see the red region in Fig.~\ref{fig:rproof}). At each fixed point, we also do linear stability analysis and find that all the fixed points under question are locally asymptotically stable. In conclusion, while 2ESS happens to be locally asymptotically stable, it need not satisfy {\color{black} inequality~(\ref{r_definition_mult})}. We observe that if $\mu$ is very small, there is no such $r>0$ such that the inequality holds true for all $({x},{y})$; however, for higher values of $\mu$ such an $r$ exists.

{\color{black}
\subsection{Information theoretical aspects}
Finally to complete the picture, now we aim to connect multiplayer hypermatrix games with information theoretical ideas. 

Let us first recall what we have already discussed in this context in previous subsections: In symmetric hypermatrix games
KL-divergence 
$D_{\mathrm{KL}}(\hat{\mathbf{x}} \| \mathbf{x})$ [see Eq.~(\ref{eqn:KL_div_lyap_sym})] 
acts as a Lyapunov function, demonstrating that the ESS is always an asymptotically stable 
fixed point of the replicator dynamics. We then extended this argument to asymmetric hypermatrix games, where we have shown that
$V_1(\mathbf{x},\mathbf{y})= D_{\mathrm{KL}}(\hat{\mathbf{x}} \| \mathbf{x})+ D_{\mathrm{KL}}(\hat{\mathbf{y}} \| \mathbf{y})$ can serve as a Lyapunov function. Consequently, ESSes under the and- and sum-definitions 
are asymptotically stable fixed points of the replicator dynamics [see Eq.~(\ref{eqn:sum_and_lyp_multi})]. 
However, the analogous stability result for the or-definition remains a conjecture (see Conjecture~1). In this section, we present more information-theoretic connections: In particular, we establish connections between ESS, replicator dynamics and Shahshahani gradient (related to Fisher information metric) in hypermatix games.

\subsubsection{$\nu$-player partnership games}
{First let us observe that if the payoff hypermatrix satisfies the following symmetry 
under the permutation of two arguments, namely the first argument $\mathbf{p}_{i}$ and 
any arbitrary argument $\mathbf{p}_{i_l}$, i.e.,
\begin{equation}\label{eq:symmetry}
	{\sf A}\left[\mathbf{p}_{i};\mathbf{p}_{i_2},\cdots,\mathbf{p}_{i_{l-1}},\mathbf{p}_{i_l},\mathbf{p}_{i_{l+1}},\cdots,\mathbf{p}_{i_\nu}\right]={\sf A}\left[\mathbf{p}_{i_l};\mathbf{p}_{i_2}, \cdots,\mathbf{p}_{i_{l-1}},  \mathbf{p}_{i}, \mathbf{p}_{i_{l+1}}, \cdots,\mathbf{p}_{i_\nu}\right]~~~\forall l\in\{2, 3, \cdots, \nu\},
\end{equation}
then one can interchange any two arguments, and the payoff remains unchanged under such a transformation. To be convinced of this claim more clearly, let us fix any two distinct indices $\mathbf{p}_{i_a}$ and $\mathbf{p}_{i_b}$ arbitrarily where $a,b\in\{1,\dots,\nu\}$. It is obvious that the transformation
that swaps positions of $\mathbf{p}_{i_a}$ and $\mathbf{p}_{i_b}$ and leaves all others arguments fixed in the payoff ${\sf A}\left[\mathbf{p}_{i};\mathbf{p}_{i_2},\cdots,\mathbf{p}_{i_\nu}\right]$ can be realized as the composition of three transformations involving the first argument. Symbolically, if $T_{\mathbf{p}_{i_a}\,\mathbf{p}_{i_b}}$ is defined as
\begin{equation}
T_{\mathbf{p}_{i_a}\,\mathbf{p}_{i_b}}{\sf A}\left[\mathbf{p}_{i};\mathbf{p}_{i_2},\cdots,\mathbf{p}_{i_a},\cdots,\mathbf{p}_{i_b},\cdots,\mathbf{p}_{i_\nu}\right]={\sf A}\left[\mathbf{p}_{i};\mathbf{p}_{i_2},\cdots,\mathbf{p}_{i_b},\cdots,\mathbf{p}_{i_a},\cdots,\mathbf{p}_{i_\nu}\right]
\end{equation}
then,
\begin{equation}
	T_{\mathbf{p}_{i_a}\,\mathbf{p}_{i_b}} \;=\; T_{\mathbf{p}_{i}\,\,\mathbf{p}_{i_a}}T_{\mathbf{p}_{i}\,\,\mathbf{p}_{i_b}}T_{\mathbf{p}_{i}\,\, \mathbf{p}_{i_a}}.\label{eq:sp}
\end{equation}
This validates our claim. {Note that this symmetry under permutation reduces to partnership game in two-player settings, which is required to establish replicator equation as a Shahshahani gradient in two-player games~\cite{hofbauer_book}. Motivated by this, we hereby generalize the idea of partnership games to the multiplayer case and call such a game a \emph{$\nu$-player partnership game} if its payoff hypermatrix satisfies the aforementioned permutation symmetry. To see why this terminology is appropriate, it is helpful to examine the special case of a three-player {two-strategy game}: The symmetry would imply ${\sf A}[\mathbf{p}_1;\mathbf{p}_2,\mathbf{p}_2]={\sf A}[\mathbf{p}_2;\mathbf{p}_1,\mathbf{p}_2]={\sf A}[\mathbf{p}_2;\mathbf{p}_2,\mathbf{p}_1]$, which means that whether a focal player chooses $\mathbf{p}_1$ or $\mathbf{p}_2$, their payoff depends only on how many players (in total) choose $\mathbf{p}_1$ and how many choose $\mathbf{p}_2$. Thus, in a $\nu$-player partnership game, all players effectively share the same payoff determined solely by the composition of strategies in the group—precisely the multiplayer analogue of a partnership.}}

When using the KL-divergence as a Lyapunov function [see Eq.~(\ref{eqn:KL_div_lyap_sym})], we previously proved that an ESS is a locally asymptotically stable fixed point---a result valid for all symmetric hypermatrix games. In contrast, now we intend to prove the converse for a special class of games:  In $\nu$-player partnership game, locally asymptotically stable fixed point corresponds to an ESS. 

For this purpose, let us define an inner product (also see Ref.~\cite{hofbauer_book}): 
\begin{equation}\label{eqn_gradient}
	\langle\bm{\xi}',\bm{\xi}\rangle=\sum_{i=1}^{m}\frac{\xi_i'\xi_i}{x_i},
\end{equation}
where $\bm{\xi}'$~and~$\bm{\xi}$ are tangent vectors tangent to $\Sigma_m$, therefore $\sum_i \xi_i=\sum_i \xi'_i=0$. Now, let's find out a potential $W(\mathbf{x})$ whose Shahshahani gradient is the replicator equation~(\ref{eq:symmetric_multi_repl}), i.e., we want a $W(\mathbf{x})$ which satisfies the following:
\begin{equation}\label{eqn:multi_gradient}
	\langle\dot{\mathbf{x}},\bm{\xi}\rangle=\sum_{i=1}^m\frac{\partial W}{\partial x_i}\xi_i.
\end{equation}
The L.H.S of Eq.~(\ref{eqn:multi_gradient}) can be found to be
\begin{equation}\label{multi_gradient}
	\langle\dot{\mathbf{x}},\bm{\xi}\rangle=\sum_{i=1}^{m}{\sf A}\left[\mathbf{p}_i;\mathbf{x}^{\nu-1}\right]\xi_i ,
\end{equation}
using Eq.~(\ref{eqn_gradient}) and the fact that $\sum_i \xi_i=0$. Owing to the above symmetry property (Eq.~\ref{eq:sp}), quick inspection reveals that the function $W (\mathbf{x})=\frac{1}{\nu}{\sf A}\left[\mathbf{x};\mathbf{x}^{\nu-1}\right]$ is the potential whose Shahshahani gradient corresponds to R.H.S of Eq.~(\ref{multi_gradient}).

We now construct a new potential \( V(\mathbf{x}) \)---which  serves as a local Lyapunov function---from \( W(\mathbf{x}) \) and go on to show that its local minimum corresponds to a fixed point of the replicator dynamic which evolves locally towards this minimum that is an ESS.  We define
\begin{eqnarray}
	V (\mathbf{x}) &&\equiv W (\hat{\mathbf{x}})-W (\mathbf{x})\equiv\frac{1}{\nu}{\sf A}\left[\mathbf{\hat{x}};\mathbf{\hat{x}}^{\nu-1}\right]-\frac{1}{\nu}{\sf A}\left[\mathbf{x};\mathbf{x}^{\nu-1}\right]\nonumber\\
	&&=\frac{1}{\nu}\sum_{{i_1},{i_2},\cdots,{i_\nu}}\hat{x}_{i_1}\cdots \hat{x}_{i_\nu}{\sf A}\left[{\mathbf{p}}_{i_1};\mathbf{p}_{i_2},\cdots,\mathbf{p}_{i_\nu}\right]-\frac{1}{\nu}\sum_{{i_1},{i_2},\cdots,{i_\nu}}x_{i_1}\cdots x_{i_\nu}{\sf A}\left[{\mathbf{p}}_{i_1};\mathbf{p}_{i_2},\cdots,\mathbf{p}_{i_\nu}\right].
\end{eqnarray}
Here $\hat{\mathbf{x}}$ corresponds to the point where the potential function $W(\mathbf{{x}})$ has the local maximum value. Therefore $V(\mathbf{{x}})$ has local minimum (which is zero) at $\mathbf{x}=\mathbf{\hat{x}}$, and $V(\mathbf{{x}})>0$ at every other $\mathbf{{x}}\ne \mathbf{\hat{x}}$. The local extremum $\mathbf{\hat{x}}$ of $V(\mathbf{x})$ appear where
\begin{equation}
	\frac{\partial V}{\partial x_i}=0 \implies{\sf A}\left[\mathbf{p}_i;\mathbf{x}^{\nu-1}\right]={\sf A}\left[\mathbf{x};\mathbf{x}^{\nu-1}\right].
\end{equation}
This implies the extremum $\mathbf{\hat{x}}$ appear exactly at the fixed point of the replicator dynamics. To find that whether this extremum is the local minimum of $V(\mathbf{{x}})$ we need to calculate the following $(m-1)\times(m-1)$-Hessian matrix of $V(\mathbf{{x}})$:
\begin{equation}
	{\sf{H}}=\left[\frac{\partial V}{\partial x_a \partial x_b}\right]=-\left[ (\nu-1)\Big\{
	{\sf A}\big[\mathbf{p}_a;\mathbf{p}_b,\mathbf{x}^{\nu-2}\big]
	-{\sf A}\big[\mathbf{p}_a;\mathbf{p}_m,\mathbf{x}^{\nu-2}\big]
	-{\sf A}\big[\mathbf{p}_m;\mathbf{p}_b,\mathbf{x}^{\nu-2}\big]
	+{\sf A}\big[\mathbf{p}_m;\mathbf{p}_m,\mathbf{x}^{\nu-2}\big]
	\Big\}\right]
\end{equation}
where we have used the notation in accordance with the remark made just after Eq.~(\ref{eq:A[xx]}), i.e.,
\begin{eqnarray}
	{\sf A}\big[\mathbf{p}_i;\mathbf{p}_j,\mathbf{x}^{\nu-2}\big]
= \sum_{i_1,\ldots,i_{\nu-2}}
x_{i_1} \cdots x_{i_{\nu-2}}\,
{\sf A}[\mathbf{p}_i;\mathbf{p}_j,\mathbf{p}_{i_1},\ldots,\mathbf{p}_{i_{\nu-2}}].
\end{eqnarray}
One can see that the Hessian matrix is symmetric, as expected. Therefore, if this matrix is positive definite, the fixed point \( \hat{\mathbf{x}} \) corresponds to a local minimum of the potential function \( V(\mathbf{x}) \). Let us now compute the time derivative of the potential \( V(\mathbf{x}) \), which determines the direction of the flow. It is given by the following:
\begin{equation}
	\dot{V}=-\sum_{i=1}^m\dot{x}_i {\sf A}\left[\mathbf{p}_i;\mathbf{x}^{\nu-1}\right]=-\sum_{i=1}^mx_i\left({\sf A}\left[\mathbf{p}_i;\mathbf{x}^{\nu-1}\right]-{\sf A}\left[\mathbf{x};\mathbf{x}^{\nu-1}\right]\right)^2.
\end{equation}
Where we have used $\sum_{i=1}^m\dot{x}_i {\sf A}\left[\mathbf{x};\mathbf{x}^{\nu-1}\right]=0$, as $\dot{\mathbf{x}}$ is a tangent vector. We can see that $\dot{V}<0$ except at $\mathbf{\hat{x}}$ where $\dot{V}=0$. This confirms that $V(\mathbf{{x}})$ is a Lyapunov function which implies that replicator dynamics asymptotically reaches the fixed point $\mathbf{\hat{x}}$ for $\nu$-player partnership games.

Furthermore, since $\mathbf{\hat{x}}$ is the local minimum of $V(\mathbf{{x}})$ which implies $V(\mathbf{{x}})>V(\mathbf{\hat{x}})$, i.e.,
\begin{equation}\label{eq:multi_ESS_info}
	{\sf A}\left[\mathbf{\hat{x}};\mathbf{\hat{x}}^{\nu-1}\right]>{\sf A}\left[\mathbf{x};\mathbf{x}^{\nu-1}\right],
\end{equation}
for every neighbouring states $\mathbf{{x}}\ne \mathbf{\hat{x}}$ of $\mathbf{\hat{x}}$. We can rewrite $\mathbf{\hat{x}}$ as ${\mathbf{x}}+(\hat{\mathbf{x}}-{\mathbf{{x}}})$ and since $(\hat{\mathbf{x}}-{\mathbf{{x}}})$ is very small, we can do Taylor expansion of L.H.S of Eq.~(\ref{eq:multi_ESS_info}) around $\mathbf{x}$ up to first order as follows: 
\begin{equation}\label{eq:taylor}
	{\sf A}\left[{\mathbf{x}}+(\hat{\mathbf{x}}-{\mathbf{{x}}});({\mathbf{x}}+(\hat{\mathbf{x}}-{\mathbf{{x}}}))^{\nu-1}\right]={\sf A}\left[{\mathbf{x}};{\mathbf{x}}^{\nu-1}\right]+\sum_{i=1}^{m-1}\frac{\partial {\sf A}}{\partial x_i}(\hat{x}_i-{x}_i)={\sf A}\left[{\mathbf{x}};{\mathbf{x}}^{\nu-1}\right]+\nu{\sf A}\left[\hat{\mathbf{x}};{\mathbf{x}}^{\nu-1}\right]-\nu{\sf A}\left[{\mathbf{x}};{\mathbf{x}}^{\nu-1}\right].
\end{equation}
Now substituting~(\ref{eq:taylor}) in the inequality~(\ref{eq:multi_ESS_info}), we ultimately get,
\begin{equation}
	{\sf A}\left[\hat{\mathbf{x}};{\mathbf{x}}^{\nu-1}\right]>{\sf A}\left[\mathbf{x};\mathbf{x}^{\nu-1}\right],
\end{equation}
which is exactly the condition of ESS in symmetric hypermatrix game (see Eq.~(\ref{eqn:mESSn})). So, the minimum of the Lyapunov function is essentially ESS where system reaches under the gradient flow in a Fisher information space or Shahshahani space.  
\subsubsection{Asymmetric hypermatrix games}\label{sec:multi_asy_imposible}
From this information-geometric perspective, situation is less rosy for asymmetric hypermatix games: Generically, it is not possible to find a smooth potential function whose Shahshahani gradient is replicator dynamics for asymmetric hypermatrix games. 

Recall the replicator dynamics [Eq.~(\ref{eqn:replicator_asym_2})] and assume that there exists a potential $W(\mathbf{{x}},\mathbf{y})$ for which the dynamics Eq.~(\ref{eqn:replicator_asym_2}) is a Shahshahani gradient, i.e., it satisfies the following relation [see Eq.~(\ref{eqn:gradient})]:
\begin{equation}\label{eq:bimatrix_innerproduct}
	\langle(\dot{\mathbf{x}},\dot{\mathbf{y}}),(\bm{\xi},\bm{\eta})\rangle=\sum_{i=1}^m\frac{\partial W}{\partial x_i}\xi_i+\sum_{i=1}^n\frac{\partial W}{\partial y_i}\eta_i,
\end{equation}
Upon doing the inner product in Shahshahani space [defined by the metric given in Eq.~(\ref{eqn:shahshahani_modified}), hence inner product is given using Eq.~(\ref{eqn:c_gradient})], where the vector $(\dot{\mathbf{x}},\dot{\mathbf{y}})$ is given by replicator dynamics Eq.~(\ref{eqn:replicator_asym_2}), we get the L.H.S of Eq.~(\ref{eq:bimatrix_innerproduct}) as 
\begin{equation}
	\langle(\dot{\mathbf{x}},\dot{\mathbf{y}}),(\bm{\xi},\bm{\eta})\rangle=\sum_{i=1}^m \left({\sf A}[\mathbf{p}_i;\mathbf{x}^{\nu_x-1}]+{\sf B}[\mathbf{p}_i;\mathbf{y}^{\gamma_x}]\right) \xi_i+\frac{1}{c}\sum_{j=1}^n \left({\sf C}[\mathbf{q}_j;\mathbf{x}^{\nu_y}]+{\sf D}[\mathbf{q}_j;\mathbf{y}^{\gamma_y-1}]\right)\eta_i.
\end{equation}
Therefore, if the replicator dynamics (\ref{eqn:replicator_asym_2}) is Shahshahani gradient under the potential $W(\mathbf{{x}},\mathbf{{y}})$, then
\begin{subequations}\label{eqn:grad_info_multi_step1}
\begin{eqnarray}
	&&\sum_i\frac{\partial W}{\partial x_i}\xi_i=\sum_i({\sf A}[\mathbf{p}_i;\mathbf{x}^{\nu_x-1}]+{\sf B}[\mathbf{p}_i;\mathbf{y}^{\gamma_x}])\xi_i,\\&&
	\sum_j\frac{\partial W}{\partial y_j}\eta_j=\sum_j\frac{1}{c}\left({\sf C}[\mathbf{q}_j;\mathbf{x}^{\nu_y}]+{\sf D}[\mathbf{q}_j;\mathbf{y}^{\gamma_y-1}]\right)\eta_j.
\end{eqnarray}
\end{subequations}
Now, if $W(\mathbf{x},\mathbf{y})$ is a smooth differentiable function, it must satisfy the following commutation relation of partial derivatives,
\begin{equation}
	\frac{\partial^2 W}{\partial y_j \partial x_i}=\frac{\partial^2 W}{\partial x_i \partial y_j}~~~~~~~\forall~i, j, (\mathbf{{x}}, \mathbf{{y}}).
\end{equation}
Since the relation must hold for all $i$ and $j$, following must hold good as well: 
\begin{equation}
	\sum_j \eta_j \sum_i \xi_i \frac{\partial^2 W}{\partial y_j \partial x_i}=\sum_i \xi_i\sum_j \eta_j \frac{\partial^2 W}{\partial x_i \partial y_j}~~~~\forall~(\mathbf{{x}}, \mathbf{{y}}).
\end{equation}
Next, using Eq.~(\ref{eqn:grad_info_multi_step1}), we get
\begin{equation}
	\sum_j \eta_j \sum_i \xi_i \frac{\partial}{\partial y_j}\left({\sf A}[\mathbf{p}_i;\mathbf{x}^{\nu_x-1}]+{\sf B}[\mathbf{p}_i;\mathbf{y}^{\gamma_x}]\right)=\frac{1}{c}\sum_i \xi_i \sum_j \eta_j \frac{\partial}{\partial x_i}\left({\sf C}[\mathbf{q}_j;\mathbf{x}^{\nu_y}]+{\sf D}[\mathbf{q}_j;\mathbf{y}^{\gamma_y-1}]\right)~~~\forall~ (\mathbf{{x}}, \mathbf{{y}}).
\end{equation}
This simplifies to the following
\begin{equation}\label{eq:integrability}
	\sum_j \eta_j \sum_i \xi_i \frac{\partial}{\partial y_j}{\sf B}[\mathbf{p}_i;\mathbf{y}^{\gamma_x}]=\frac{1}{c}\sum_i \xi_i \sum_j \eta_j \frac{\partial}{\partial x_i}{\sf C}[\mathbf{q}_j;\mathbf{x}^{\nu_y}]~~~\forall~ (\mathbf{{x}}, \mathbf{{y}}).
\end{equation}
As we can see, the L.H.S. of Eq.~(\ref{eq:integrability}) is a function of \( \mathbf{y} \) alone, while the R.H.S is a function of \( \mathbf{x} \) alone, and since \( \mathbf{x} \) and \( \mathbf{y} \) are independent variables, it is not always possible to satisfy this integrability condition for all possible \( (\mathbf{x}, \mathbf{y}) \). Hence, {we cannot always find a potential whose Shahshahani gradient is the replicator dynamics for asymmetric hypermatrix games}. 

However, there may exist certain non-generic cases where this integrability constraint is bypassed; below, we discuss two such possibilities:
\begin{enumerate}
	\item[]\textbf{Case 1:} If \( {\sf B}[\mathbf{p}_i;\mathbf{y}^{\gamma_x}]\) and \({\sf C}[\mathbf{q}_j;\mathbf{x}^{\nu_y}]\) are some constants, then the condition~(\ref{eq:integrability}) is trivially satisfied; and therefore, we cannot rule out the existence of a potential in such cases---indeed, we have been able to find a potential for symmetric hypermatrix in the immediately preceding subsection, where only intra-specific interactions are considered (${\sf B}={\sf C}=0$).
	
	\item[] \textbf{Case 2:} The last possibility is that of the two-player games for which Eq.~(\ref{eq:integrability}) can be cast as
	\begin{eqnarray}
		&&\sum_j \eta_j \sum_i \xi_i \frac{\partial({\sf B}\mathbf{y})_i}{\partial y_j}=\frac{1}{c}\sum_i \xi_i \sum_j \eta_j \frac{\partial({\sf C}\mathbf{x})_j}{\partial x_i},\\
		\implies &&\sum_j \eta_j \sum_i \xi_i(b_{ij}-b_{in})=\frac{1}{c}\sum_i \xi_i \sum_j \eta_j (c_{ji}-c_{jm}),\nonumber\\
		\implies &&\sum_{ij} \eta_j \xi_ib_{ij}=\frac{1}{c}\sum_{ij} \xi_i \eta_j c_{ji}.\label{eqn:two_ply_potential_cond}
	\end{eqnarray}
	Since \( \bm{\xi} \) and \( \bm{\eta} \) are arbitrary tangent vectors, the matrix elements of \( {\sf B} \) and \( {\sf C} \) must satisfy the relation \( {\sf B} = \frac{1}{c} {\sf C}^{T} \) for~(\ref{eqn:two_ply_potential_cond}) to hold. As discussed in Section~\ref{section_information}, this is precisely the condition for rescaled partnership games. This observation explains why we have been able to find a potential for two-player asymmetric games but not for more general multiplayer asymmetric hypermatrix games.
\end{enumerate}

}

\section{Conclusions} \label{sec:DnC}
The primary objective of this study has been to critically revisit various definitions of ESS, specifically in the context of asymmetric (involving both inter and intra-specific interactions) and bimatrix games (with solely inter-specific interactions), and to unravel their suitability. Through our analyses, we have brought forward intricacies inherent in the definitions and have suggested their corresponding resolutions by examining them through the lens of dynamical stability. The insights thus gained help resolve the prevailing ambiguities surrounding the ESS concept in asymmetric interaction scenarios.

As we have described in this paper, the definitions of ESS in a population that involves two subpopulations can be broadly classified into three classes, which we may appositely call (i) and-definitions, (ii) sum-definitions, and (iii) or-definitions. In the case of and-definitions (Definitions 1a to 1e),  the fitness of host individuals in each subpopulation is compared against that of the corresponding mutant. Whereas, in the case of sum-definitions (Definitions 2a to 2e) one compares the added fitnesses of residents in both the subpopulations with the combined fitness of corresponding mutants. Finally, in the case of or-definitions (Definitions 3a to 3e), a state is said to be ESS if at least one of the two subpopulations of residents are fitter against the mutants. The ESS under or-definition {\color{black}(face-value perspective)} has a special name in the literature: 2ESS~\cite{Cressman_Tao_2014}.

If no intra-specific interaction is present, one has so called bimatrix or bihypermatrix games. One finds that ESSes of and- and sum-definitions can only be pure, {\color{black}whereas the case of or-definitions is a little subtle, we find that it can be mixed if one adopts the simultaneous deviation perspective but in case of face-value perspective, it is always pure.} However, when it comes to the desirable property that an ESS should be locally asymptotically stable fixed point of the appropriate replicator equation, it may not possess the property if it is a non-strict NE. {\color{black} This does not imply that mixed ESS in bimatrix games lacks evolutionary significance---one can show that some evolutionary dynamics (e.g., adjusted replicator dynamics) other than the replicator dynamics can render the mixed ESS (under the or-definition) stable. We remark that there exists another definition of ESS, namely the Nash--Pareto pair (see Appendix~\ref{sec:Nash_Pareto}) which allows for mixed ESS and can be Lyapunov stable under replicator dynamics~\cite{hofbauer_book}.} When intra-specific interactions are present in the so-called asymmetric matrix or hypermatrix games, ESSes under all three classes of definitions can be either pure or mixed. While it seems that mixed ESSes are locally asymptotically stable fixed points of the corresponding replicator equation, pure ESS need not be so if it corresponds to non-isolated fixed points.

We bring again to the readers' attention that apart from using information-theoretic idea of {\color{black}KL-divergence} in finding the Lyapunov function in {\color{black}two-player and multiplayer games}, we have used the Fisher information metric in the form of the Shahshahani metric to decipher an interpretation of the weight factor ($r$) present in an equivalent definition of 2ESS: $r$ is the inverse of the rescaling factor in a distinct class of two-player games known as rescaled partnership. {\color{black} Interestingly, we find that a potential function whose Shahshahani gradient is replicator dynamics can be found for $\nu$-player partnership games; however, this is, in general, not possible for asymmetric hypermatrix games.}

The most important take-home message, we would like to convey is that while a strict NE is an ESS under all three classes of definitions, other ESSes need not simultaneously satisfy all the classes. Not only that, the latter ESSes may not always be locally asymptotically stable under replicator equation.  In a given situation, it may be practical to adopt that definition which renders ESS locally asymptotically stable; after all, if an ESS is not reached dynamically starting from a mutant-invaded state, then it doesn't seem to serve its purpose. The  asymptotic approach towards ESS state is also compatible with the principle of minimum relative entropy---a rather general principle of nature.

We conclude by highlighting potential avenues for extending the findings of this paper. While we have provided an illustrative example, the rigorous proof of results concerning mixed 2ESS in asymmetric hypermatrix games remains an open question. Establishing the asymptotic stability of a pure 2ESS in multiplayer games using techniques beyond linear stability analysis is also missing. Moreover, since the study in this paper has been confined to matrix games, the implications of nonlinear payoffs~\cite{Archetti2012,Patra2022} in non-matrix population games remains unexplored. Exploring the realm of asymmetric games in finite~\cite{Nowak2004,Sekiguchi2015} populations to locate an appropriate definition of ESS is also an interesting problem. Finally, extending the concept of ESS to non-convergent outcomes, which may manifest as oscillatory orbits rather than fixed points~\cite{Bhatacharjee_etal_2023,Dubey2024}, opens up another avenue for investigation.

\acknowledgements
Authors thank Ross Cressman for helpful remarks on the initial idea of this work. 
Arunava Patra thanks CSIR (India) for the financial support in the form of Senior Research Fellowship. Sagar Chakraborty acknowledges the support from SERB (DST, govt. of India) through project no. MTR/2021/000119. 
\appendix
\section{Alternative definitions of ESS in bimatrix games}\label{sec:alternate_defs}
In this section, we present alternative ESS definitions that, although not equivalent to those in the main text, are motivated by the same underlying intent.
\subsection{Alternative and-definitions (primed and-definition)}\label{sec:alternate_defs_and}
	We write three equivalent alternative and-definitions, {\color{black}which we termed as primed and-definition}, in the context of bimatrix games as follows:
	\begin{enumerate}
		\item[]\textbf{Definition 1a$^\prime$}: A state $(\mathbf{\hat{x}},\mathbf{\hat{y}})$ is called ESS if for any mutant state $(\mathbf{{x}},\mathbf{{y}}) \in\Sigma_m \times \Sigma_n$ there exists an invasion barrier, $\epsilon_{(\mathbf{x},\mathbf{y})} >0$ such that, $\forall(\epsilon_x,\epsilon_y)$,  with $0 < \epsilon_x <\epsilon_{(\mathbf{x},\mathbf{y})} $ and $0 < \epsilon_y <\epsilon_{(\mathbf{x},\mathbf{y})}$,
		\begin{equation}\label{eq:ess1_fp'}
			\mathbf{\hat{x}}\cdot{\sf B}(\epsilon_y \mathbf{y}+(1-\epsilon_y) \mathbf{\hat{y}}) > \mathbf{x}\cdot {\sf B}(\epsilon_y \mathbf{y}+(1-\epsilon_y) \mathbf{\hat{y}}),~\mathbf{{x}}\neq\mathbf{\hat{x}}~\text{\emph{and}}~~
			\mathbf{\hat{y}}\cdot {\sf C}(\epsilon_x \mathbf{x}+(1-\epsilon_x) \mathbf{\hat{x}}) >\mathbf{y}\cdot {\sf C}(\epsilon_x \mathbf{x}+(1-\epsilon_x) \mathbf{\hat{x}}),~\mathbf{{y}}\neq\mathbf{\hat{y}}.
		\end{equation}
		Here, $\epsilon_x$ and $\epsilon_y$ are the sufficiently small fractions of mutants inside the subpopulations X and Y respectively. 
		\item[] \textbf{Definition 1b$^\prime$}: A state $(\mathbf{\hat{x}},\mathbf{\hat{y}})$ is an ESS if, for any mutant  $(\mathbf{{x}},\mathbf{{y}})$ the following conditions hold,
		\begin{equation}\label{eqn:equlibrium_1b'}
			\text{(i)}~\mathbf{\hat{x}}\cdot{\sf B}\mathbf{\hat{y}}\geq \mathbf{x}\cdot{\sf B}\mathbf{\hat{y}}~\forall\mathbf{{x}}\neq\mathbf{\hat{x}}~\text{and}~~\mathbf{\hat{y}}\cdot{\sf C}\mathbf{\hat{x}}\geq\mathbf{y}\cdot{\sf C}\mathbf{\hat{x}}~\forall~\mathbf{{y}}\neq\mathbf{\hat{y}};
		\end{equation}
		(ii) and if $\mathbf{\hat{x}}\cdot{\sf B}\mathbf{\hat{y}}= \mathbf{x}\cdot{\sf B}\mathbf{\hat{y}}$, then for all $\mathbf{y}$,
		\begin{equation}\label{eqn:and_definiton1b2'}
			\mathbf{\hat{x}}\cdot\sf{B}\mathbf{y}> \mathbf{x}\cdot{\sf B}\mathbf{y};
		\end{equation}
		(iii) and if $\mathbf{\hat{y}}\cdot{\sf C}\mathbf{\hat{x}}=\mathbf{y}\cdot{\sf C}\mathbf{\hat{x}}$, then, for all $\mathbf{x}$,
		\begin{equation}
			\mathbf{\hat{y}}\cdot{\sf C}\mathbf{x}>\mathbf{y}\cdot{\sf C}\mathbf{x}.
		\end{equation}
	As written, conditions (ii) and (iii) are extremely restrictive: They can never be satisfied. This is because, if $\mathbf{\hat{x}}\cdot{\sf B}\mathbf{\hat{y}}= \mathbf{x}\cdot{\sf B}\mathbf{\hat{y}}$, then a mutant of the form $(\mathbf{x},\mathbf{\hat{y}})$ renders inequality (\ref{eqn:and_definiton1b2'}) invalid. Therefore, the above definition effectively reduces to the following stricter and simpler form, as given in~\cite{hofbauer_sigmund_1988}.
 
			\item[] \textbf{Definition 1b$^{\prime\prime}$}: A state $(\mathbf{\hat{x}},\mathbf{\hat{y}})$ is an ESS if, for any mutant  $(\mathbf{{x}},\mathbf{{y}})$ the following conditions hold,
		\begin{equation}\label{eqn:equlibrium_1b''}
			\text{(i)}~\mathbf{\hat{x}}\cdot{\sf B}\mathbf{\hat{y}}>\mathbf{x}\cdot{\sf B}\mathbf{\hat{y}}~\forall\mathbf{{x}}\neq\mathbf{\hat{x}}~\text{and}~~\mathbf{\hat{y}}\cdot{\sf C}\mathbf{\hat{x}}>\mathbf{y}\cdot{\sf C}\mathbf{\hat{x}}~\forall~\mathbf{{y}}\neq\mathbf{\hat{y}}.
		\end{equation}
		Finally, we write the neighborhood version of this definition: 
		\item[] \textbf{Definition 1c$^\prime$}: A state $(\mathbf{\hat{x}},\mathbf{\hat{y}})$ is an ESS if for every population state $(\mathbf{x},\mathbf{y})$ that are sufficiently close to  $(\mathbf{\hat{x}},\mathbf{\hat{y}})$ such that
		\begin{equation}\label{eqn:ess1c'_neighbourhood}
			\mathbf{\hat{x}}\cdot{\sf B}\mathbf{y}> 	\mathbf{x}\cdot{\sf B}\mathbf{y},~\mathbf{{x}}\neq\mathbf{\hat{x}}~\text{\emph{and}}~\mathbf{\hat{y}}\cdot{\sf C}\mathbf{x}> \mathbf{y}\cdot{\sf C}\mathbf{x},~\mathbf{{y}}\neq\mathbf{\hat{y}}.
		\end{equation}
	\end{enumerate}
{\color{black} From (\ref{eqn:equlibrium_1b''}) it is obvious that this definition is nothing but the definition of strict NE itself, hence, ESS under these definitions can only be pure.} 
\subsection{Nash-Pareto pair}\label{sec:Nash_Pareto}
Hofbauer and Sigmund proposed an alternative definition of ESS to accommodate mixed ESS in bimatrix games~\cite{hofbauer_book,hofbauer_2003}. They refer to this concept as the Nash--Pareto pair:
\begin{enumerate}
\item[]\textbf{Definition 3b$^\prime$}: A state $(\hat{\mathbf{x}},\hat{\mathbf{y}})$ is a Nash--Pareto pair if, for any mutant $(\mathbf{{x}},\mathbf{{y}})\neq(\mathbf{\hat{x}},\mathbf{\hat{y}})$, the following conditions hold:
\begin{equation}\label{eq:3b'i}
	\text{(i)}~~	\mathbf{\hat{x}}\cdot{\sf B}\mathbf{\hat{y}}\geq	\mathbf{x}\cdot{\sf B}\mathbf{\hat{y}}~~\text{and}~~\mathbf{\hat{y}}\cdot{\sf C}\mathbf{\hat{x}}\geq\mathbf{y}\cdot{\sf C}\mathbf{\hat{x}};
\end{equation}
(ii) and  $\forall(\mathbf{x},\mathbf{y})\in\Sigma_m\times\Sigma_n$ for which both inequalities in (i) hold as equalities, following condition holds:
\begin{equation}\label{eq:3b'ii}
	\text{if}~	\mathbf{\hat{x}}\cdot\sf{B}\mathbf{y}<\mathbf{x}\cdot{\sf B}\mathbf{y} ~\text{then}~\mathbf{\hat{y}}\cdot{\sf C}\mathbf{x}>\mathbf{y}\cdot{\sf C}\mathbf{x}\text{ and if}~\mathbf{\hat{y}}\cdot{\sf C}\mathbf{x}<\mathbf{y}\cdot{\sf C}\mathbf{x}~\text{then}~~\mathbf{\hat{x}}\cdot\sf{B}\mathbf{y}> \mathbf{x}\cdot{\sf B}\mathbf{y}.\\
\end{equation}
\end{enumerate}
We would like to bring it to the fore that the Nash--Pareto pair is different from 2ESS (equivalently, Definition 3b). Observe that, we can rewrite~(\ref{eq:3b'ii}) as
`either $\mathbf{\hat{x}}\cdot{\sf B}\mathbf{y} \geq \mathbf{x}\cdot{\sf B}\mathbf{y}$ or  $\mathbf{\hat{y}}\cdot{\sf C}\mathbf{x} > \mathbf{y}\cdot{\sf C}\mathbf{x}$ and 
either $\mathbf{\hat{y}}\cdot{\sf C}\mathbf{x} \geq \mathbf{y}\cdot{\sf C}\mathbf{x}$ or $\mathbf{\hat{x}}\cdot{\sf B}\mathbf{y} > \mathbf{x}\cdot{\sf B}\mathbf{y}$'.
This follows from the material implication rule~\cite{hurley2014concise}: `if $P$ then $Q$' is equivalent to `either not P or Q' (i.e., symbolically, $P \rightarrow  Q\equiv\neg P \lor Q $).  
Thus, even if the resident state satisfies inequalities in~(\ref{eq:3b'i}) as equalities and furthermore,  
$\mathbf{\hat{x}}\cdot{\sf B}\mathbf{y} = \mathbf{x}\cdot{\sf B}\mathbf{y}$ and $\mathbf{\hat{y}}\cdot{\sf C}\mathbf{x} = \mathbf{y}\cdot{\sf C}\mathbf{x}$ hold true, then $(\mathbf{\hat{x}},\mathbf{\hat{y}})$ still qualifies as a Nash-Pareto pair; however, clearly, such an $(\mathbf{\hat{x}},\mathbf{\hat{y}})$ is not an ESS under or-definition.

{\color{black} Before we conclude this appendix, we remark that a mixed Nash--Pareto pair can be Lyapunov stable in certain classes of games, as discussed in Ref.~\cite{hofbauer_book}.
}

\section{Non-applicability of Lyapunov function~(\ref{eqn:lyapunov_asym_DKL}) in bimatrix games}\label{appendix_cressman_proof}
The Lyapunov function given in Eq.~(\ref{eqn:lyapunov_asym_DKL}) appears quite general, and one might think that it applies to bimatrix games as well. However, as we already know that the internal fixed point in bimatrix games cannot be asymptotically stable under replicator dynamics. Therefore, the Lyapunov function in Eq.~(\ref{eqn:lyapunov_asym_DKL}) is not applicable in the case of bimatrix games. In this section, we discuss why this is the case. To do so, we use the proof of Theorem 1 provided in Appendix A of Ref.~\cite{Cressman1996}. While the elaborate proof is best seen in the original reference, let us briefly point out at which point of the argument in the proof ceases to be valid for bimatrix games.

We begin by referring to a statement present in the proof: ``\emph{When $\theta=0$, the point is$-((p^*-p_1)\cdot(Ap_1+Bq_1),(q^*-q_1)\cdot(Cp_1+Dq_1))$ which is in quadrant IV. Similarly, when 	$\theta=\pi/2$ the point is in quadrant II.}'' Here $(p^*,q^*)$ is the resident state and $(p_1,q_1)$ is the mutant state, which in our notation denoted by $(\mathbf{\hat{x}},\mathbf{\hat{y}})$ and $(\mathbf{x},\mathbf{y})$, respectively. $A,\,B,\,C,\,D$ are payoff matrices ${\sf A,\,B,\,C,\,D}$. In this proof, one tries to show that Definition 3d implies inequality~(\ref{r_definition}); for this purpose, one introduces $\theta$ as a parameter to explore in $\mathbb{R}^2$ space spanned by $(p^*-p_1)\cdot(Ap_1+Bq_1)$ and $(q^*-q_1)\cdot(Cp_1+Dq_1))$ as the two coordinates. The idea is to basically find a point in this $\mathbb{R}^2$ space at which Definition 3d is not satisfied and hence, to  contradict an initial hypothesis that there is no such $r$ for which inequality~(\ref{r_definition}) is satisfied.
	
A subsequent sentence in the scheme of the proof is: ``\emph{Thus, for some $\theta_0$ between $0$ and $\pi/2$, the point has zero first component and, by Definition 4.3, negative second. Then for $\theta_0+\pi/2$, the point has zero first component and positive second. This gives our contradiction.}'' Here Definition 4.3 is Definition 3d of this paper. For bimatrix games, however, $A=0$ and $D=0$, and since this proof is valid for interior fixed point (which is also a mixed NE), we can add $(p^*-p_1)\cdot Bq^*$ (which is zero) to the coordinate, $(p^*-p_1)\cdot(Ap_1+Bq_1)$, and $(q^*-q_1)\cdot Cp^*$ (which also is zero) to the second coordinate, $(q^*-q_1)\cdot(Cp_1+Dq_1))$. Subsequently, on redefining $(p^*-p_1)$ as $\bm{\xi}$ and $(q^*-q_1)$ as $\bm{\eta}$, the corresponding point in $\mathbb{R}^2$ takes the following form: $(\bm{\xi}\cdot B\bm{\eta},\bm{\eta}\cdot C\bm{\xi})$. To complete the proof, one needs to argue that the point has zero first component and negative second component---this is exactly what is not valid because if the first component is zero then the second component is automatically zero as well. Let us see explicitly why this is so.

We know that any $2 \times 2$ bimatrix game can be expressed in the form of either a rescaled zero-sum game or a rescaled partnership game~\cite{hofbauer_book}. For the rescaled zero-sum game, ${\sf C}= -c{\sf B}^T-c{\sf \Omega}^T+{\sf \Omega}'$; whereas for the rescaled partnership game, we have ${\sf C}= c{\sf B}^T-c{\sf \Omega}^T+{\sf \Omega}'$ (see Section~\ref{section_information}). For a rescaled zero-sum game or a rescaled partnership game, easy calculations show that if $\bm{\xi}\cdot B\bm{\eta}=0$ then  $\bm{\eta}\cdot C\bm{\xi}=0$ \cite{hofbauer_book}. This is because $\sum_i\xi_i=\sum_i\eta_i=0$ and hence, we can trivially show: $\bm{\xi}\cdot \Omega\bm{\eta}=\bm{\eta}\cdot\Omega'\bm{\xi}=0$. 

In conclusion, due to the above arguments, the equivalence between inequality~(\ref{r_definition}) and 2ESS does not hold in bimatrix games, which prevents Eq.~(\ref{eqn:lyapunov_asym_DKL}) from being a valid Lyapunov function in such games.	 	
Finally, we remark that the above arguments extends beyond $2 \times 2$ case, simply because it is known that asymptotic stability of an interior fixed point (of corresponding replicator equation) is not possible for bimatrix games in general~\cite{hofbauer_book}.
\section{ESS under Definition 2e for bihypermatrix games can only be pure}
\label{sec:app:2eESS}
We now demonstrate that there cannot be a mixed ESS under Definition 2e in bihypermatrix games (${\sf A=D=0}$). While we focus on the 2-strategy case for simplicity, the arguments can be easily generalized for any number of strategies. We begin by noting that for mixed ESS state, $(\mathbf{\hat{x}}, \mathbf{\hat{y}})$, the following condition holds [see Eq.~(\ref{eqn:bihyper_matrix_int})]:
\begin{eqnarray}\label{eqn:mixed_ESSb}
		{\sf B}[\mathbf{p}_1;\mathbf{\hat{y}}^{\gamma_x}]={\sf B}[\mathbf{p}_2;\mathbf{\hat{y}}^{\gamma_x}]={\sf B}[\mathbf{\hat{x}};\mathbf{\hat{y}}^{\gamma_x}]~\text{and}~
		{\sf C}[\mathbf{q}_1;\mathbf{\hat{x}}^{\nu_y}]={\sf C}[\mathbf{q}_2;\mathbf{\hat{x}}^{\nu_y}]={\sf C}[\mathbf{\hat{y}};\mathbf{\hat{x}}^{\nu_y}].
\end{eqnarray}
Note that for the two-strategy case, we have $(\mathbf{\hat{x}}, \mathbf{\hat{y}})=((\hat{x},1-\hat{x}),(\hat{y},1 -\hat{y}))$. Hence the neighbourhood vector can be written as, $(\mathbf{x},\mathbf{y})\equiv(\mathbf{\hat{x}}+\bm{\delta}_x,\mathbf{\hat{y}}+\bm{\delta}_y)\equiv((\hat{x}+\delta_x,1-\hat{x}-\delta_x),(\hat{y}+\delta_y,1-\hat{y}-\delta_y)) $. We want to cover the entire neighborhood around the internal fixed point $(\mathbf{\hat{x}},\mathbf{\hat{y}})$; therefore, $\delta_x$ and $\delta_y$ can be varied independently within the infinitesimal range, and their signs can be anything---either positive or negative. Also, since the payoff is linear in its arguments, the following property---which can be easily proved using the expansion, ${\sf B}\left[\mathbf{\hat{x}};\mathbf{y}^{\gamma_x}\right]=\sum_{i,j_1,\cdots,j_{\gamma_x}}\hat{x}_iy_{j_1}\cdots y_{j_{\gamma_x}}{\sf B}[\mathbf{p}_i;\mathbf{q}_{j_1}\cdots\mathbf{q}_{j_{\gamma_x}}]$---comes in handy:
\begin{eqnarray}\label{eqn:payoff_hyp_porp}
	{\sf B}\left[\mathbf{\hat{x}};\mathbf{y}^{\gamma_x}\right]\equiv{\sf B}\left[\mathbf{\hat{x}};(\mathbf{\hat{y}}+\bm{\delta}_y)^{\gamma_x}\right]=\sum_{i=0}^{\gamma_x}\binom{\gamma_x}{i}{\sf B}\left[\mathbf{\hat{x}};\bm{\delta}_y^{i},\mathbf{\hat{y}}^{\gamma_x-i}\right].
\end{eqnarray}

To show that Definition 2e cannot be satisfied by a mixed state, we need to argue that ${\sf B}\left[\mathbf{\hat{x}};\mathbf{y}^{\gamma_x}\right]-{\sf B}\left[\mathbf{x};\mathbf{y}^{\gamma_x}\right]+{\sf C}\left[\mathbf{\hat{y}};\mathbf{x}^{\nu_y}\right]-{\sf C}\left[\mathbf{y};\mathbf{x}^{\nu_y}\right]$ cannot be positive everywhere in the entire neighbourhood. To this end, we expand the terms in Definition 2e to arrive at
\begin{eqnarray}
{\sf B}\left[\mathbf{\hat{x}};\mathbf{y}^{\gamma_x}\right]-{\sf B}\left[\mathbf{x};\mathbf{y}^{\gamma_x}\right]+{\sf C}\left[\mathbf{\hat{y}};\mathbf{x}^{\nu_y}\right]-{\sf C}\left[\mathbf{y};\mathbf{x}^{\nu_y}\right]&=&(\hat{x}-x)	\left({\sf B}\left[\mathbf{p}_1;\mathbf{y}^{\gamma_x}\right]-{\sf B}\left[\mathbf{p}_2;\mathbf{y}^{\gamma_x}\right]\right)+
	(\hat{y}-y)\left({\sf C}\left[\mathbf{q}_1;\mathbf{x}^{\nu_y}\right]-{\sf C}\left[\mathbf{q}_2;\mathbf{x}^{\nu_y}\right]\right),\nonumber\\
	&=&-\delta_x	\left({\sf B}\left[\mathbf{p}_1;\mathbf{y}^{\gamma_x}\right]-{\sf B}\left[\mathbf{p}_2;\mathbf{y}^{\gamma_x}\right]\right)
	-\delta_y\left({\sf C}\left[\mathbf{q}_1;\mathbf{x}^{\nu_y}\right]-{\sf C}\left[\mathbf{q}_2;\mathbf{x}^{\nu_y}\right]\right)\label{eqn:sum_ESS},\nonumber\\
	&=&-\delta_x\left(b_1\delta_y+b_2\delta_y^2+\cdots+b_{\gamma_x}\delta_y^{\gamma_x}\right)-\delta_y\left(c_1\delta_x+c_2\delta_x^2+\cdots+c_{\nu_x}\delta_x^{\nu_y}\right).\nonumber\\&&\label{eqn:bihyp_ESS_s}
\end{eqnarray}
To get to the final form Eq.~(\ref{eqn:bihyp_ESS_s}), we have used Eq.~(\ref{eqn:mixed_ESSb}) and Eq.~(\ref{eqn:payoff_hyp_porp}). The coefficients $b_1,b_2,\cdots,b_{\gamma_x}$ depend on elements of ${\sf B}$ and $\hat{y}$; similarly, $c_1,c_2,\cdots, c_{\nu_y}$  depend on elements of ${\sf C}$ and  $\hat{x}$.

Now, to the leading order, the R.H.S. of Eq.~(\ref{eqn:bihyp_ESS_s}) is $-(b_1+c_1)\delta_x\delta_y$ which for any given values of $b_1$ and $c_1$ cannot be positive for all infinitesimal (positive or negative) values of $\delta_x$ and $\delta_y$. For similar reason, in case $-(b_1+c_1)\delta_x\delta_y=0$, the next order term in $\delta_x$ and $\delta_y$ of the R.H.S. of Eq.~(\ref{eqn:bihyp_ESS_s}) leads to same conclusion, and so on. In conclusion, mixed states cannot be ESS under Definition 2e for bihypermatrix games. We remark the line of the proof adopted herein can be extended to the case of mixed ESS under Definition 1e.
\bibliography{Dubey_etal_bibliography.bib}
\end{document}